\documentclass[11pt, oneside]{article}   	
\pdfoutput=1 
\usepackage{geometry}                		
\usepackage[dvipdfmx]{graphicx}				
\usepackage[blocks, affil-it]{authblk}
\usepackage{subcaption}
\usepackage{amssymb}
\usepackage{amsmath}
\usepackage{color}
\usepackage{authblk}
\usepackage{lineno}
\usepackage{todonotes}
\usepackage[amssymb]{SIunits}
\usepackage{microtype}
\usepackage{hyperref}
\usepackage{booktabs}

\hypersetup{
    colorlinks,
    citecolor=black,
    filecolor=black,
    linkcolor=black,
    urlcolor=black
}

\graphicspath{{figures/}}

\newcounter{mycomment}

\newcommand{\insertgraph}[4]{
	\begin{figure}
		\centering		\includegraphics*[width=#1\textwidth]{figures/#2}
		\caption{#3}
		\label{#4}
	\end{figure}
	}
	
\newcommand{\inserttwographs}[6]{
	\begin{figure}
	\centering
	\includegraphics*[width=#1\textwidth]{figures/#2}
	\hfill
	\includegraphics*[width=#3\textwidth]{figures/#4}
	\caption{#5}
	\label{#6}
	\end{figure}
}

\providecommand{\keywords}[1]
{
  \small	
  \textbf{\textit{Keywords---}} #1
}

\title{The SuperFGD Prototype Charged Particle Beam Tests}

\author[a,*]{A.\,Blondel}
\author[b]{M.\,Bogomilov}
\author[c]{S.\,Bordoni}
\author[a]{F.\,Cadoux}
\author[a]{D.\,Douqa}
\author[d]{K.\,Dugas}
\author[e]{T.\,Ekelof}
\author[a]{Y.\,Favre}
\author[f]{S.\,Fedotov}
\author[e]{K.\,Fransson}
\author[g]{R.\,Fujita}
\author[h]{E.\,Gramstad}
\author[i]{A.K.\,Ichikawa}
\author[b]{S.\,Ilieva}
\author[g]{K.\,Iwamoto}
\author[j]{C.~Jes\'{u}s-Valls}
\author[k]{C.K.\,Jung}
\author[d]{S.P.\,Kasetti}
\author[f]{M.\,Khabibullin}
\author[f]{A.\,Khotjantsev}
\author[a]{A.\,Korzenev}
\author[f]{A.\,Kostin}
\author[f,l,m]{Y.\,Kudenko}
\author[d]{T.\,Kutter}
\author[j]{T.\,Lux}
\author[a]{L.\,Maret}
\author[n]{T.\,Matsubara}
\author[f]{A.\,Mefodiev}
\author[o]{A.\,Minamino}
\author[f]{O.\,Mineev}
\author[p]{G.\,Mitev}
\author[c]{M.\,Nessi}
\author[a]{L.\,Nicola}
\author[a]{E.\,Noah}
\author[a]{S.\,Parsa}
\author[b]{G.\,Petkov}
\author[a]{F.\,Sanchez}
\author[c,@]{D.\,Sgalaberna}
\author[q]{W.\,Shorrock}
\author[r]{K.\,Skwarczynski}
\author[f]{S.\,Suvorov}
\author[k]{A.\,Teklu}
\author[b]{R.\,Tsenov}
\author[q]{Y.\,Uchida}
\author[b]{G.\,Vankova-Kirilova}
\author[f]{N.\,Yershov}
\author[g]{M.\,Yokoyama}
\author[r]{J.\,Zalipska}
\author[e]{Y.\,Zou}
\author[r]{W.\,Zurek}

\affil[a]{University of Geneva, Section de Physique, DPNC, Geneva, Switzerland}

\affil[b]{University of Sofia, Sofia, Bulgaria}

\affil[c]{European Organization for Nuclear Research (CERN), Geneva, Switzerland}

\affil[d]{Louisiana State University, Baton Rouge, United States}

\affil[e]{Uppsala University, Uppsala, Sweden}

\affil[f]{Institute for Nuclear Research of RAS, Moscow, Russia}

\affil[g]{University of Tokyo, Tokyo, Japan}

\affil[h]{University of Oslo, Oslo, Norway}

\affil[i]{Kyoto University, Kyoto, Japan}

\affil[j]{Institut de Fisica d'Altes Energies (IFAE), Bellaterra Spain}

\affil[k]{Stony Brook University, New York, United States}

\affil[l]{Moscow Institute of Physics and Technology (MIPT), Moscow, Russia}

\affil[m]{National Research Nuclear University MEPhI, Moscow, Russia}

\affil[n]{High Energy Accelerator Research Organization (KEK), Tsukuba, Japan}

\affil[o]{Yokohama National University, Yokohama, Japan}

\affil[p]{Institute for Nuclear Research and Nuclear Energy of BAS, Sofia, Bulgaria}

\affil[q]{Imperial College London, Department of Physics, London, United Kingdom}

\affil[r]{National Centre for Nuclear Research, Warsaw, Poland}

\affil[*]{Now at IN2P3 Paris-Sorbonne, Paris, France}

\affil[@]{Now at ETH Zurich, Zurich, Switzerland}

\date{}							

\begin{document}
\maketitle
\newpage
\begin{abstract}
A novel scintillator detector, the SuperFGD, has been selected as the main neutrino target for an upgrade of the T2K experiment ND280 near detector. The detector design will allow nearly 4$\pi$ coverage for neutrino interactions at the near detector and will provide lower energy thresholds, significantly reducing systematic errors for the experiment. The SuperFGD is made of optically-isolated scintillator cubes of size $10\times10\times10$ mm$^3$, providing the required spatial and energy resolution to reduce systematic uncertainties for future T2K runs. The SuperFGD for T2K will have close to two million cubes in a $1920 \times 560 \times 1840$ mm$^3$ volume. A prototype made of $24 \times 8 \times 48$ cubes was tested at a charged particle beamline at the CERN PS facility. The SuperFGD Prototype was instrumented with readout electronics similar to the future implementation for T2K. Results on electronics and detector response are reported in this paper, along with a discussion of the 3D reconstruction capabilities of this type of detector. Several physics analyses with the prototype data are also discussed, including a study of stopping protons.

\end{abstract}

\newpage
\section{Introduction}

Long plastic scintillator bars arranged perpendicularly to the neutrino beam direction have been used extensively in two Fine Grained Detector's (FGDs) \cite{Amaudruz:2012esa}, which are part of the near detector suite (ND280) of T2K \cite{Abe:2011ks, Abe:Nature} and in other experiments such as MINOS \cite{Michael:2008bc} and MINERvA \cite{Aliaga:2013uqz}. This geometry is very effective to measure long ranged tracks such as high momentum leptons and hadrons. For lower momentum tracks the limitations in acceptance and position resolution of this traditional plastic scintillator bar geometry quickly become apparent: a particle traveling along the scintillator bar or short tracks in events with multiple outgoing particles often cannot be tracked or its momentum reconstructed accurately.

A novel scintillator detector concept with 3D fine granularity and quasi-3D read-out capabilities based on complementary 2D read-outs was proposed for T2K-II, a program of precision measurements of oscillation parameters aiming to observe CP violation with more than a 3$\sigma$ Confidence Limit (CL) \cite{Abe:2019whr}. The T2K-II proposal consists of a beam power upgrade from 485~kW to 1.3~MW~\cite{Friend:2017oav,Abe:2019fux}, along with an upgrade of the current ND280 near detector complex~\cite{Abe:2019whr}.

The new scintillator detector design addresses acceptance and tracking limitations by adopting a highly segmented layout. The basic unit is a cube covered on all sides by an optical reflector. This unit is replicated as many times as required to fill the detector volume. Scintillation light emitted by the cubes is collected by wavelength-shifting fibers along three orthogonal directions. With such a scheme, the location of each energy deposition event can be determined with a precision related to the cube size by combining information from two or more readout views. The third view helps to resolve hit ambiguities due to multiple tracks.

With its quasi-3D readout scheme, the Super Fine-Grained Detector (SuperFGD) is a key component of the T2K ND280 near detector upgrade \cite{Sgalaberna:2017khy}. Secondary particles originating from neutrino interactions in the SuperFGD and exiting its volume will be tracked by existing Time Projection Chambers (TPCs) in the forward direction, and by two new High-Angle TPCs (HA-TPCs) located above and below the SuperFGD~\cite{attie2019performances}. This setup enhances capabilities for measuring final state leptons and nuclear residuals resulting from neutrino interactions by providing 3D tracking with close to $4\pi$ acceptance and lower energy thresholds. It addresses the increasingly demanding requirements to measure with better precision the outgoing lepton, which is emitted more isotropically at lower energies, and any additional hadrons from the initial neutrino interactions or nuclear breakup, leading to better neutrino energy reconstruction. Improvements in antineutrino energy reconstruction are expected with a new method based on precise measurements of the outgoing neutron on an event-by-event basis~\cite{Munteanu:2019llq}.

The 10~mm cube size chosen for the SuperFGD is a natural granularity scale corresponding to the range of 200~MeV/$c$ protons in plastic, a high probability momentum for protons arising from neutrino interactions at T2K. This cube size also strikes an acceptable balance between position resolution and number of readout channels. At one end of each fiber, a silicon photomultiplier (SiPM) will detect the light signal carried by the fiber. The current design foresees the SuperFGD dimensions to be 192 (width) $\times$ 56 (height) $\times$ 184 (length) cubes along the $x$-$y$-$z$ detector axes respectively, consisting of 1,978,368 cubes and 56,384 channels, with the neutrino beam along the $z$-axis. The expected improvements are reported in the ND280 upgrade technical design report~\cite{Abe:2019whr}. 

A $5\times 5\times 5$ array of scintillator cubes was tested at CERN in autumn 2017~\cite{Mineev:nim}. These first positive results motivated the assembly and testing of a larger prototype, a $24\times 8\times 48$ array ($x$-$y$-$z$ detector axes) composed of 9,216 cubes. This prototype, referred to as the ``SuperFGD Prototype", was tested extensively at the T9 beamline of the Proton Synchrotron (PS) accelerator at CERN. It consists of 1,728 instrumented readout channels. It is the first of its type and size, capable of providing relevant information for the design of the SuperFGD.

The construction and testing of the SuperFGD Prototype was motivated by a need to test the scintillator cube assembly method, components in a B-field representative of operational conditions at T2K ND280, readout electronics similar to the final detector and general performance such as calibration, light yields, hit efficiencies, crosstalk, fiber attenuation, time resolution. We report on the SuperFGD Prototype assembly and instrumentation, analysis methods and results from charged particle beam tests.

\section{The SuperFGD Prototype}

\label{sectionSFGDproto}

The physical dimensions of the SuperFGD Prototype were imposed by the requirement to fit the detector inside the MNP17 magnet, a general purpose dipole magnet made available to users at CERN. This magnet was used to generate a 0.2~T magnetic field across the prototype during the charged particle beam test, in order to mimic the magnetic field in the ND280 detector~\cite{Abe:2011ks}.

\subsection{Scintillator Cubes}
\label{sec:cubes}

The scintillator cubes were produced at UNIPLAST Co.\ (Vladimir, Russia). The scintillator composition is polystyrene doped with 1.5\% of paraterphenyl (PTP) and 0.01\% of 1,4-bis benzene (POPOP).

More than 10,000 cubes were fabricated between the end of 2017 and the beginning of 2018 to assemble mechanical mock-ups and detector prototypes. These $10\times 10\times 10$~mm$^3$ cubes were cut to size from long 10~mm-thick extruded slabs. The variation in the cube sizes using this method was relatively large. The standard deviation, $\sigma_\mathrm{w}$, of the cube width roughly equaled 100~{\micro\meter}. Cubes produced with this method were used to construct the SuperFGD Prototype.

To reduce the variation in cube size, the cube production has been updated since the SuperFGD Prototype construction. The new method is based on injection molding, which improves cube size tolerance threefold ($\sigma_\mathrm{w}<30$~{\micro\meter}). This method will be used for the production of SuperFGD cubes to be installed in ND280.

A reflective layer was etched onto each of the cubes' surfaces with a chemical agent, resulting in the formation of a 50--80~{\micro\meter}-thick white polystyrene micropore deposit~\cite{Kudenko:2001qj}. 

Each cube was placed on a jig designed to hold it in place during the drilling of the three orthogonal through holes of 1.5~mm diameter. A Computer Numerical Control (CNC) milling machine was used to drill the holes. The average distance between the hole center and the cube side was measured to be $3.11\pm0.08$~mm, which is slightly above the specified value of 3.00~mm. 

Random deviations of the hole position less than 0.2~mm should not significantly increase the cube position uncertainty because of the free gap between the 1~mm thick WLS fiber and the 1.5~mm thick hole. It is the cube size stability that is the key factor for the SuperFGD assembly.

\subsection{WLS Fibers and Photosensors}
\sloppy{
For the SuperFGD, we are using the \mbox{Y-11(200)MS} WLS fiber, produced by Kuraray Co.~\cite{Kuraray}. This is the same fiber that is used in ND280's current FGDs. It is a multi-clad, round shape fiber of S-type (increased flexibility) with a diameter of 1.0~mm. The performance and quality of this fiber are very well established by the T2K project and many large experiments~\cite{Mineev:jinst}.
}

A custom plastic optical coupler (ferrule) developed for the Baby MIND detector was used in conjunction with the WLS fibers and the photosensors~\cite{Noah:2017ati}. It consists of two main parts shown in Fig.~\ref{fig:prototype_photosensor_assembly}. One end of each WLS fiber was glued into one part, the fiber ferrule, using EJ-500 optical cement produced by Eljen Technology. The protruding end was cut off from the tip of the ferrule with a sharp blade. The connector end was then polished using a small drilling machine with a fine polishing wheel. The other end of the fiber was left open, unpolished and without a reflective coating. Altogether, 1,728 fibers were produced. The fibers attached to the ferrules can be seen in Fig.~\ref{fig:prototype_photosensor_assembly}.

The second part of the optical coupler is a container for a single SiPM. Both parts of the plastic optical coupler latch together by means of a locking groove inside the container and a matching ring protrusion on the ferrule. A small foam spring inside the container provides reliable optical contact between the photosensor face and the fiber end. The photosensor pins protrude from the container and attach to a mini-PCB into which an ultra-miniature low-profile coaxial connector is soldered. An ultra-small coaxial cable (U.FL-2LP-088) of 1.5~m length sends the photosensor signal to the front-end electronics.

\inserttwographs{0.48}{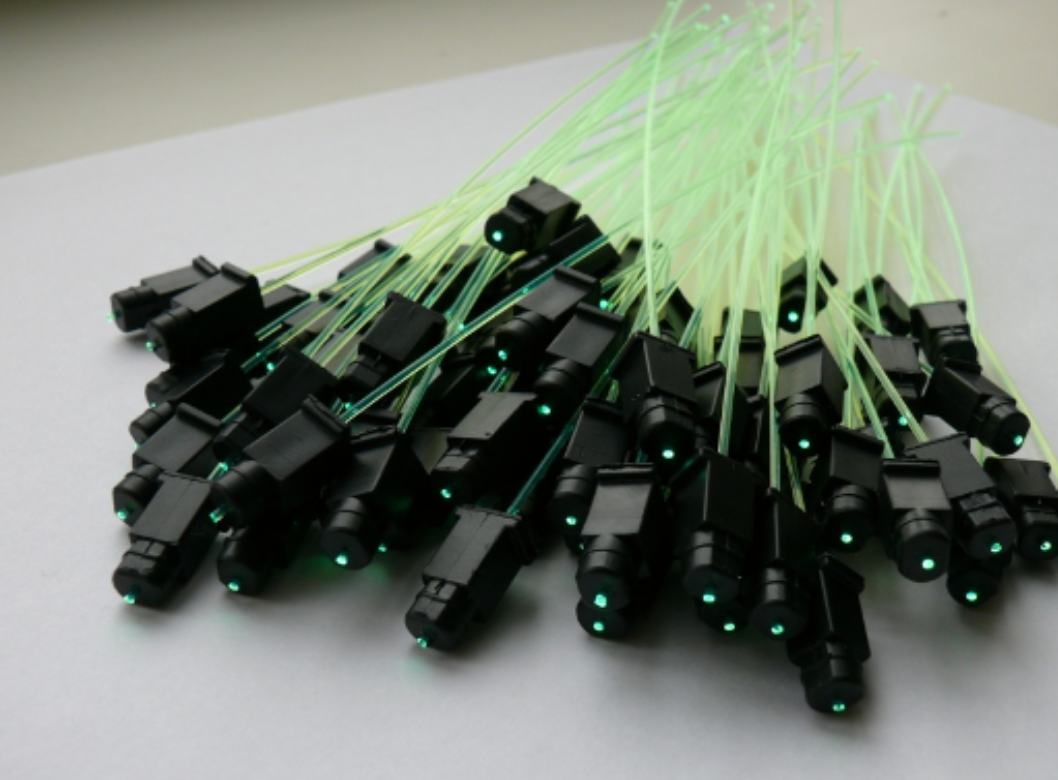}{0.48}{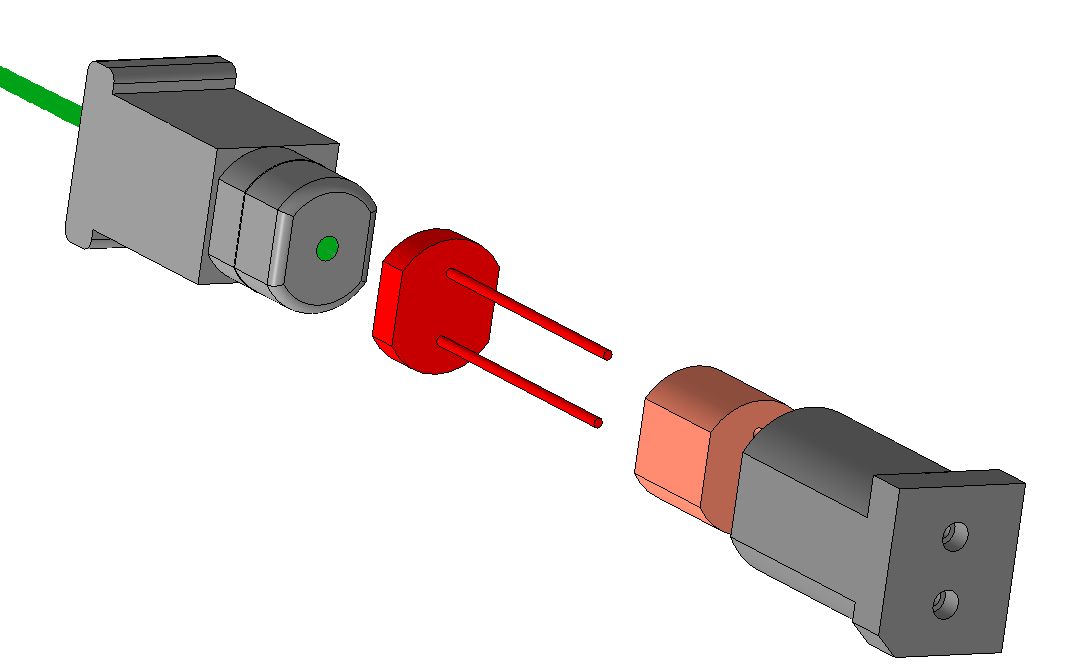}{Left: WLS fibers with optical connectors, prepared for installation. Right: The optical coupling interface between the WLS fiber (shown in green) and an SiPM (shown in red). The fiber ferrule (left) latches onto the photosensor's plastic housing (right). A foam spring (shown in light red) ensures contact between the fiber end and the SiPM.}{fig:prototype_photosensor_assembly}

The SuperFGD Prototype is equipped with three types of SiPM referred to as Multi-Pixel Photon Counters (MPPCs) by the manufacturer, Hamamatsu. The majority were the same model that has been chosen for the SuperFGD, the \mbox{S13360-1325CS}. The MPPCs used on the prototype have a ceramic casing, whereas the ones that will eventually be used on the SuperFGD are surface-mounted, providing an improved integrated solution. Two other types of MPPCs were also installed on the detector in order to save on instrumentation costs (they were available as spares) and for comparison with the proposed model of MPPC. The three MPPC types are referred to as Type I, II and III. Their specifications are listed in Table~\ref{tab:mppcs_on_prototype}.

\begin{table}

\centering

\begin{tabular}{@{}llll@{}cccc}
\toprule
\textbf{Description}  	&	\textbf{Type I}  &	\textbf{Type II}   & \textbf{Type III}	\\
\midrule
Manufacturer ref.	        &	S13360-1325CS	        &	S13081-050CS                &   S12571-025C 			    \\
No.\ in Prototype	        &	1152	        	    &	384                     & 192 			        \\
Pixel pitch	[{\micro\meter}]&	25			            &	50                      & 25			        \\
Number of pixels	        &	2668		            &	667                     & 1600 			        \\
Active area	[mm$^2$]        &	$1.3\times1.3$		    &	$1.3\times1.3$          & $1.0\times1.0$	    \\
Operating voltage [V]	    &	$56$--$58$	    &	$53$--$55$       & $67$--$68$     \\
Photon detection eff.\ [\%]		            &	25		        	    &	35                      & 35 			        \\
Dark count rate [kHz]	    &	70		        	    &	90                      & 100 			        \\
Gain			            &	$7\times10^5$		    &	$1.5\times10^6$         & $5.15\times10^5$ 	    \\
Crosstalk probability [\%]  &	1		        	    &	1                       & 10 			        \\
\bottomrule
\end{tabular}
\caption{Main parameters for the three MPPC types installed on the SuperFGD Prototype.}
\label{tab:mppcs_on_prototype}
\end{table}

\subsection{Readout Electronics}
\label{sec:readout_elec}

The readout electronics scheme developed for the Baby-MIND detector of the WAGASCI experiment was used \cite{Noah:2266598, Basille:2019mcp} for the SuperFGD Prototype. It is based on the CITIROC (Cherenkov Imaging Telescope Integrated Read Out Chip) front end ASIC (Application-Specific Integrated Circuit), which is engineered for the readout of SiPMs, and has been chosen for the final SuperFGD design. The main component in this scheme is the Front End Board (FEB) which houses three CITIROC chips. The detector is equipped with 18 FEBs, distributed across four minicrates. 

There are 32 input channels per CITIROC and hence 96 channels per FEB. As well as the three CITIROCs, each FEB contains a Field Programmable Gate Array (FPGA) for timing and data flow control, an 8-channel Analogue-to-Digital Converter (ADC) used for digitizing the CITIROC analogue outputs, and a USB 3.0 micro-controller for data transmission to a DAQ PC.

To avoid damage to the circuitry, the readout electronics were kept outside of the magnetic field within which the SuperFGD Prototype was placed during the beam test. To achieve this, a custom frame was constructed from aluminum bars that were several meters long, and the electronics were mounted on each corner of the frame, with the scintillator cube array fixed in the center.

\subsection{Mechanical Preassembly of the Cubes}

For the SuperFGD, the ``fishing line'' assembly method has been developed and validated to ensure good coupling and alignment of all detector components: cubes, fibers, mechanics and photosensors.

The basic concept is to preassemble the cube arrays using a flexible plastic cord of calibrated diameter, which is threaded through the holes drilled into the cubes. A fishing line of 1.3~mm diameter was the natural choice for this purpose. First, the cube array is assembled using the fishing lines, which forms the 3D skeleton structure of specified geometry. Then, the fishing lines are removed and the WLS fibers are inserted in place one by one. The fishing line diameter allows smooth insertion through the 1.5~mm cube holes, even with the variation in hole position described in Sec.~\ref{sec:cubes}, while leaving some tolerance for the subsequent installation of the 1.0~mm fibers. 

\insertgraph{0.5}{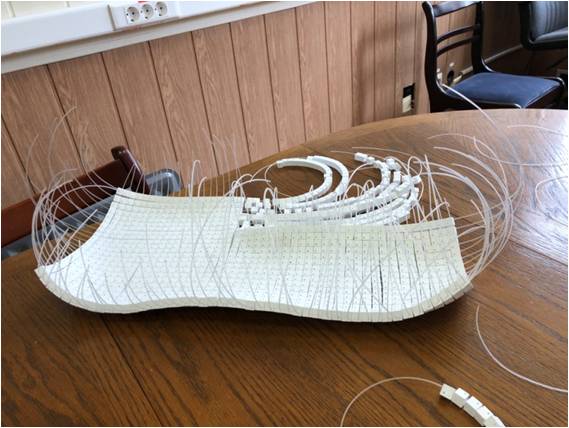}{A plane of scintillator cubes formed with fishing lines.}{fig:fishingline_plane}

A linear chain of cubes is the basic element of more complicated arrays. The linear chains are sewn together into 2D flat planes, also using fishing lines. An example of a plane prepared for a detector prototype is shown in Fig.~\ref{fig:fishingline_plane}. During the last step, the cube planes are merged into a 3D body.

Mock-ups of differing size were preassembled prior to the prototype construction to test possible assembly methods. One of the mock-ups, an array of $8\times 30\times 30$ cubes (7,200 cubes), was close in size to SuperFGD Prototype. 

The tests demonstrated that the fibers can be inserted through all the aligned cubes over a 2~m length, even though the mock-ups were made with the first batch of extruded cubes of relatively variable size ($\sigma_x=100$~{\micro\meter}).

\subsection{Assembly of the SuperFGD Prototype}

\label{subsec:SuperFGD_cubes_assembly}

The SuperFGD Prototype for the beam test was assembled in several stages using the fishing line method. The first sub-unit is a chain of 48 cubes (one line along the $z$-axis), strung together on a fishing line. Another set of fishing lines was used to bind 24 such individual chains to form a 2D plane ($x$-$z$ axes) of $24\times48$ cubes. This produced eight separate planes of $24\times48$ cubes. These 2D planes were stacked on top of each other to build up the height of the detector, with each 2D ($x$-$z$ axes) layer separated by a sheet of 150~{\micro\meter} thick Tyvek paper reflector and threaded with fishing lines along the vertical $y$-axis. This separation using Tyvek sheets is only for research and development purposes and not envisioned for the final detector design. The final array of $24\times8\times48$ cubes ($x$-$y$-$z$ detector axes) is shown in Fig.~\ref{fig:prototype_parts}.

The cubes of the prototype are held together by the pressure applied from an external box. The box is made of 8~mm-thick acrylic plates, into which a matrix of holes was drilled to accommodate the fibers and the connectors. The panels which make up the box are shown in Fig.~\ref{fig:prototype_parts} next to the SuperFGD Prototype volume.

\inserttwographs{0.416}{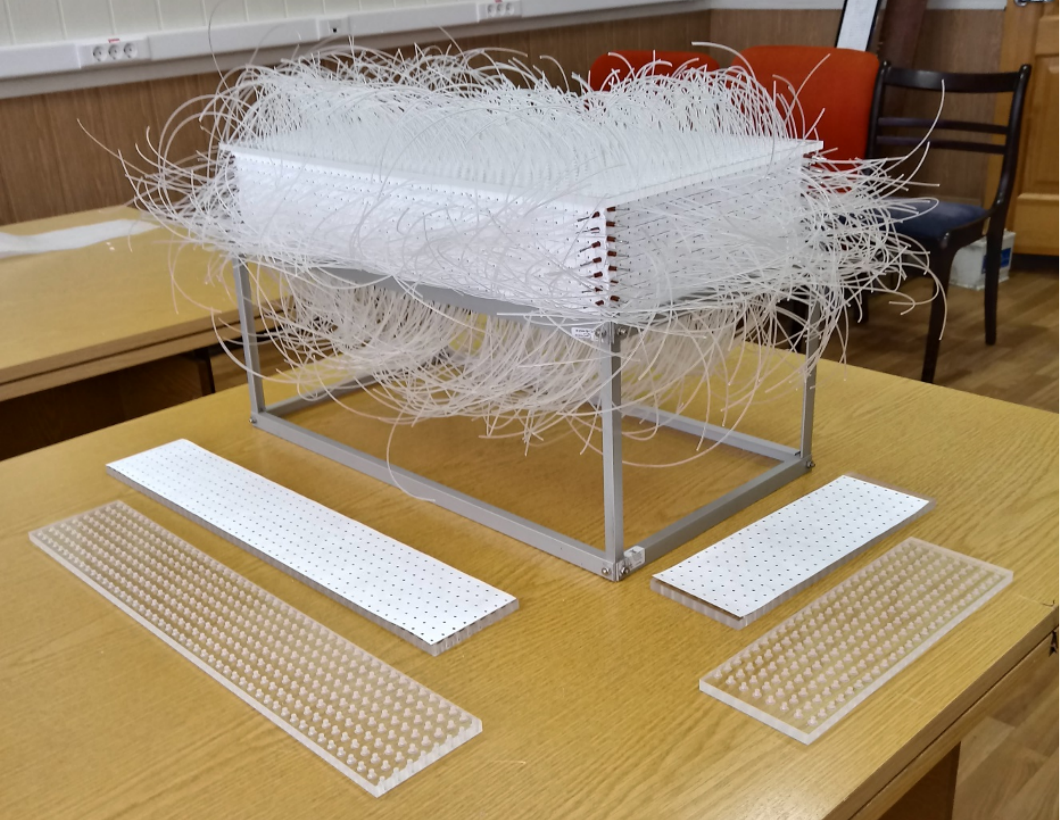}{0.574}{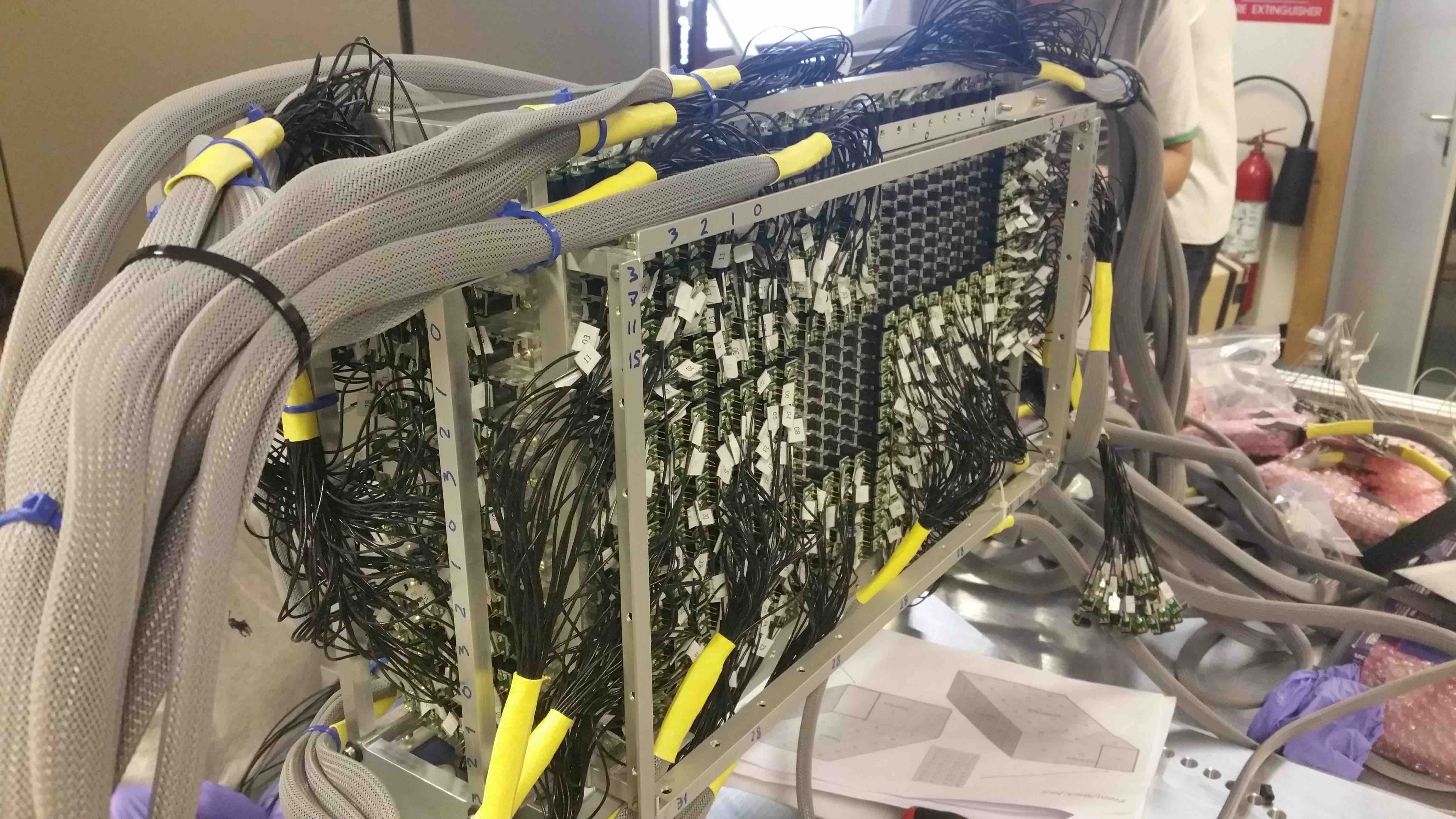}{Left: The SuperFGD Prototype volume assembled with the fishing line method, before the insertion of WLS fibers and the attachment of the acrylic box. Right: The partially instrumented SuperFGD Prototype bottom face. The detector is rotated by 90$^\mathrm{o}$ about the $z$-axis to enable access to the bottom face.}{fig:prototype_parts}

The box panels were positioned on the surfaces of the cube assembly with the fishing lines in place, and secured together with brass screws at their edges. Then, the fishing lines were replaced one-by-one with the WLS fibers. The size of the ferrules did not allow them to be connected to adjacent fibers. Instead, one row of fibers was inserted from one side of the prototype, then fibers were inserted into the neighboring row from the opposite side so that the optical couplers were distributed over all sides of the box uniformly.

To keep the optical couplers in place, they are pressed against the walls of the box using~6mm-thick acrylic strips with cut-outs and held in place by brass screws. One strip is used for each row of 8 or 24 couplers, depending on which side of the detector the couplers are mounted. The protruding ends of the fibers were cut off using a sharp blade.

\begin{figure}
  \begin{center}
    \includegraphics*[width=0.8\textwidth]{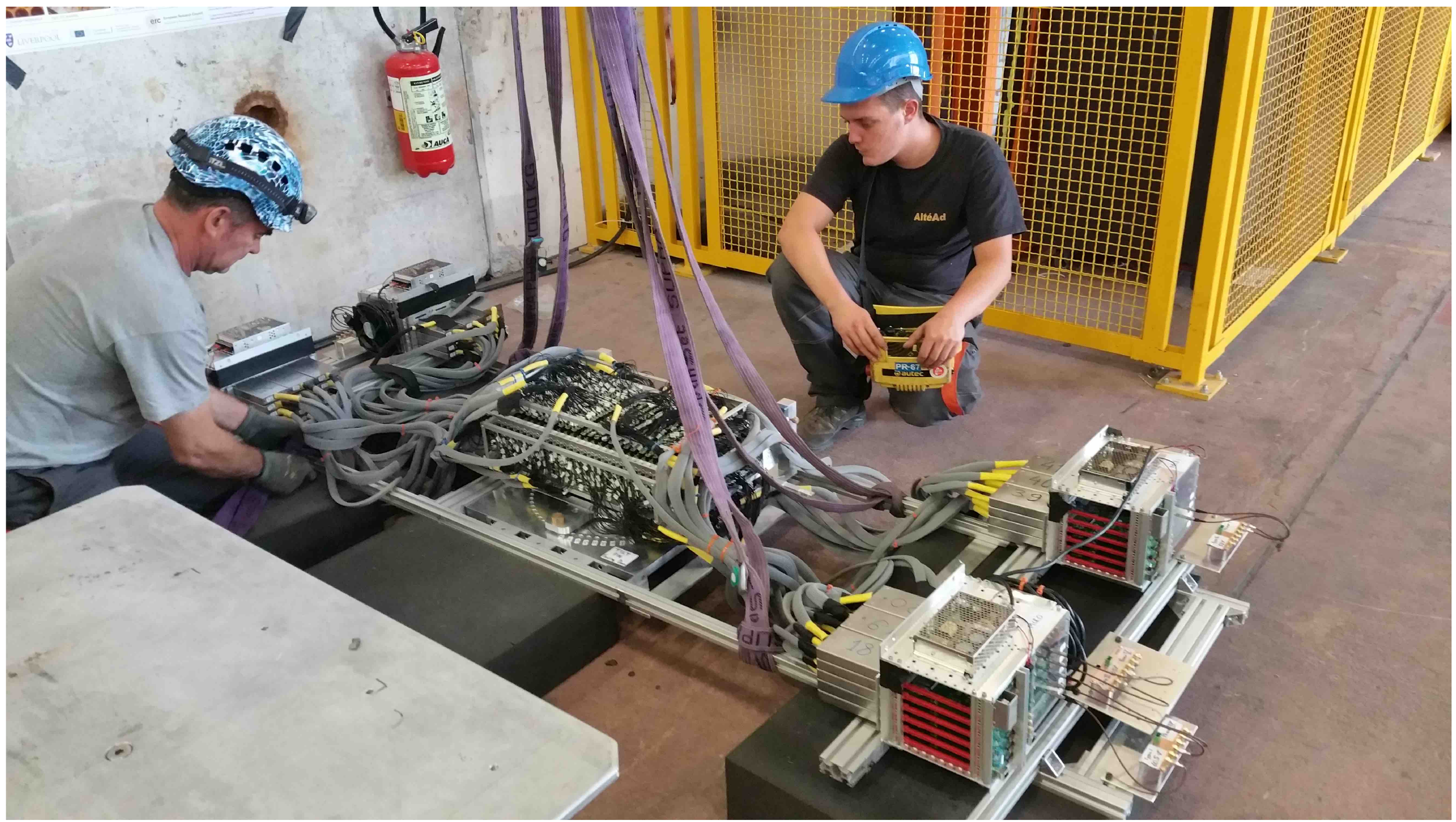}
  \end{center}
\caption{The SuperFGD Prototype in the CERN-PS East Hall during transport to the T9 beamline. The minicrates housing the front end electronics can be seen on each of the four corners of the SuperFGD Prototype mounted in the center of the mechanical assembly.}
\label{fig:Target-proto_inside_mnp17}
\end{figure}

Finally, a frame constructed from aluminum profiles was mounted around the detector to serve as a support for readout cables. This provided mechanical protection for the photosensors and prevented light isolation sheets wrapped around the detector from interfering with the mounted MPPCs and cables. A photograph of the SuperFGD Prototype mounted in the aluminum frame and connected to the readout electronics is shown in Fig.~\ref{fig:Target-proto_inside_mnp17}.

The distribution of MPPC types around the six faces of the detector is shown in Fig.~\ref{fig:Target-mppc_distribution}. Fibers along the $z$-axis collect light from many cubes simultaneously, as the particle beam is directed along this axis, so we chose to instrument all $z$-fibers with the MPPCs with the largest dynamic range (Type I). We instrumented the $y$-$z$ plane fully with Type I MPPCs to have the same gain and calibration for all readout channels detecting through-going tracks. The $x$-$z$ plane was instrumented with the three different types of MPPCs, keeping the same MPPC Type I towards the back of the detector where possible to have all three projections equipped with the same MPPCs for the detector volume corresponding to the stopping point of 0.8~GeV/$c$  protons. Each MPPC's operating voltage was set to the manufacturer recommended voltage corresponding to the MPPC gain value reported in Table~\ref{tab:mppcs_on_prototype}. Given the variation in operating voltages of up to 2~V across the whole sample of MPPCs of a given type, the MPPCs were preselected and sorted into batches of 32 to achieve an operating voltage spread no greater than $\pm 100$~mV per batch. 

\insertgraph{0.7}{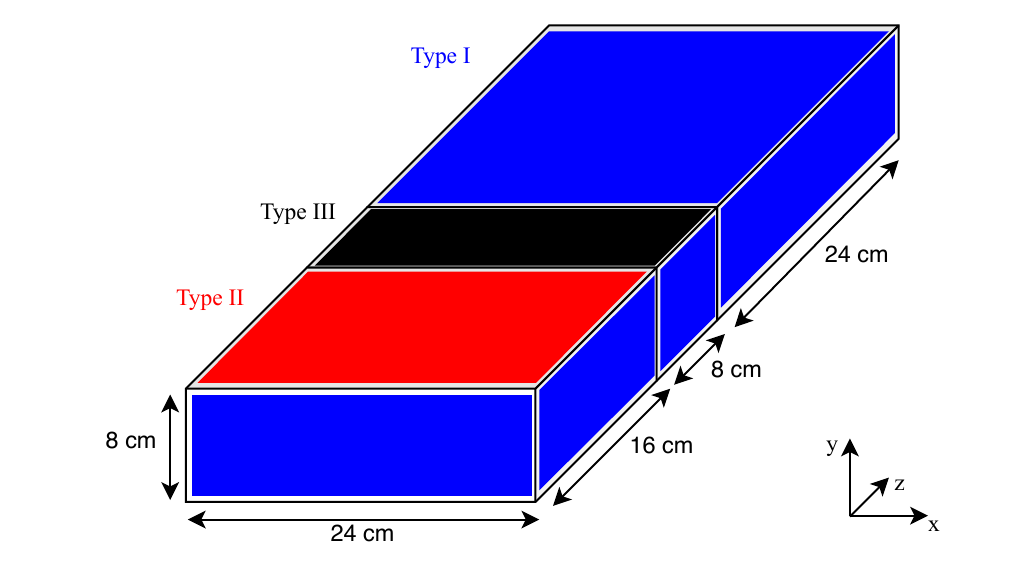}{Distribution of the three types of MPPCs around the six faces of the SuperFGD Prototype: 1,152 Type I (S13360-1325CS), 384 Type II (S13081-050CS), and 192 Type III (S12571-025C) MPPCs represented by blue, red and black respectively.}{fig:Target-mppc_distribution}

\label{sec:cube_production_assembly}

\section{Signal Processing and Calibration}

\subsection{Signal Processing}
\label{sec:signal}

Each MPPC is connected to a single input channel on a CITIROC that splits the incoming signal down two paths, the High Gain (HG) path and the Low Gain (LG) path (Fig.~\ref{fig:Target-3signal_outputs}). The HG and LG paths each start with an independent, tunable preamplifier followed by a slow shaper (SSH). The HG and LG slow-shaper outputs are sampled either using a built-in peak detector or by applying an externally-controlled delay. These analogue HG and LG signals are then multiplexed in the CITIROC and digitized outside the CITIROC on the FEB 12-bit ADC. The CITIROC also provides independent trigger lines with a fast shaper that can either be connected to the HG or LG preamplifier outputs (for the CERN beam test it was the HG preamplifier). The fast trigger output signal is processed through an adjustable threshold discriminator and sampled directly by the FEB FPGA at 400~MHz, assigning timestamps to the rising and falling edges of the trigger line output signals.

\insertgraph{0.7}{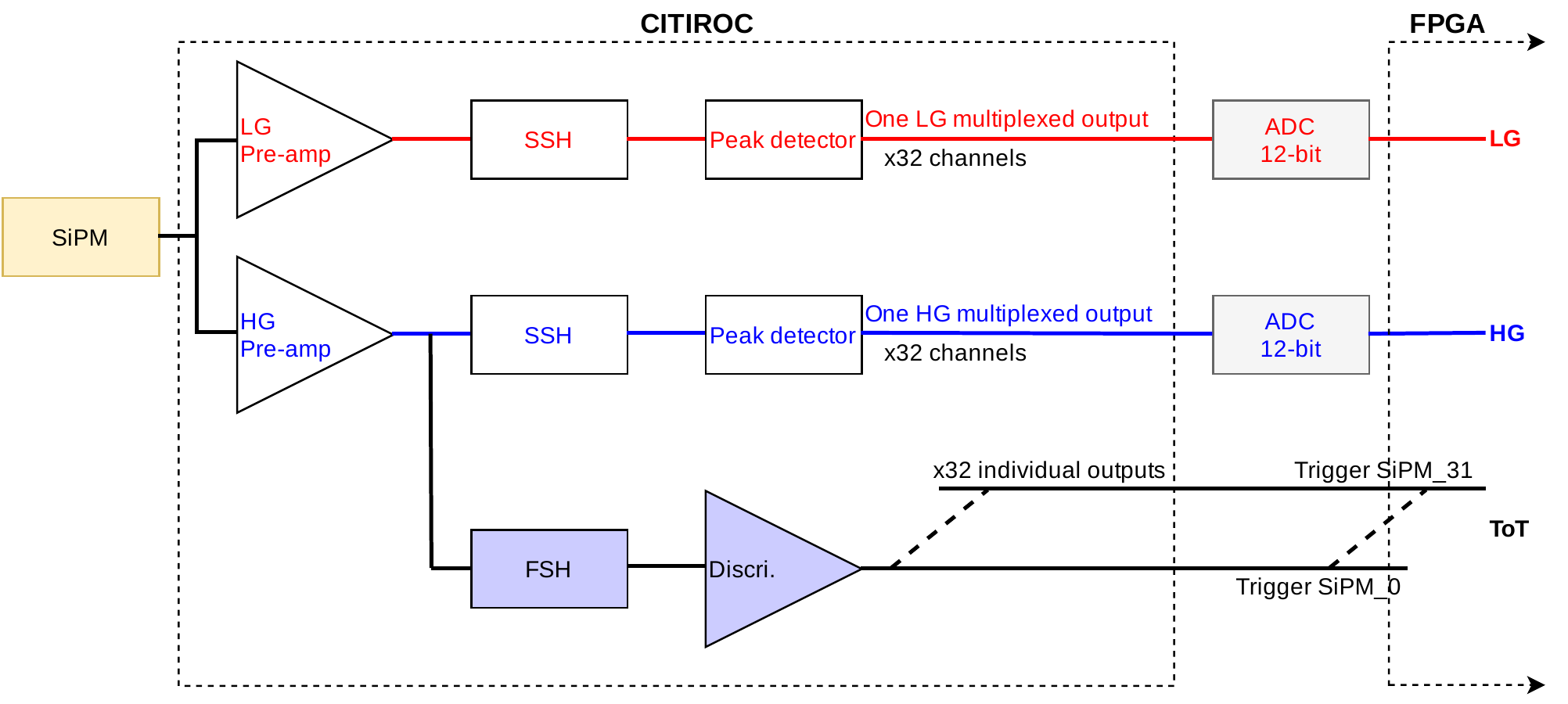}{Sketch of the three signal outputs providing amplitude information for an event. A different calibration process is required for the amplitude reading from each path. HG: High Gain. LG: Low Gain. SSH: Slow Shaper. FSH: Fast Shaper.}{fig:Target-3signal_outputs}

The fast trigger output is also used to deduce charge amplitudes. The difference in time between the falling and rising edge timestamps yields the Time-over-Threshold (ToT) of the signal, which is a function of the signal amplitude. This is critical for the registration of hits that would otherwise not be recorded by the HG and LG signal paths. For a given readout cycle, the start of the acquisition gate for HG and LG signals is triggered by the first channel on the FEB that produces a fast trigger output. The hit multiplicity is limited to one HG and one LG hit per channel for this acquisition gate, the duration of which was set to 10~{\micro\second} for the beam tests. Moreover, the deadtime introduced by the analogue output stages---caused by multiplexing the 32 channels into one single output each for HG and LG, and digitization of these outputs---is 9.12~{\micro\second}. As a result, only one HG and one LG hit can be recorded per channel every 19.12~{\micro\second}, whereas ToT hits can be recorded continuously and without any deadtime. 

\subsection{Calibration}
\begin{figure}
	\centering
	\includegraphics*[trim=0 -52 0 0,clip,width=0.495\textwidth]{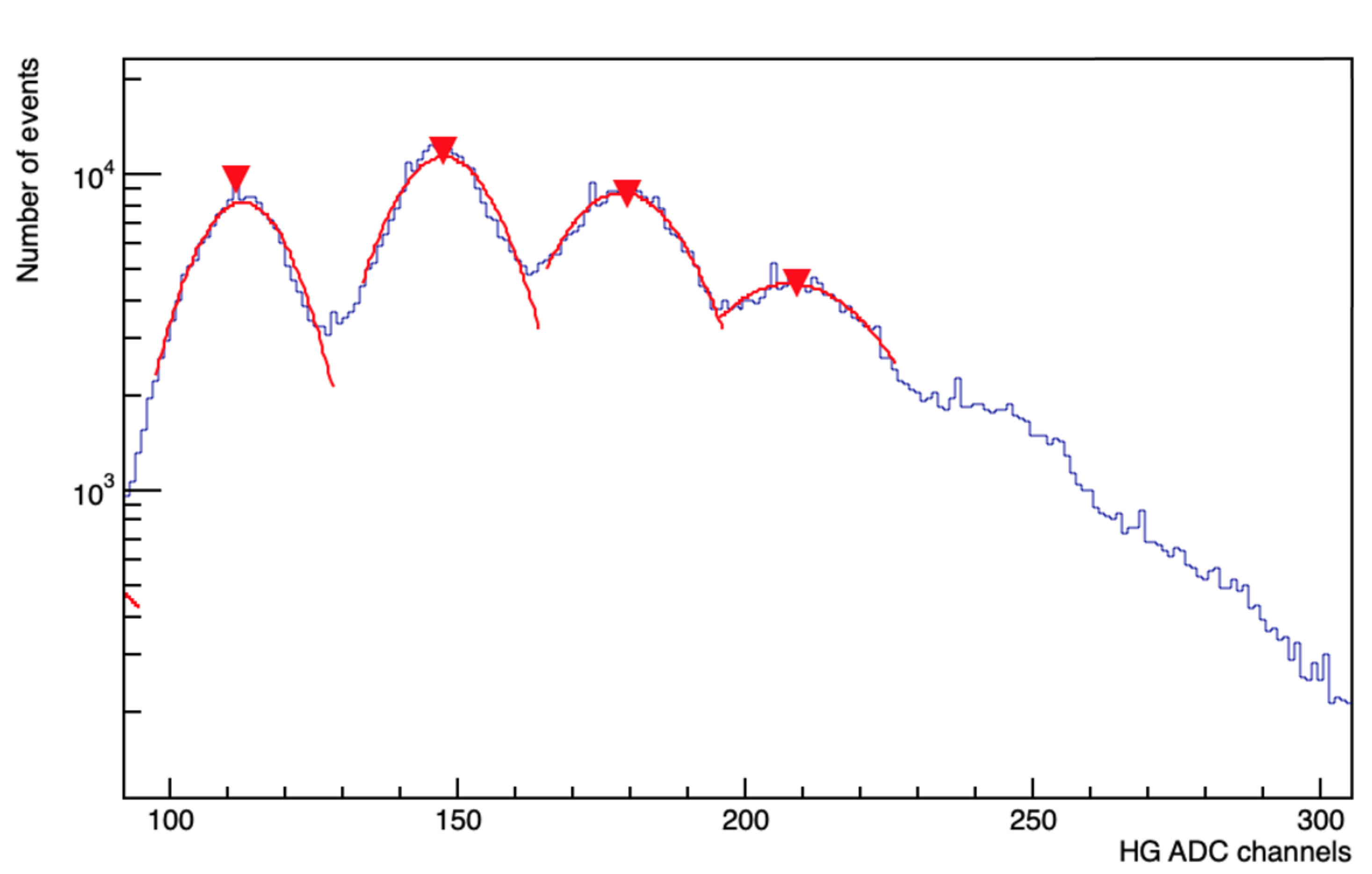}
	\hfill
	\includegraphics*[width=0.495\textwidth]{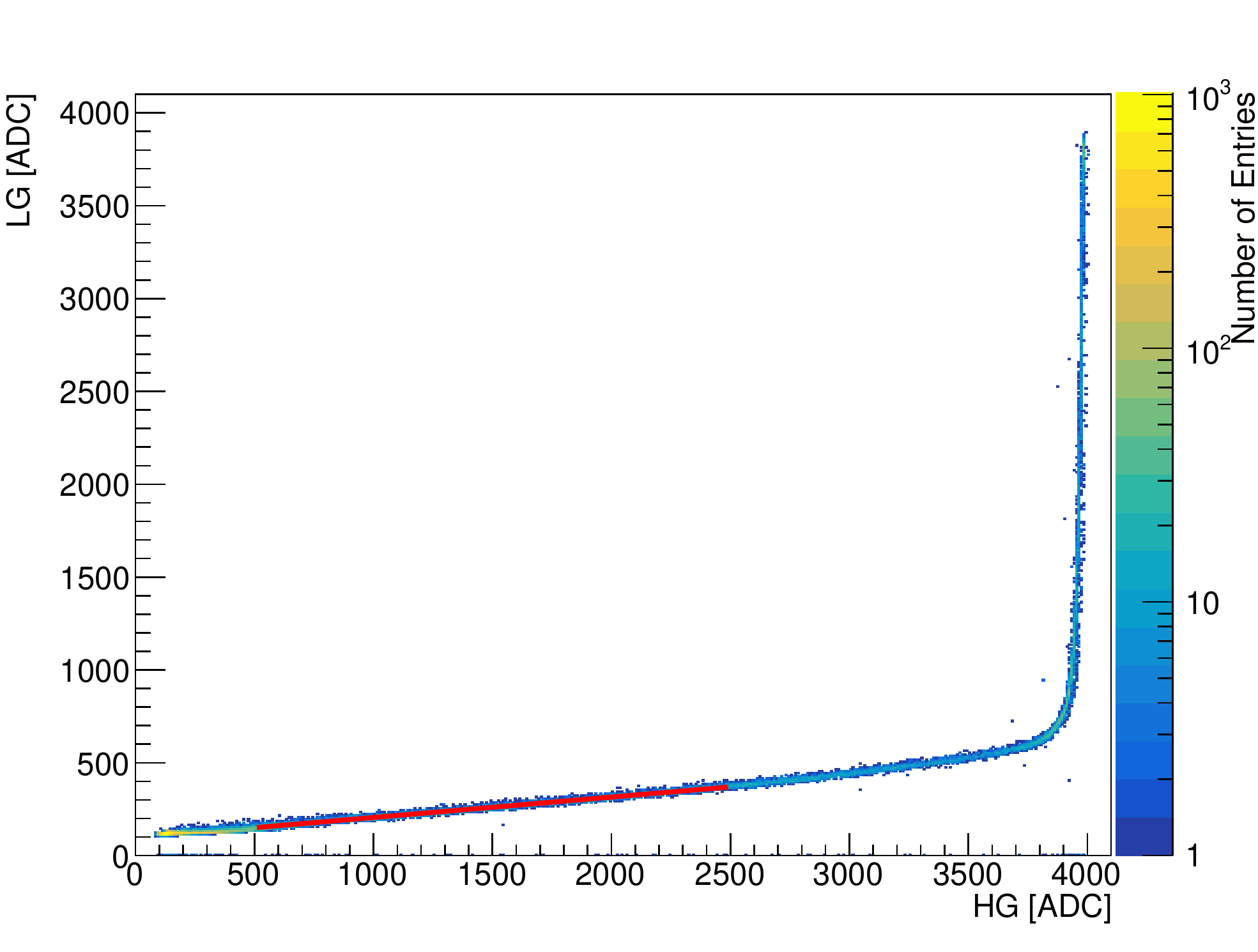}
	\caption{Left: typical calibration plot used to extract the ADC/pe ratio for the HG signal path. Right: Calibration of LG signal path against the HG signal path. The linear fit used for the LG calibration is also shown in red.}
	\label{fig:hg_lg_calibration}
\end{figure}

As outlined in Sec.~\ref{sec:signal}, there are three different signal readout paths that provide a measurement of amplitude: the HG, LG and ToT. HG and LG amplitudes are returned in ADC units, and ToT amplitudes are returned in timestamp units of 2.5~ns. All three signal types are calibrated in units of photoelectrons (p.e.). The calibration method for the light yield is specific to HG, LG and ToT:
\begin{itemize}
    \item HG calibration - obtained the ADC/p.e.\ gain ratio from either dark counts (Type III) or LED signals for each MPPC channel using a custom LED system~\cite{7753500}. Clearly distinguishable individual photoelectron peaks are required (Fig.~\ref{fig:hg_lg_calibration}); 
    \item LG calibration - a linear fit of LG data against HG data where the latter is below about 3600 ADC counts, for a given channel, provided its LG calibration parameters (Fig.~\ref{fig:hg_lg_calibration});
    \item ToT calibration - compared ToT data against HG data for signals up to 100~p.e.\ and against LG data for signals above 100~p.e. The nonlinear relationship between ToT and HG/LG was described using 5$^\mathrm{th}$-order polynomial fits (Fig.~\ref{fig:tot_calibration}).
\end{itemize}

\inserttwographs{0.495}{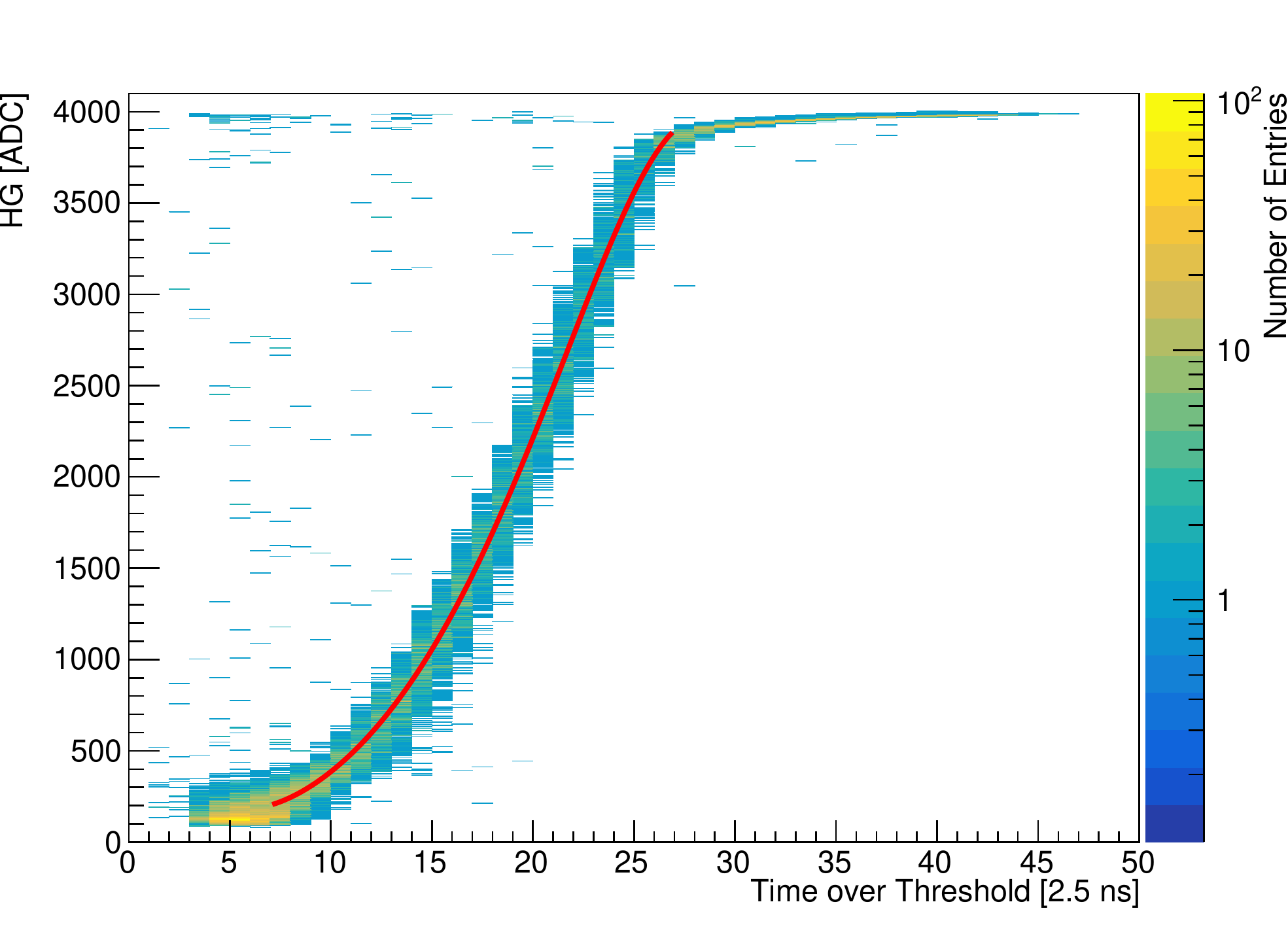}{0.495}{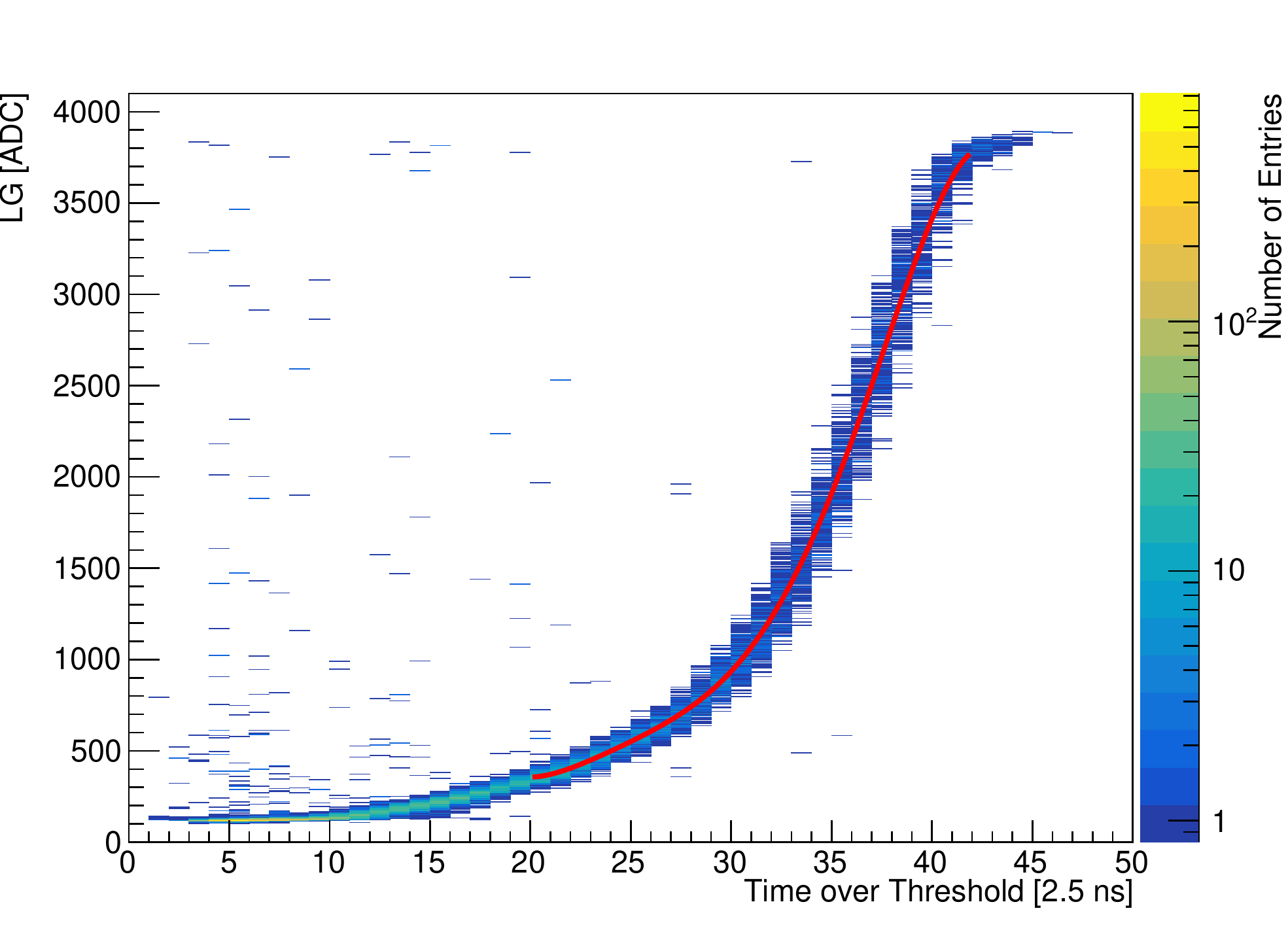}{Calibration of the Time over Threshold (ToT) amplitude measurement using the HG and LG signal paths. Left: HG vs ToT. Right: LG vs ToT. 5$^\mathrm{th}$-order polynomial fits used for the ToT calibration are also shown in red.}{fig:tot_calibration}

Dedicated LED calibration runs were carried out at the beamline location to extract the HG calibration factors. The LG and ToT calibration analysis was performed using beamline data, which provided the required dynamic range. Different HG and LG preamplifier gain settings were used throughout the test campaign. The three HG gain ratios used were 42, 29 and 23~ADC/p.e. The process of choosing between those three gain ratios was driven by the requirement to measure at least 1 MIP and possibly up to 2 MIPs without saturating the HG signal, and by the requirement to calibrate efficiently all channels. Gain distributions from HG calibration runs with the most commonly used gain ratio, around 29~ADC/p.e. are shown in Fig.~\ref{fig:Target-calib_gain_spread}. 
 
These distributions show large variations. In future, the gain could be further tuned on an individual channel basis by fine adjustment of the CITIROC 10-bit DAC that trims each MPPC operating voltage independently, rather than applying the same 10-bit DAC value to groups of 32 MPPCs as was done for this study.

\begin{figure}
    \centering
    
    \includegraphics[trim=22 0 0 0, clip, width=0.535\textwidth]{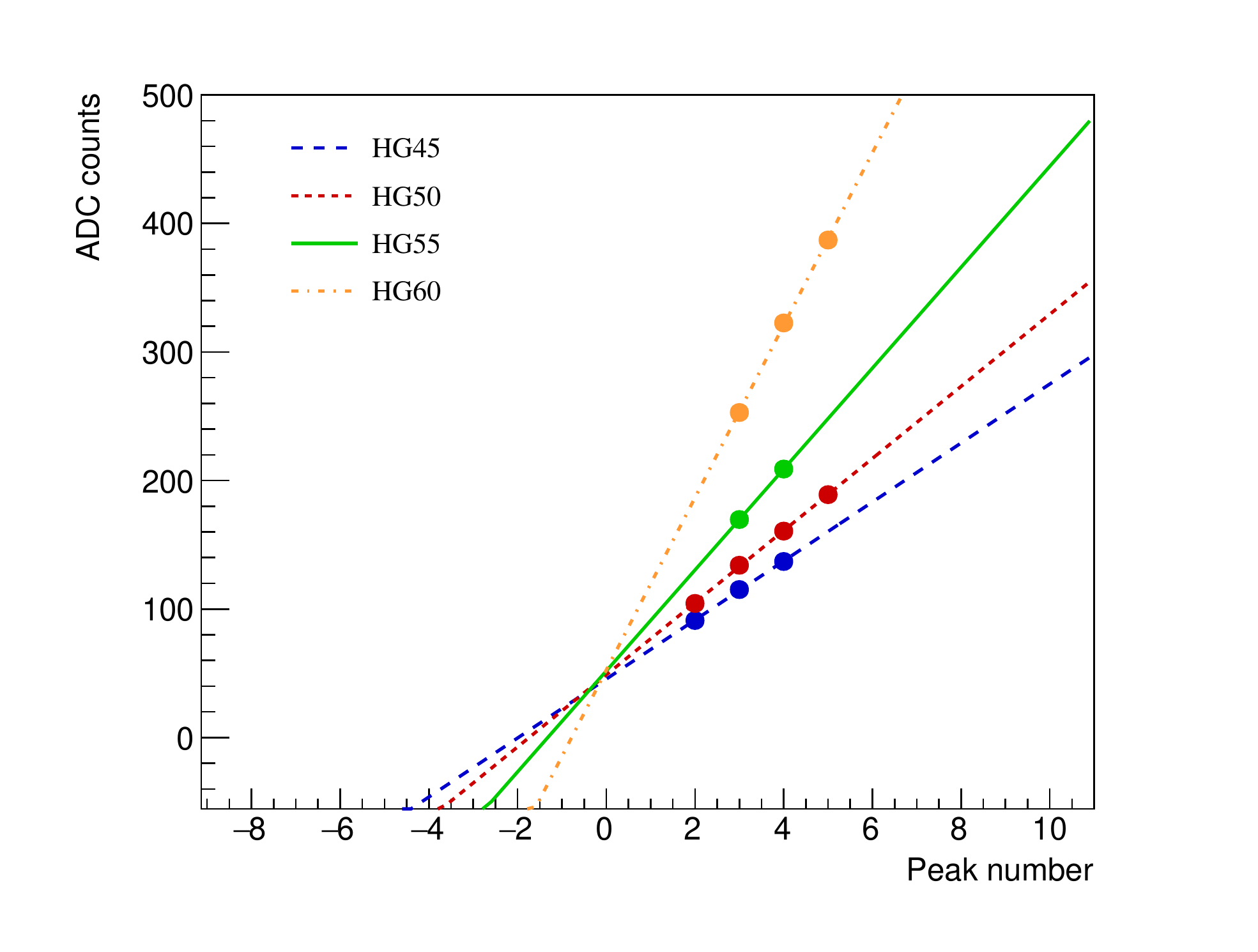}
    \caption{Pedestal finding method for a single channel with four different HG values. The ADC values of each photoelectron peak are plotted and extrapolated. The point of intersection is taken as the pedestal position.}
    \label{fig:pedestal_finder}
\end{figure}

The variation in pedestal values across all channels is large, at several hundred ADC counts. Establishing pedestal values was therefore a significant part of the calibration process, especially for studies of low signal amplitudes such as detector thresholds or optical crosstalk. Because the signal thresholds were set above 0.5 p.e., the electronics baseline was not recorded. Pedestal estimations therefore required dedicated calibration runs with up to four different HG preamplifier gain settings per channel. A linear fit through the location of the first two to five individual photo-electron peaks was applied per channel per HG preamplifier setting. The location of the pedestal was taken to be the point of intersection of lines (each from one HG preamplifier value) reconstructed from fit parameters and extrapolated to lower ADC values (Fig.~\ref{fig:pedestal_finder}).

\insertgraph{0.495}{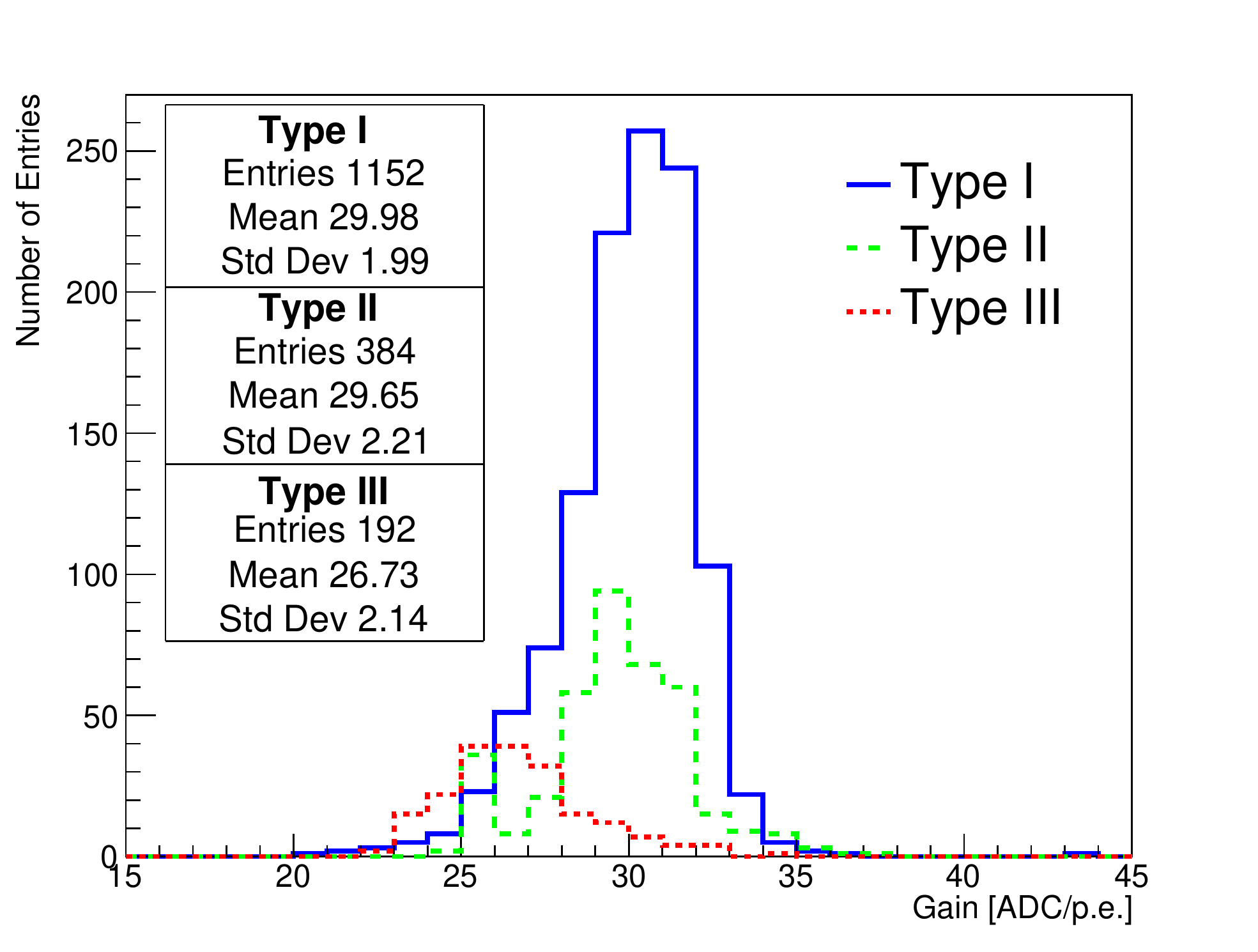}{Gain distributions in units of ADC/p.e.\ for all channels with Type I (green solid), II (blue dashed) and III (red dotted) MPPCs with the ``lower" CITIROC HG preamplifier settings of 50, 43 and 45 respectively.}{fig:Target-calib_gain_spread}

Finally, the three resulting values of the amplitude in p.e. are combined to create one value that is referred to as the hit \textit{PE}. To create this value, the charge of the hit is considered equal to the HG signal if it exists and is below saturation. If the HG signal fails one or both of these conditions, the hit PE is equal to the LG signal (also if it exists and is below saturation). Otherwise, the ToT signal is used for the hit PE. For the analysis of the data, hit PE is used as the standard value of the light yield in photoelectrons.

\section{Beamline Setup}

In this section we report on the layout of the beamline used in the 2018 CERN beam tests, on the properties of the beam and on the particle trigger system that was used to provide information about the different particle types.

\subsection{The Beamline Layout}
\label{subsec:beamline}
\begin{figure}[htbp]
  \begin{center}
    \includegraphics*[width=0.8\textwidth]{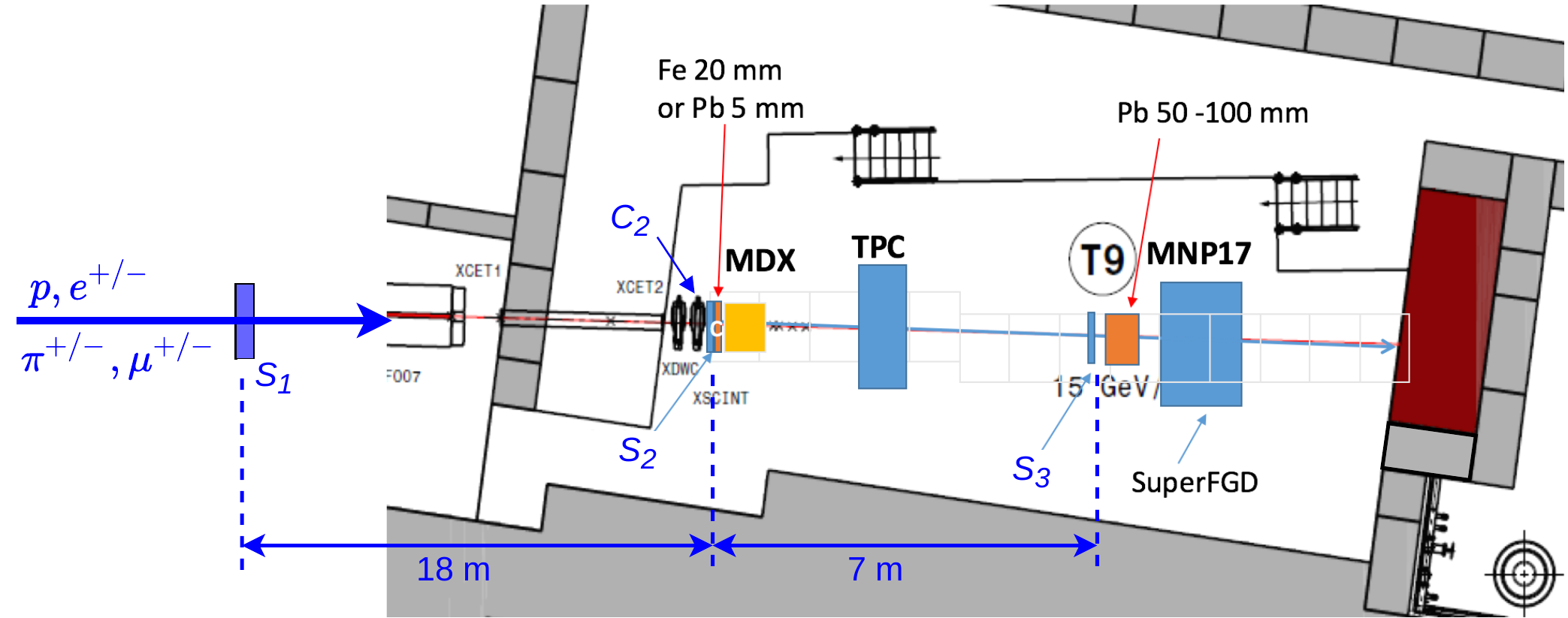}
  \end{center}
\caption{Layout showing the main components on the PS T9 beamline platform. Scintillators ($S_1$, $S_2$, $S_3$) and a Cherenkov detector ($C_2$) were used in the particle trigger system.}
\label{fig:Target-t9_layout}
\end{figure}

The SuperFGD Prototype was tested at the CERN-PS T9 beamline with a TPC placed upstream as shown in Fig.~\ref{fig:Target-t9_layout}. This TPC was part of a separate study~\cite{attie2019performances}.

Time-of-flight counters were installed along the beamline to provide particle identification. The beam momentum was fixed for each run, chosen within a range of $\pm400$~MeV/c to $\pm8$~GeV/c, by setting beamline magnet currents to predefined values. We operated two main beam modes. The ``hadron" beam mode was the standard beam mode, with a mix of charged particle types including positive beams (negative beams in parentheses) of protons, $\pi^+$ ($\pi^-$), $\mu^+$ ($\mu^-$) and $e^+$ ($e^-$). The resulting relative fraction of particle species in the beam could be further modified by the insertion of beam stoppers and paraffin, Fe, Cu or Pb converters to enhance or suppress a particular particle type. The ``muon" beam mode operated with a thick beam stopper in the beamline that suppressed the hadronic component of the beam and most $e^+$ ($e^-$).

\subsection{Particle Triggers}
\label{subsec:triggers}

A trigger system based on scintillator detectors ($S_1$, $S_2$, $S_3$) and a Cherenkov detector ($C_2$) was used to discriminate different particle types. The position of the detectors in the beamline can be seen in Fig.~\ref{fig:Target-t9_layout}. In order to achieve good time synchronisation between the particle triggers and all photosensors, the system used delays and coincidence logic to form electrical signals that were attenuated and routed to a dedicated FEB that processed these signals through four CITIROC channels. The CITIROC processed these trigger signals as it would photosensor signals.

Examples of signal combinations are described in Table~\ref{tab:trigger_combinations}, though these combinations frequently evolved throughout the beam tests. Sample purity was good for the $e^{+/-}$ trigger because it was the only particle type detectable with the Cherenkov detector. All other particles $\pi/\mu/p/K$ were below detection threshold for the particle momentum ($<8$~GeV/c) and Cherenkov gas pressure ($<3$~bar) chosen. The proton sample purity was \textbf{$> 99\%$} at momenta $<1$~GeV/c, due to the large difference in time of flight with respect to the other particle types. The $\pi/\mu$ trigger was less efficient, as it included $e^{+/-}$ that were not detected by the Cherenkov detector. 

\begin{table}
 \centering
 \begin{tabular}{@{}llll@{}cccc} 
  \toprule
  \textbf{Beam mode} & \textbf{Particle trigger} & \textbf{Trigger setup} & \textbf{Purity} \\ 
  \midrule
  Hadrons & All & $S_{2L} \times S_3 \times S_1$ & N/A\\
  Hadrons & $p$ & $S_{2S} \times S_3 \times S_1$ & $>90\%$ \\
  Hadrons & $e^+$ & $S_{2L} \times S_3 \times C_2$ & $>90\%$ \\ 
  Hadrons & $\pi / \mu$ & All $\times\ \overline{e}\times\overline{p}$ & [$e^+$ ($40-50\%$)]\\
  Muons & $\mu$ & All $\times\ \overline{e}\times\overline{p}$ & [$e^+$ ($10-20\%$)]\\
  \bottomrule
 \end{tabular}
 \caption{Combination of trigger signals to form particle identification triggers, and typical purity. Percentages of unwanted particles are expressed in square parentheses.}
 \label{tab:trigger_combinations}
\end{table}

\section{Detector Response}
\label{detectorResponse}

The detector response to an energy deposition event by a charged particle in one scintillator cube leads to several features that were studied in detail. We start the discussion with a review of the hit amplitude thresholds and hit time structure, to assess limitations of the readout system. One of the features of this type of detector is the amount of scintillation light that leaks to adjacent cubes from the main cube where it was produced. This cube-to-cube optical crosstalk was evaluated using two methods which are outlined.

We continue the detector response discussion with a study of channel response, defined as the response of an individual readout channel to a given event. The cube position is neither known before the channel response study nor extracted from such a study. However the channel response study is important in evaluating channel uniformity and the quality of the calibration. This response is directly related to all photons collected by the WLS fiber on that channel. 

By combining hits in two projections we can determine the position of the cube that was hit. This enables the study of the attenuation of light signals as they transit down the WLS fibers from the point of incidence and wavelength conversion in the fiber at the level of the cube to the photosensor. The cube response subsection presents attenuation-corrected cube light yields for several thousand cubes.

Finally, the time resolution of the SuperFGD Prototype detector is reported.

\subsection{Hit Amplitude Thresholds}
\label{sec:thresholds}
Hardware hit amplitude thresholds were set prior to the beam-test runs at two levels: at the level of the CITIROC discriminator, which sets thresholds for trigger line pulses; and at the level of the FPGA to zero-suppress digitized HG/LG values sent to the DAQ PC. The first threshold was applied as a general setting with the same value for all CITIROCs connected to all FEBs. The second value was loosely determined on a CITIROC-by-CITIROC basis by acquiring some dummy runs and registering a rough threshold for each group of 32 channels.

\begin{figure}
    \centering
    \includegraphics[width=\linewidth]{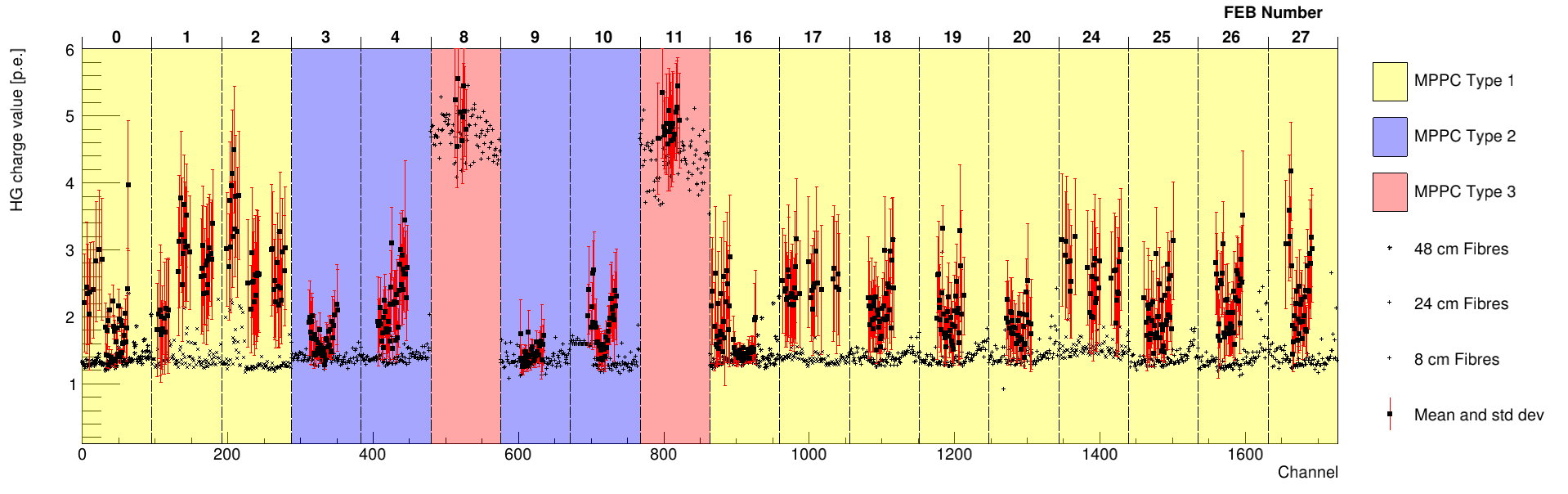}
    \caption{Measured values of the hit thresholds for each channel. Colors represent different MPPC types, and the fiber length of the channel is represented by the marker style (see legend). Two different representations of the hit threshold are shown. The minimum HG value measured over the whole data run for each channel is shown by either a plus, cross or star marker. The mean value of the minimum HG values measured for each spill for each channel is represented by a solid black square, and the standard deviation in this value is shown by the red error bars. Data points with standard deviations above 1~p.e.\ are not shown.}
    \label{fig:thresholds}
\end{figure}

These thresholds on hit amplitudes set a limit on the minimum light yield that can be measured by each readout channel. A study was carried out that found the minimum value of the HG light yield measurement for each channel, as well as the mean value of the lowest HG light yield for each spill in each channel. The results for all channels are shown in Fig.~\ref{fig:thresholds}.

The minimum thresholds for Type I and II MPPCs are very similar, with values around 1.2~p.e. The Type III MPPCs have higher values around 4.8~p.e. This is not an unexpected result, as the CITIROC thresholds were set differently for each MPPC type, and thresholds for the MPPC Type III were set deliberately high to reduce dark count hits given this type of MPPC has a crosstalk probability 10 times higher than the other MPPCs.  

The mean value and standard deviation of the lowest hit charge per spill differ significantly between channels, as opposed to the minimum values across all spills, which are relatively consistent between channels. Such variations can be attributed to the different number of hits seen by each channel per spill. Channels that saw many hits, for example some channels in FEBs 0, 3, 9, and 16, have very small standard deviations and tend to have mean values around 1.4~p.e. Channels with lower statistics varied between 2--3~p.e, and channels that had a standard deviation over 1~p.e.\ were ignored. The Type III MPPCs again gave different results to the other MPPCs, with average values around 5~p.e.

\subsection{Hit Time Structure}

For the SuperFGD Prototype, because the hit amplitude is the result of processing along a signal path that includes both the SiPM and the CITIROC readout ASIC, and because the readout ASIC includes a shaper with a time constant that is set to any of eight values in steps of 12.5~ns, the separation of the different time-dependent components of a hit is not possible. Moreover, the CITIROC shaper output can display additional peaks beyond the main peak. 

The left-hand side plot in Fig.~\ref{fig:AP} shows a display of hits recorded within a time window of $\pm1.2$~{\micro\second} of a particle trigger signal for a 2~GeV/c muon beam. The main trigger is recorded at -250~ns and is surrounded on both sides by a repeating structure every 350~ns. This structure is correlated with the proton beam and is a sub-structure of the PS accelerator revolution frequency of 477~kHz, which can also be observed with higher peaks every 2~{\micro\second}. This structure is less distinct in MPPCs of Type III, which suffer from a larger percentage of dark counts.

\inserttwographs{0.49}{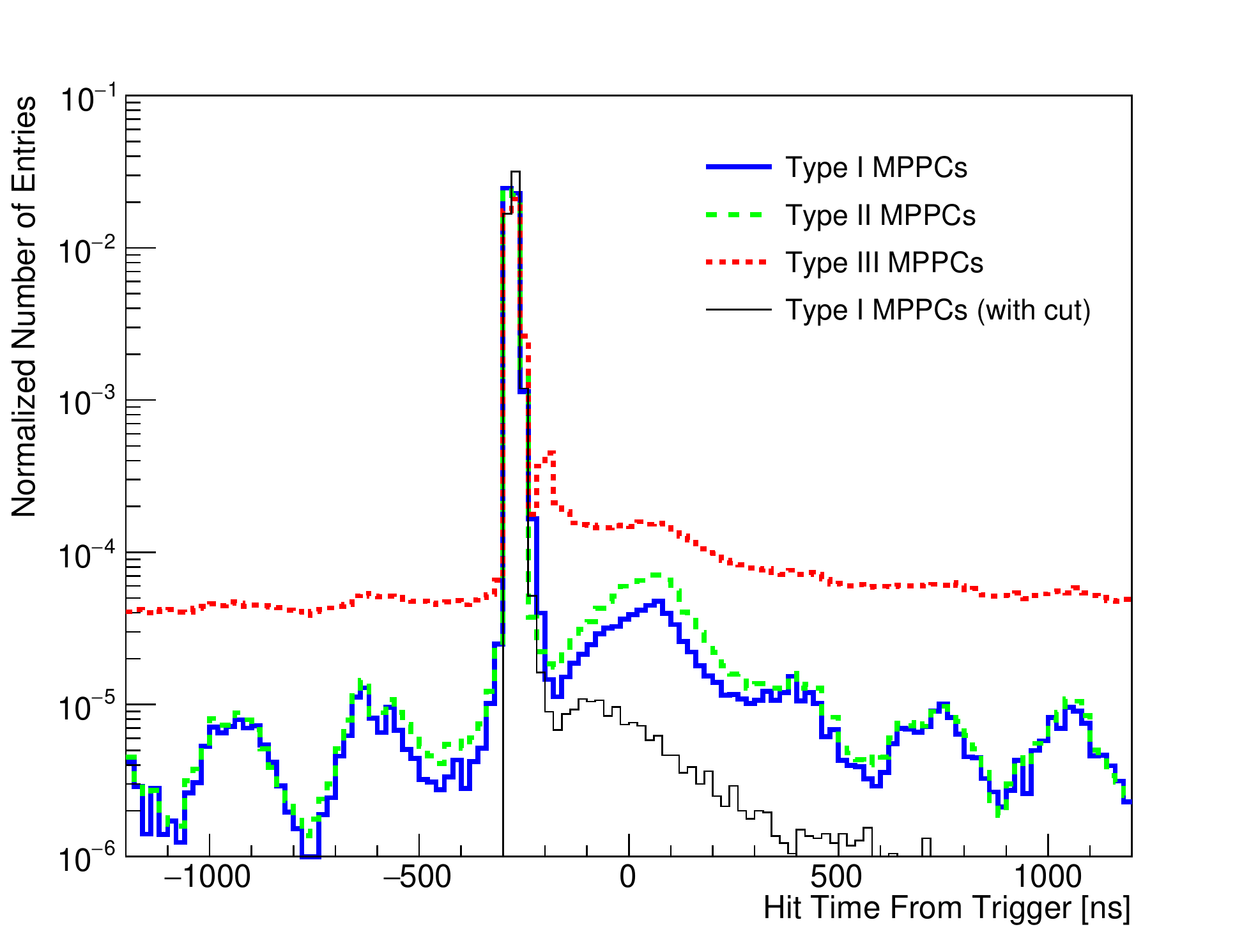}{0.49}{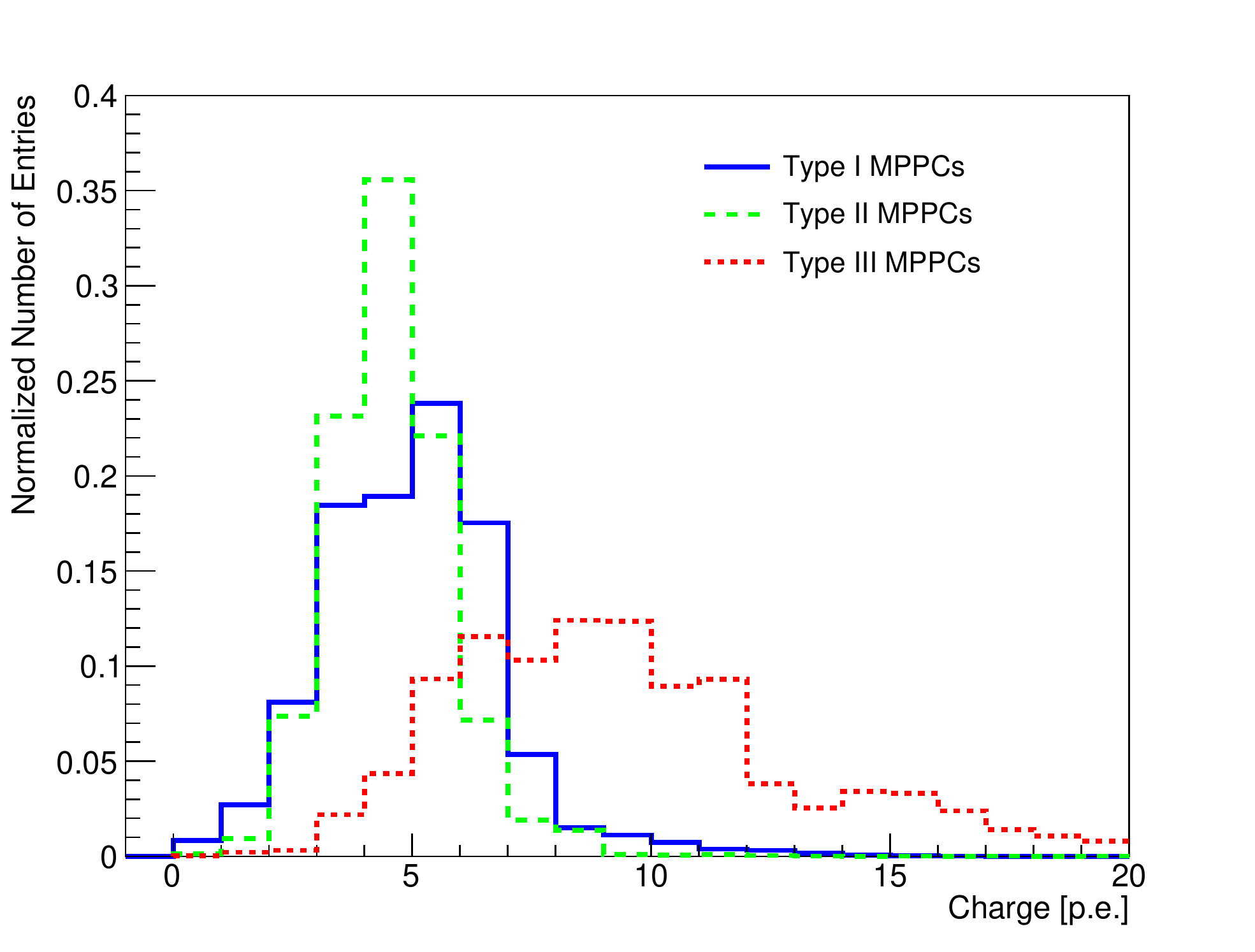}{Left: The time distribution of hits w.r.t. the external trigger for the three types of MPPCs used in the prototype, and a distribution for MPPCs of type I with a cut eliminating the beam structure. Right: The average charge recorded in the time windows of interest for the three types of MPPCs.}{fig:AP}

Apart from this beam structure, we observe a secondary peak, less than 100~ns following the main trigger, for MPPCs of Type III. Whereas MPPCs of Types I and II exhibit a wide peak around 0~ns. In both cases, these peaks are comprised of low energy hits with an average charge of 4.5~p.e. The charge distribution of these hits is shown on the right-hand side plot in Fig.~\ref{fig:AP}, where a time cut was used to select a time window around the secondary peak for each MPPC type, along with a charge cut at 20~p.e. used to eliminate events containing a MIP (with a typical charge of around 50~p.e.) in order to eliminate the contribution from the beam structure.

This effect is quantified by calculating the percentage of hits in these peaks per recorded MIP (minimum ionizing particle) hits for each FEB. In order to remove the component of the beam structure, a cut was used to eliminate events where a MIP was recorded outside the triggered event time window. The black curve in the left-hand side plot in Fig.~\ref{fig:AP} shows the remaining hits for MPPCs of Type I after this cut is applied. Table~\ref{tab:afterpulse_per_MPPC} shows the percentage of occurrence for each type of MPPC, as well as the average charge for the peak hits. A noticeable difference is observed between the three types of MPPCs used in the prototype as MPPCs of Type III exhibit a much larger percentage of occurrence and average charge compared to Types I and II.

\begin{table}[htbp]
\centering
\begin{tabular}{llll}
\toprule
\textbf{MPPC} & \textbf{\% of occurrence per MIP} & \textbf{Mean charge [p.e.]}\\
\midrule
Type I & 0.29 & $4.41\pm1.42$ \\
Type II & 0.51 & $4.67\pm1.36$ \\
Type III & 11.27 & $8.93\pm2.60$ \\
\bottomrule
\end{tabular}
\caption{Details on the peak structures in the time distribution of hits for the three types of MPPCs.}
\label{tab:afterpulse_per_MPPC}
\end{table}

The origin of these peaks is still unknown and under investigation. Therefore, the following studies employ time cuts that eliminate these hits along with the beam structure by carefully choosing a time window that only contains the main triggered events.

Finally, this time structure can be used to estimate the recovery time required for a channel to be ready to receive a second hit. From Fig.~\ref{fig:AP}, the sharp peak around -250~ns corresponding to the triggered event hits has a width of about 100~ns. Therefore, we estimate it to be of order 200~ns, a safe period of time following the triggered event.

\subsection{Channel Response}
Channel response across the whole detector provides detail on readout channel uniformity. The data collected also allow for a direct comparison between the different signal processing paths of the CITIROC. This is a very important step in validating the calibration procedure.

\inserttwographs{0.49}{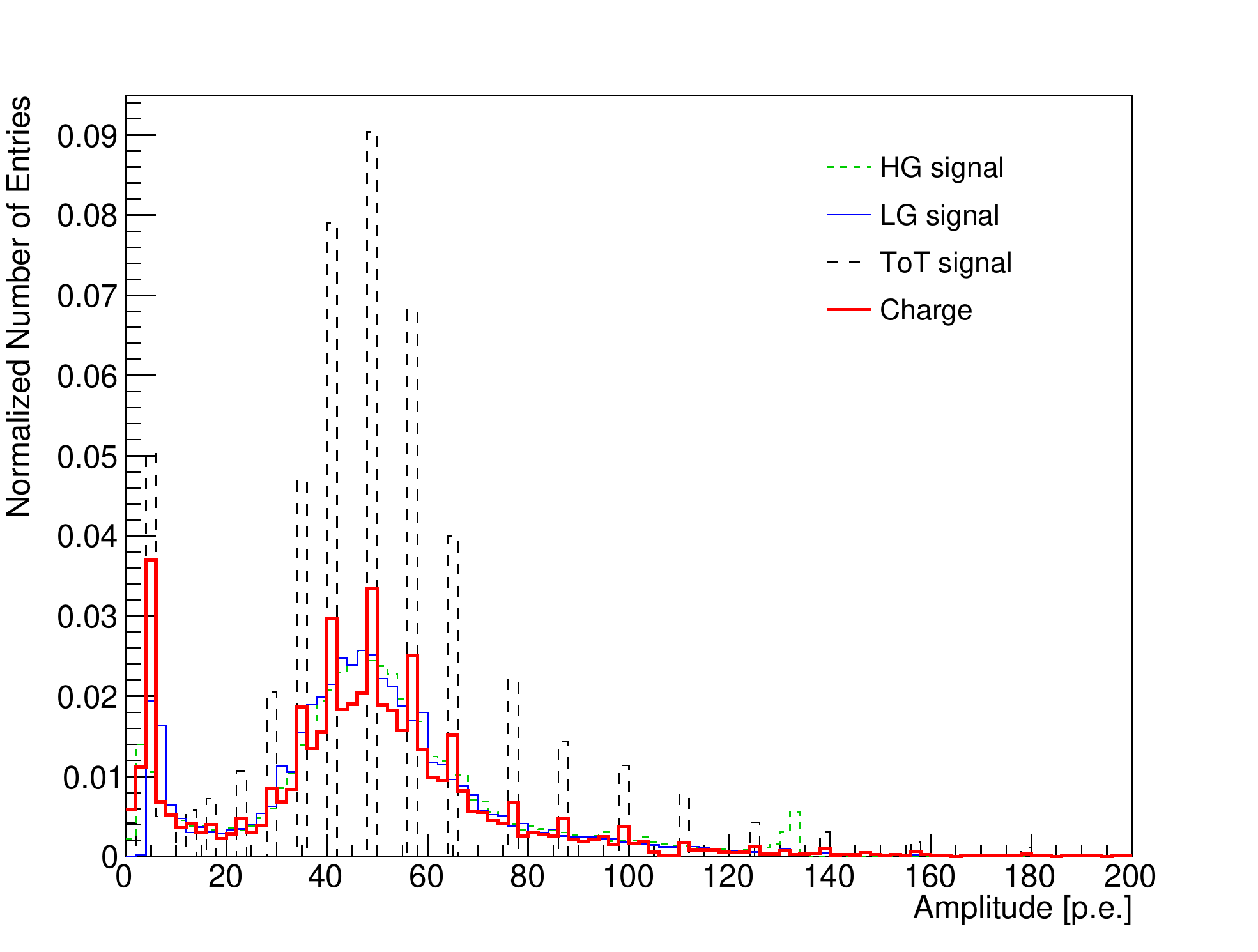}{0.49}{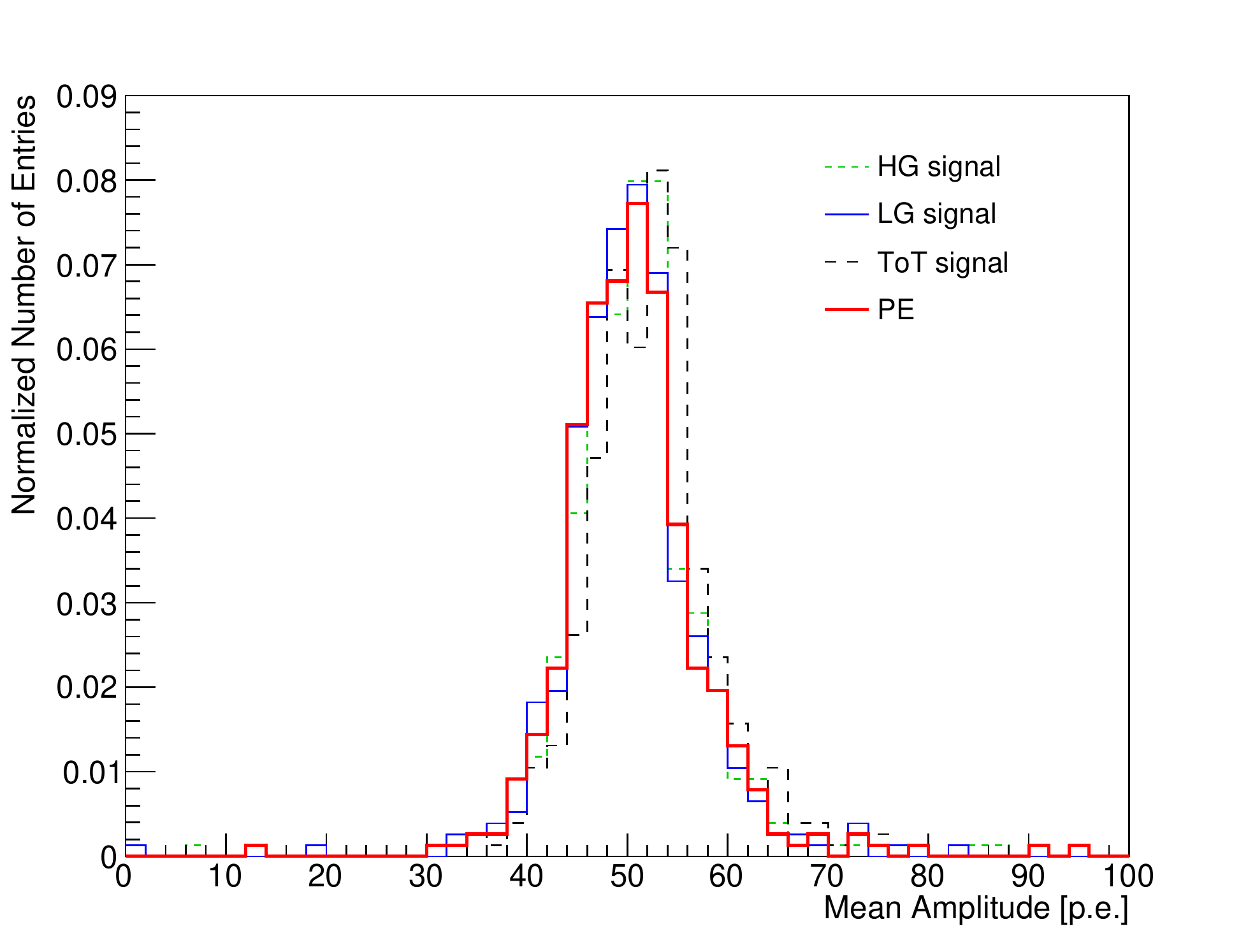}{Left: Response of one readout channel to minimum ionizing particles. This readout channel is connected to a 24~cm fiber, Type I MPPC. Right: Mean light yield for 384 readout channels connected to 24~cm fibers, Type I MPPCs. The three CITIROC signal outputs are shown, HG, LG and ToT, along with PE which is a combination of these.}{fig:channel_response}

A study into channel uniformity is shown in Fig.~\ref{fig:channel_response}. The left plot in the figure shows amplitude distributions as recorded by the HG, LG, ToT signals as well as the combined hit PE for one readout channel connected to a 24 cm fiber, Type I MPPC, illustrating the typical response of all channels. There is no correction for attenuation in the fiber, so the distributions integrate different hit positions in different cubes along the fiber. The discrete nature of the ToT distribution is due to the ToT hit sampling with 2.5~ns resolution. 

The right-hand plot of Fig.~\ref{fig:channel_response} shows the distribution of mean amplitudes for each channel of the 384 channels connected to 24 cm fibers. The four different signal types show similar mean values and standard deviations, which is an indication that calibration parameters were correctly applied to the data. A summary of light yields for different MPPC types and fiber lengths is given in Table~\ref{tab:channel_light_yield}. 
\begin{table}[htbp]
\centering
\begin{tabular}{@{}lllllll@{}ccccccc}
\toprule
\textbf{Fiber} & \textbf{MPPC} & \textbf{\# of MPPCs} & \textbf{HG [p.e.]} & \textbf{LG [p.e.]} & \textbf{ToT [p.e.]} & \textbf{PE [p.e.]}\\
\midrule
24 cm & Type I & 384 & 50.40 & 50.15 & 51.93 & 50.16\\
8 cm & Type I & 192  & 52.78 & 52.02 & 54.26 & 52.53\\
8 cm & Type II & 128 & 51.61 & 51.12 & 52.23 & 51.56\\
8 cm & Type III & 64 & 43.23 & 40.18 & 42.84 & 42.14\\
\bottomrule
\end{tabular}
\caption{Mean light yield obtained from the different readout electronics signal paths for the different fiber lengths and photosensor types instrumenting the SuperFGD Prototype.}
\label{tab:channel_light_yield}
\end{table}
The MPPC Type I connected to 24~cm fibers shows a lower light yield compared to those connected to 8~cm fibers due to higher attenuation in the WLS fiber, as expected. The Type III MPPCs record the lowest light yields, possible due to worse geometrical matching to the WLS fiber area given they are smaller (Table~\ref{tab:mppcs_on_prototype}).

\subsection{Optical Crosstalk Between Adjacent Cubes}
\label{subsec:crosstalk}
Cube-to-cube optical crosstalk occurs when light produced in a particular cube travels to a neighboring cube. This phenomena happens because the cubes' surfaces are not completely opaque. Measuring how much light is shared among neighboring cubes is crucial. A large amount of cube-to-cube optical crosstalk significantly complicates the 3D reconstruction because of the increasing ambiguities in the 2D to 3D matching. If the spread of light among cubes is too large, the fine granularity of the detector can be compromised. On top of that, optical crosstalk measurements are key to describe and simulate the detector response.

\begin{figure}
	\centering
	\includegraphics*[width=0.495\textwidth]{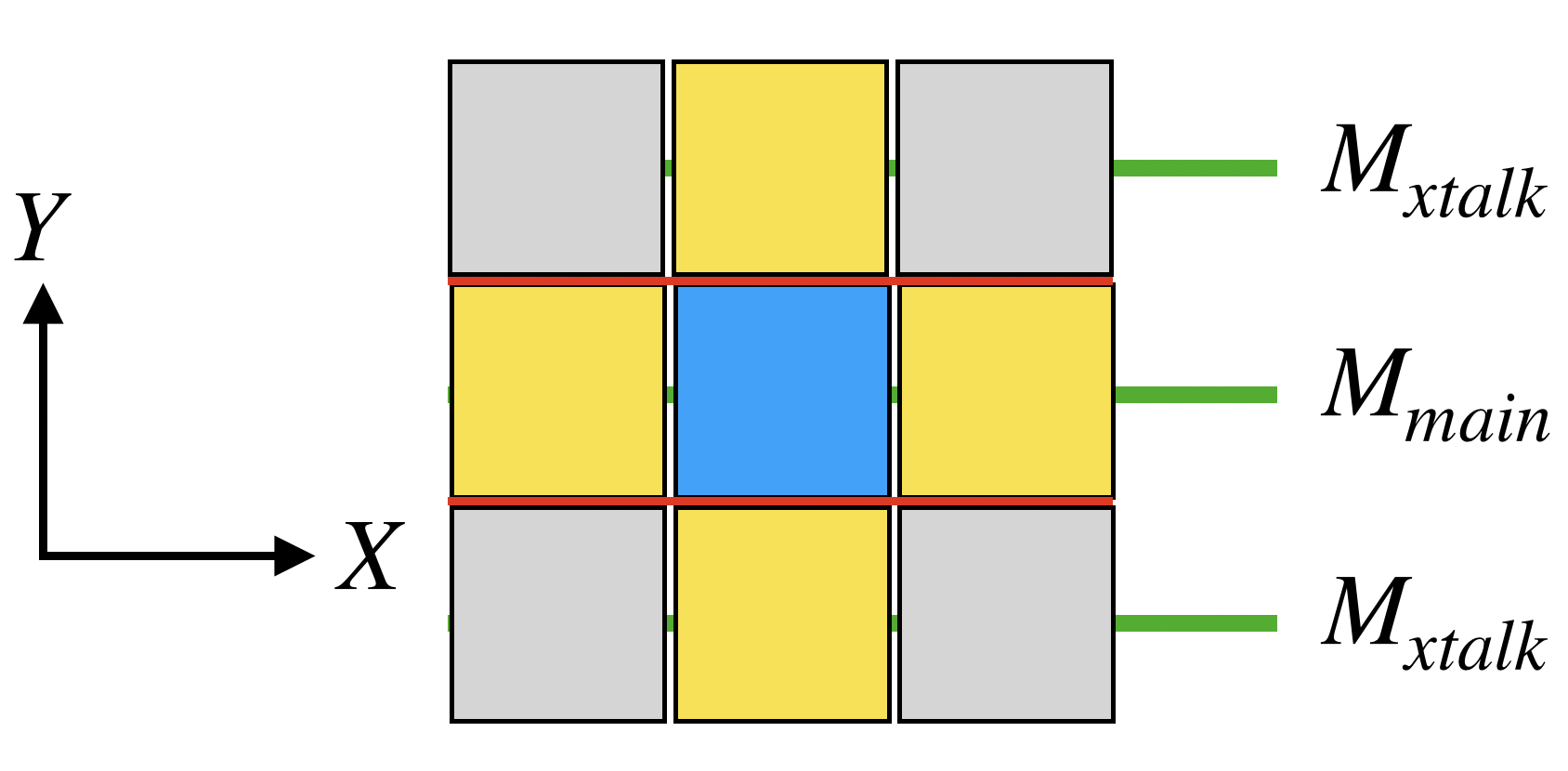}
	\caption{Sketch illustrating the main cube hit (blue) and 4 crosstalk cubes (green) in the $x$-$y$ plane, $x$-fibers are shown, $y$-fibers are not shown. The optical fiber $M_{main}$ that collects light from the main cube hit also collects light from 2 adjacent crosstalk cubes.}
	\label{fig:xtalk_sketch}
\end{figure}

\begin{figure}
	\centering
	\includegraphics*[width=0.495\textwidth]{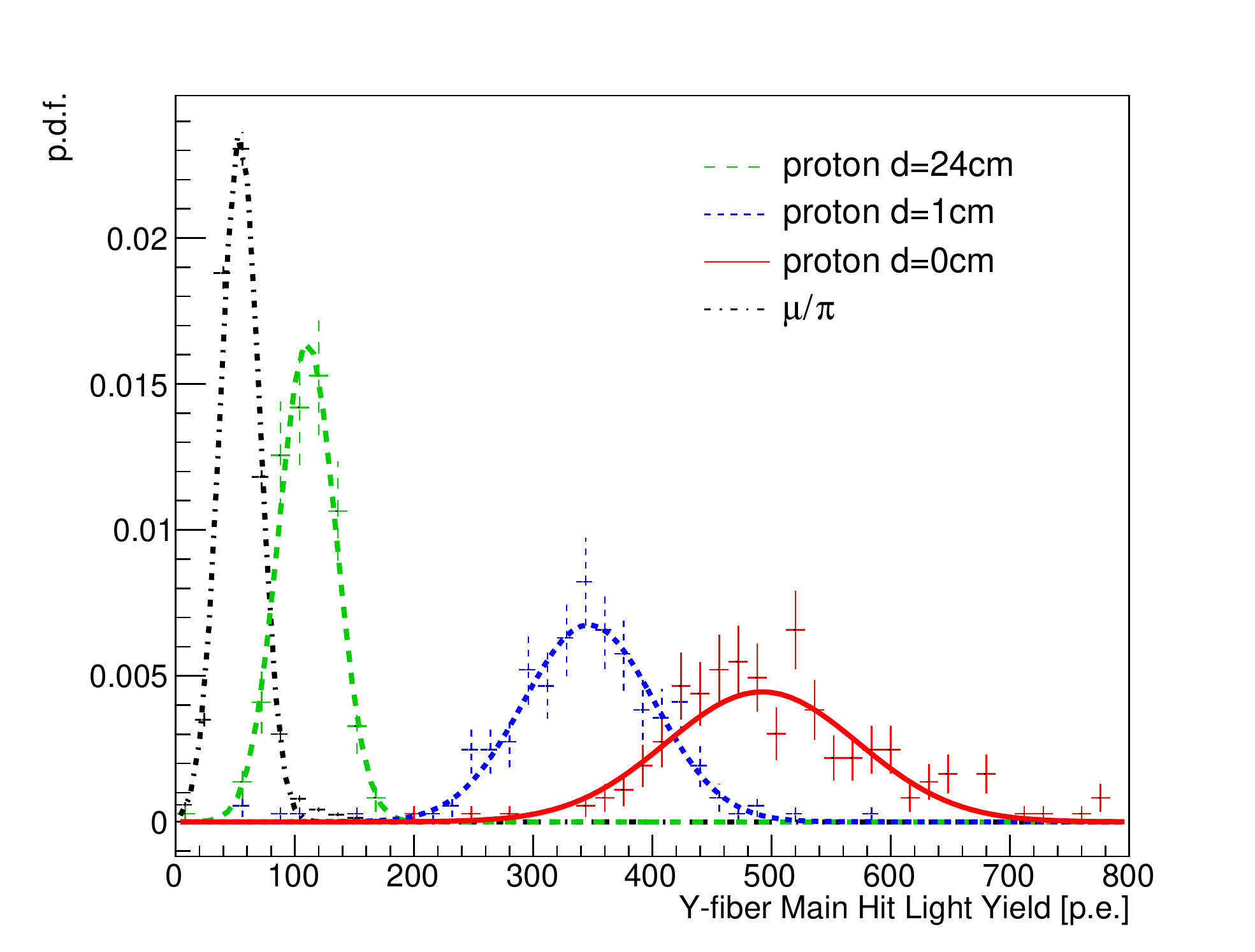}
	\hfill
	\includegraphics*[width=0.495\textwidth]{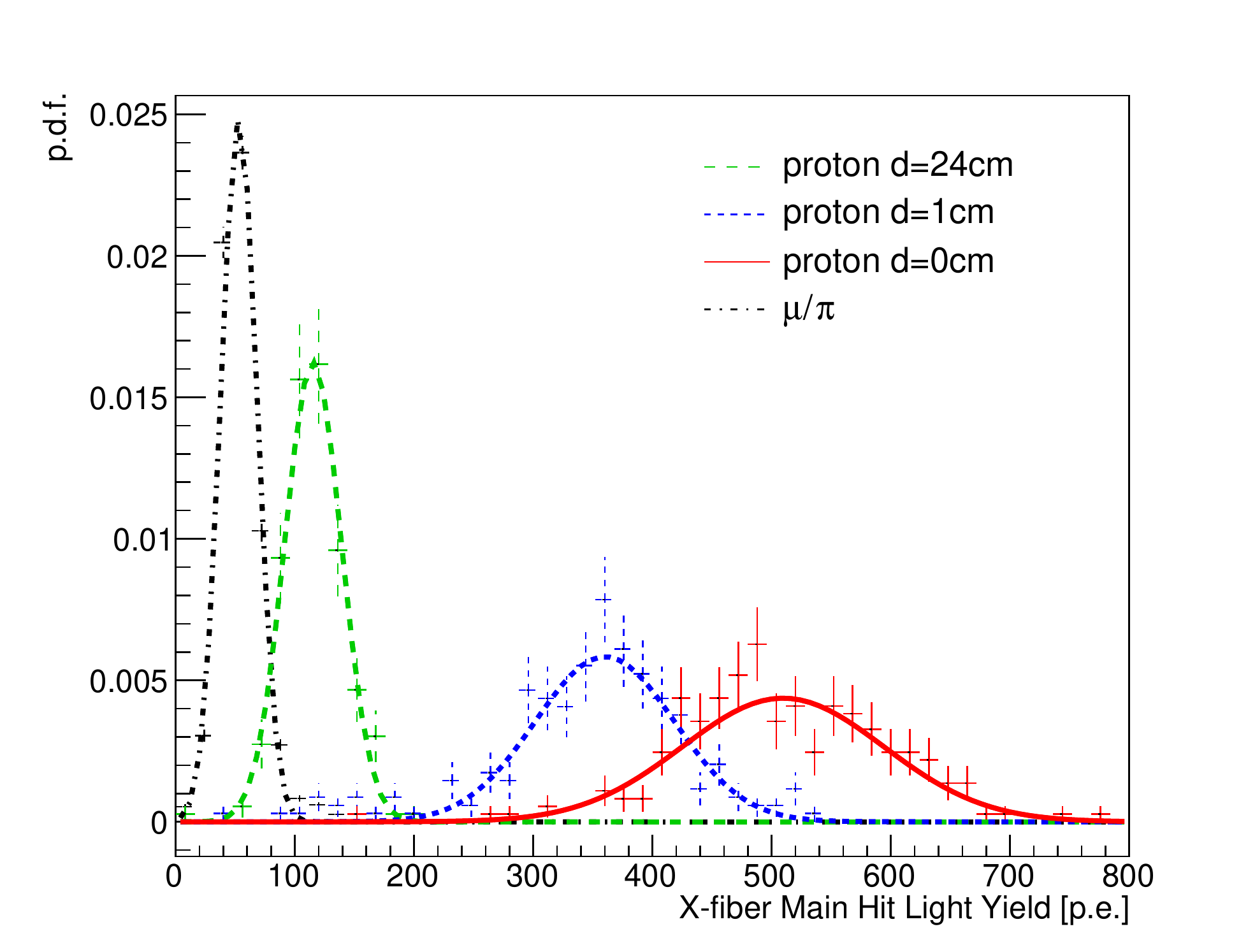}\\
	\includegraphics*[width=0.495\textwidth]{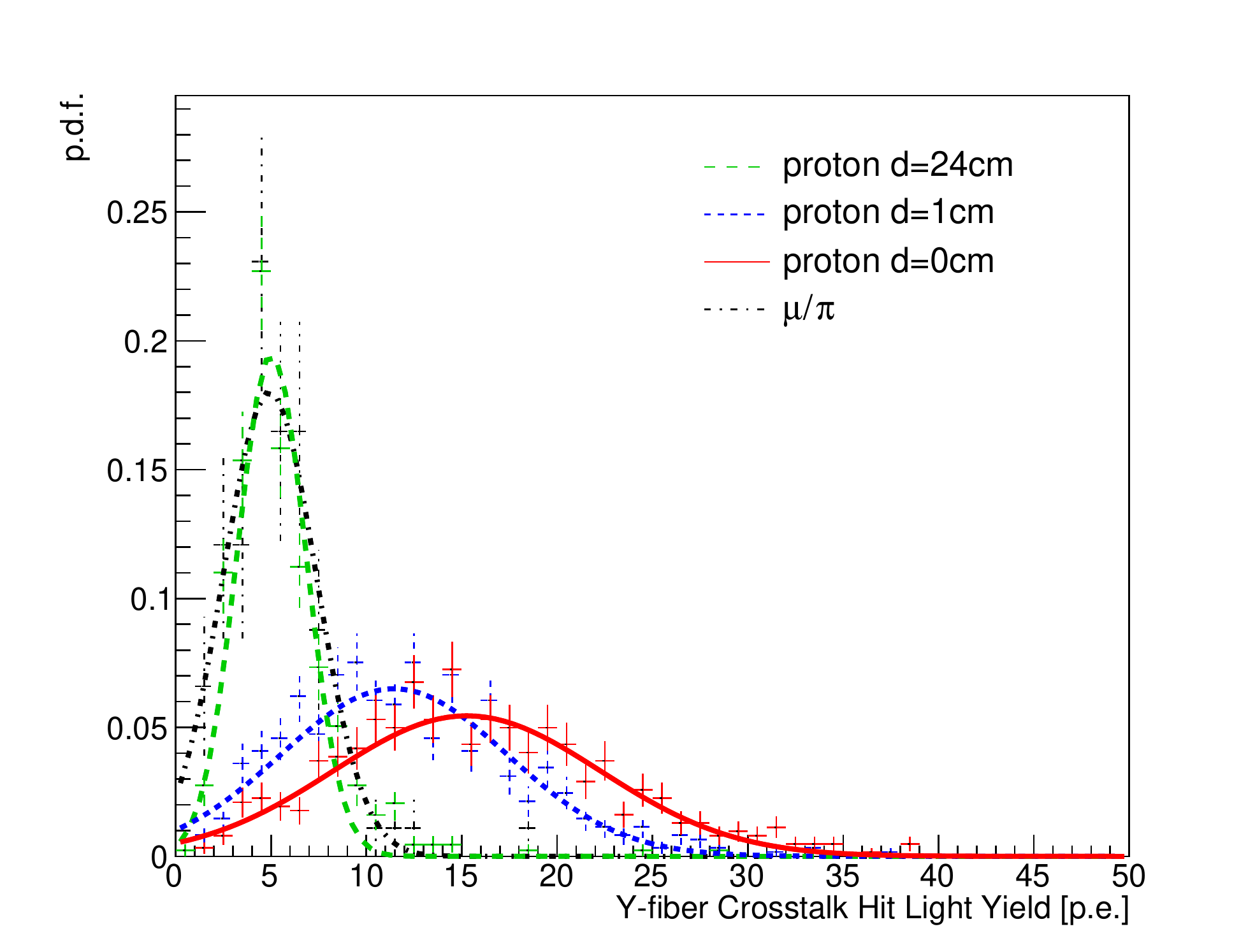}
	\hfill
	\includegraphics*[width=0.495\textwidth]{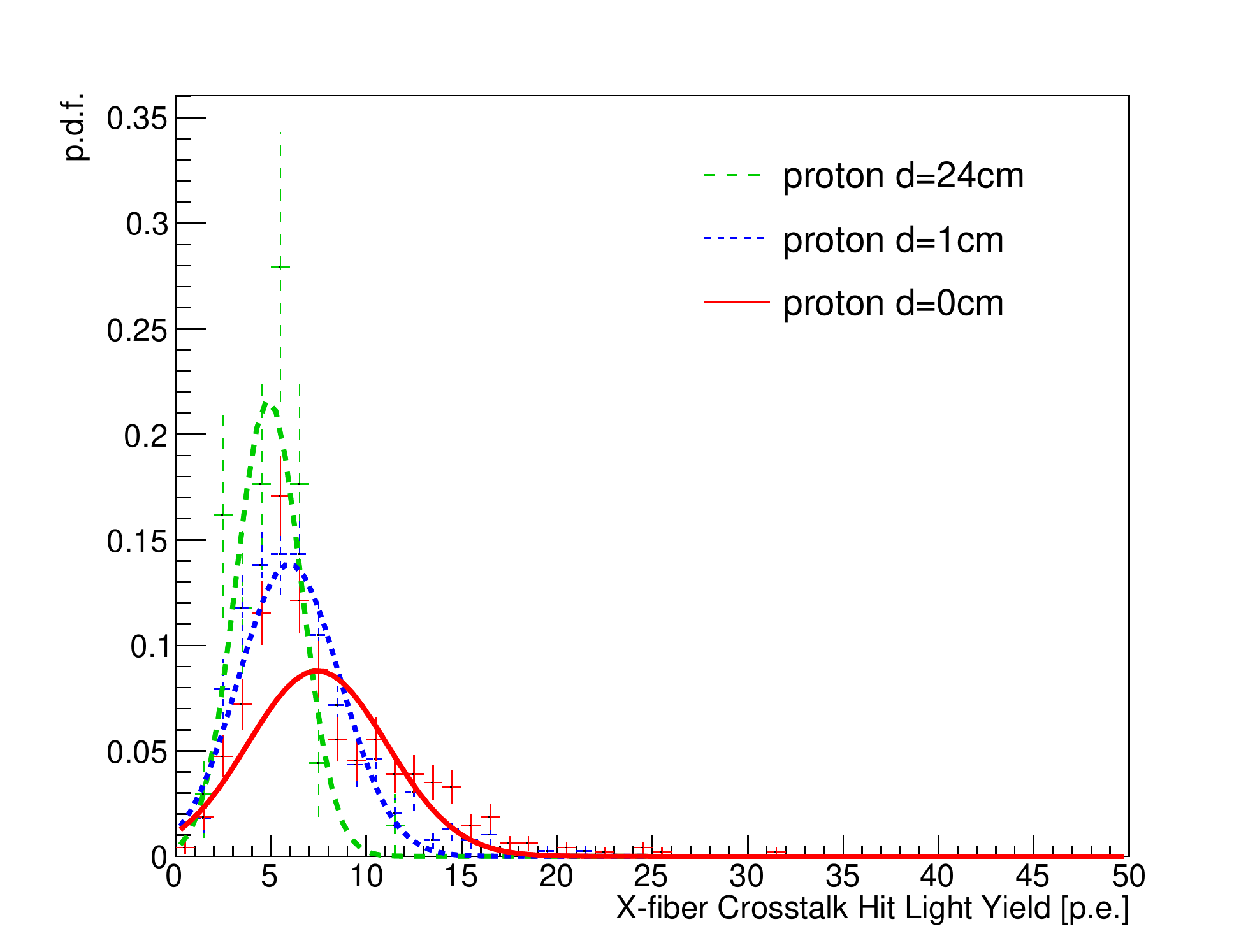}
	\caption{Light yields measured using fibers perpendicular to the track propagation for main hits and crosstalk hits. Distributions from both the 0.8~GeV/$c$ proton and 2.0~GeV/$c$ $\mu$/$\pi$ triggers are shown. The selected protons stopped within the prototype volume, leaving deposits of different energies along their tracks. The light yield for these protons is measured at three different distances $d$ from the stopping point: at 0~cm, 1~cm and 24~cm. The bottom right plot does not include $\mu$/$\pi$ crosstalk p.d.f., since very few crosstalk hits were recorded for this sample.}
	\label{fig:hit_LY}
\end{figure}

As mentioned in Sec.~\ref{subsec:SuperFGD_cubes_assembly}, a layer of Tyvek light insulator was placed between each $x$-$z$ plane of cubes. The beam was directed parallel to the $z$-axis of the detector, hence the detector sees an asymmetric optical crosstalk response along $x$ and $y$ fibers. In our study of optical crosstalk, only perfectly straight tracks are selected allowing hits with more than 20~p.e.\ (more than 40~p.e.\ for protons) to be found only in a single $x$ and $y$ coordinates.  In this way, the $x$ and $y$ coordinates of the cubes crossed by the beam tracks are known. For simplicity, these cubes will be referred to as ``main cubes" and cubes only containing energy deposits from optical crosstalk will be named ``crosstalk cubes", an sketch of this is provided in Fig.~\ref{fig:xtalk_sketch}. 
The light yield distributions for crosstalk and main cubes measured along $x$ and $y$ fibers is presented in Fig.~\ref{fig:hit_LY}. 

To compute the percentage fraction of light, $\kappa$, that exits the main cube through each side, we need to compare the main cube measurement, $M_{\textup{main}}$, to its neighboring crosstalk cube measurement, $M_{\textup{xtalk}}$. $\kappa$ is not simply the ratio $M_{\textup{xtalk}}/M_{\textup{main}}$, since it is necessary to correct the main cube light yield so that it also includes the light that escaped into neighboring cubes. In total, and neglecting $\mathcal{O}(\kappa^2)$ contributions, the light flows to six adjacent cubes and is recovered a total of four times: two times from the crosstalk cubes in the fiber direction and two times from the crosstalk contributions from the surrounding main cube neighbors. Therefore,
\begin{align}
\label{eq:kappa}
\kappa = \frac{M_{\textup{xtalk}}}{M_{\textup{main}}+2M_{\textup{xtalk}}}.
\end{align}
Here we assume that the crosstalk light yields originating from a single cube are all equal. To estimate $\kappa$ in an unbiased way, it is necessary that $M_{\textup{main}}$ and $M_{\textup{xtalk}}$ sample the real probability distribution function (p.d.f.) of the light deposits. As discussed in Sec.~\ref{sec:thresholds}, the different electronic thresholds limit the minimum number of photons that we can detect from the MPPCs. This limit deforms the light yield p.d.f. in the low light yield region and therefore to accurately compute $\kappa$ we need to use the highest possible crosstalk deposits. Hence, the best estimator for $\kappa=2.94\pm0.05\%$, shown in Fig.~\ref{fig:kappa},  is obtained for the stopping point of protons given that its light yield is generally above the electronics thresholds, as it is clear by comparing Fig.~\ref{fig:thresholds} w.r.t Fig.~\ref{fig:hit_LY}.

The case is not the same for $y$-fiber measurements. To quantify the optical crosstalk suppression achieved by the use of Tyvek sheets, we measured $\kappa$ along $z$-fibers where the crosstalk light is accumulated along the full length of the detector, producing signals far above the threshold, as shown in Fig.~\ref{fig:LY_XY}.
\inserttwographs{0.49}{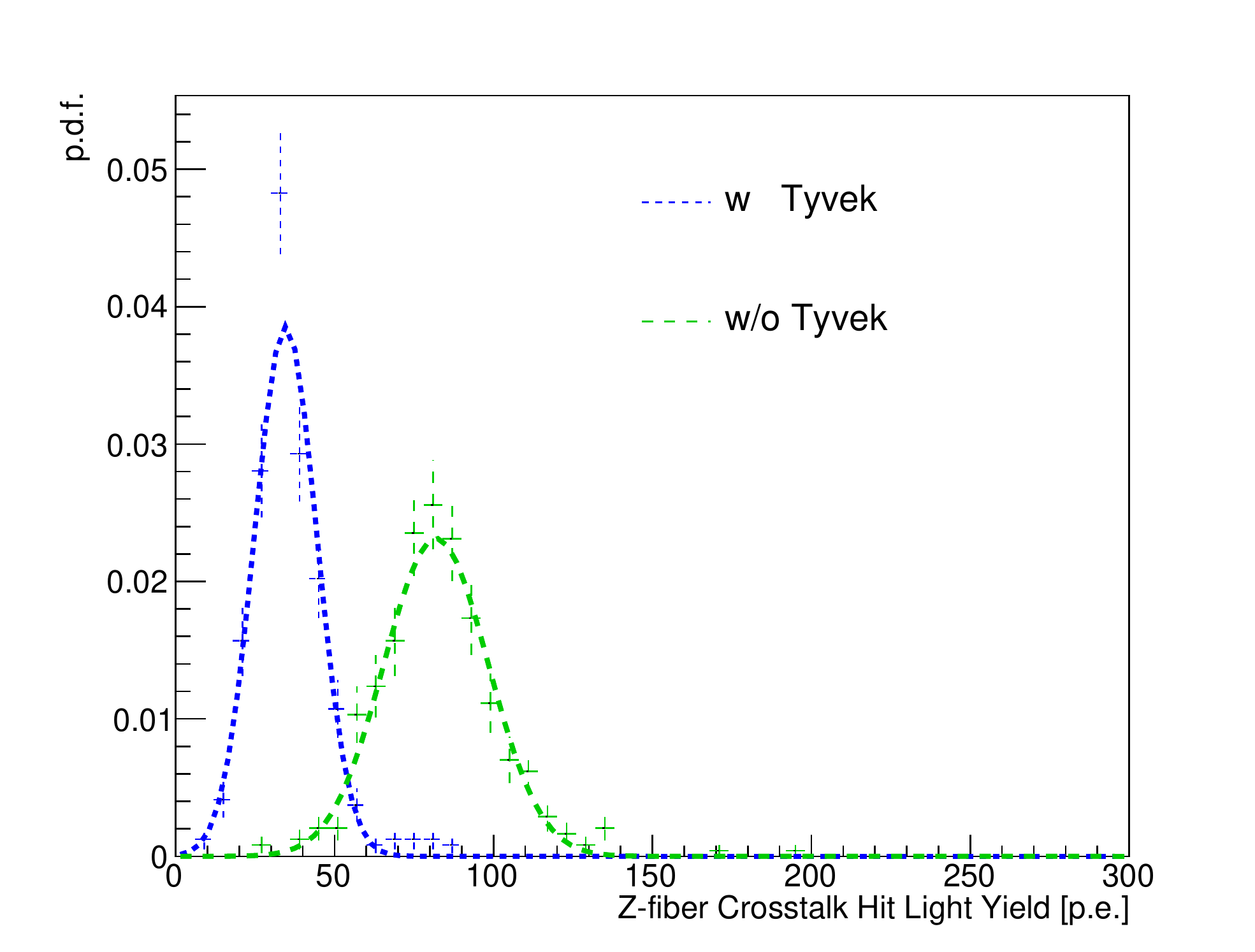}{0.49}{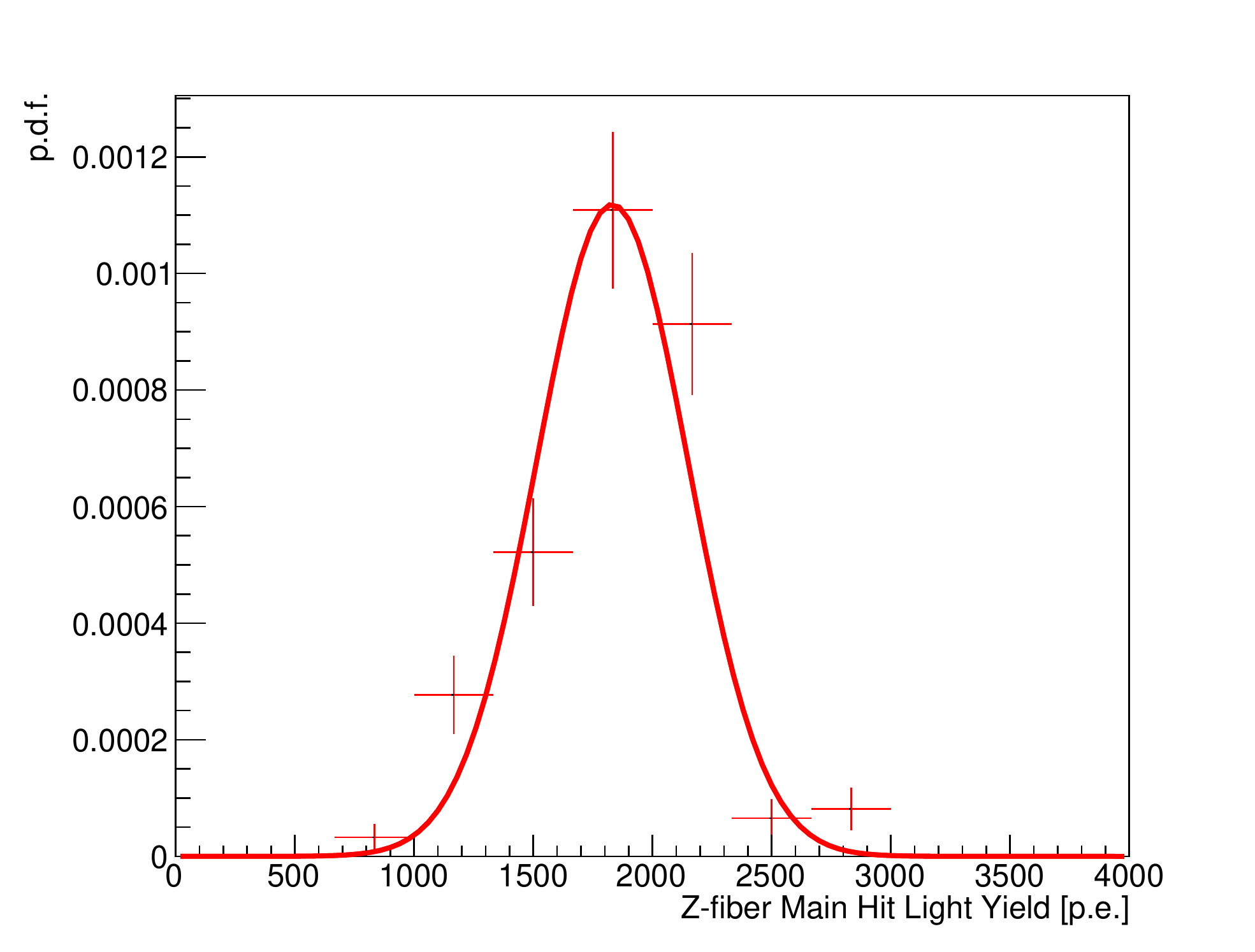}{Light yield for $z$-fiber main and crosstalk hits using 2.0~GeV/c data with $\mu$/$\pi$ trigger. For the crosstalk samples, the inputs have been classified as coming from cubes insulated by Tyvek or with no Tyvek insulation with respect to the main cube.}{fig:LY_XY}
This does present a limitation: along $z$-fibers the accumulated collected light in all the main deposits creates a very high light yield signal. The response of the sensitive elements saturates for such large deposits, making the main hit measurement smaller than it should be and biasing $\kappa$ towards larger values. Correcting for this effect is not possible without a dedicated characterization of the photosensors and electronics response at very high light yields, which is not the goal of this study. Instead, we compare the value of $\kappa$ using the saturated MPPC hits and the vertical or horizontal crosstalk hits in the $x$-$y$ plane. The ratio between both $\kappa$ estimates, presented in Fig.~\ref{fig:kappa}
\inserttwographs{0.49}{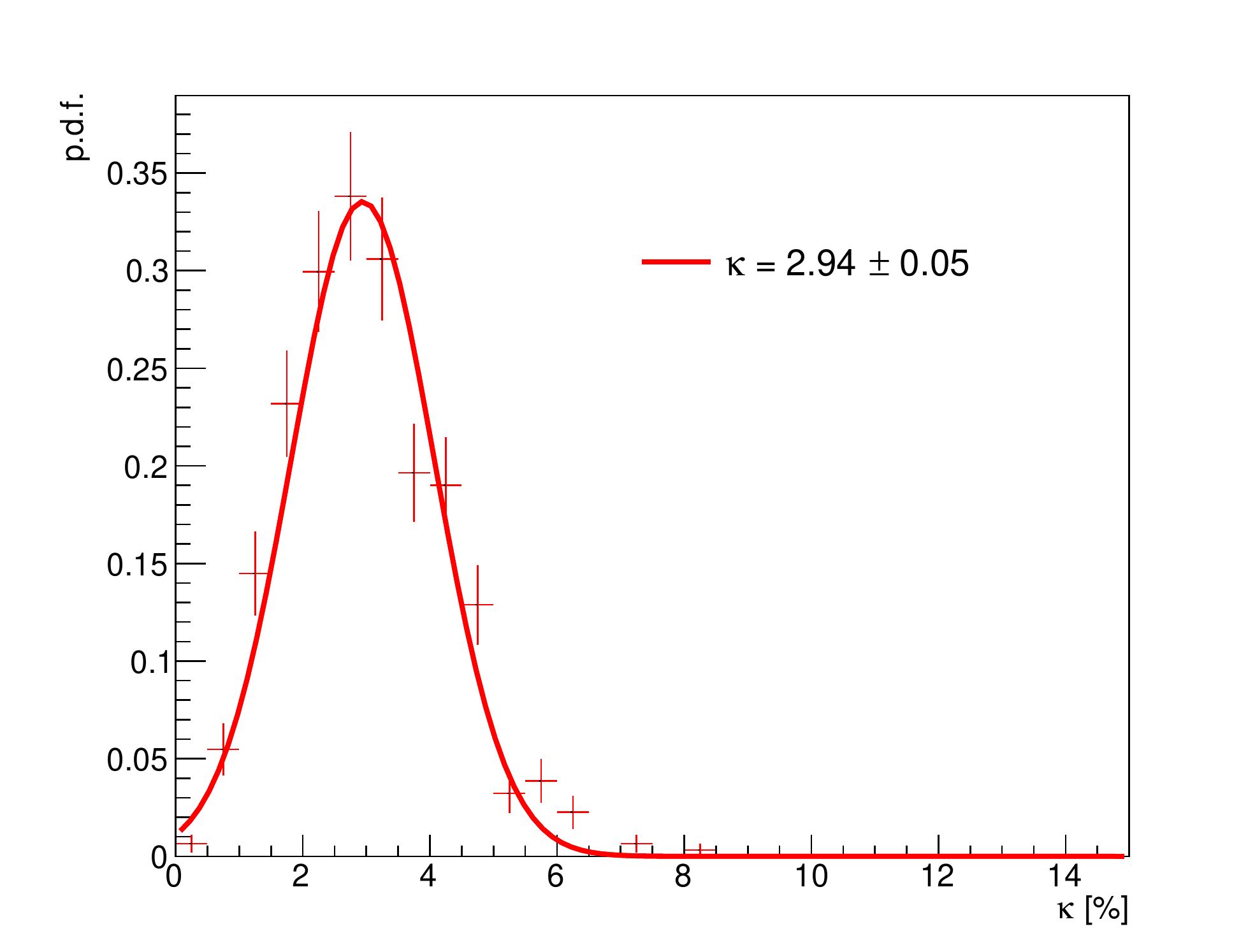}{0.49}{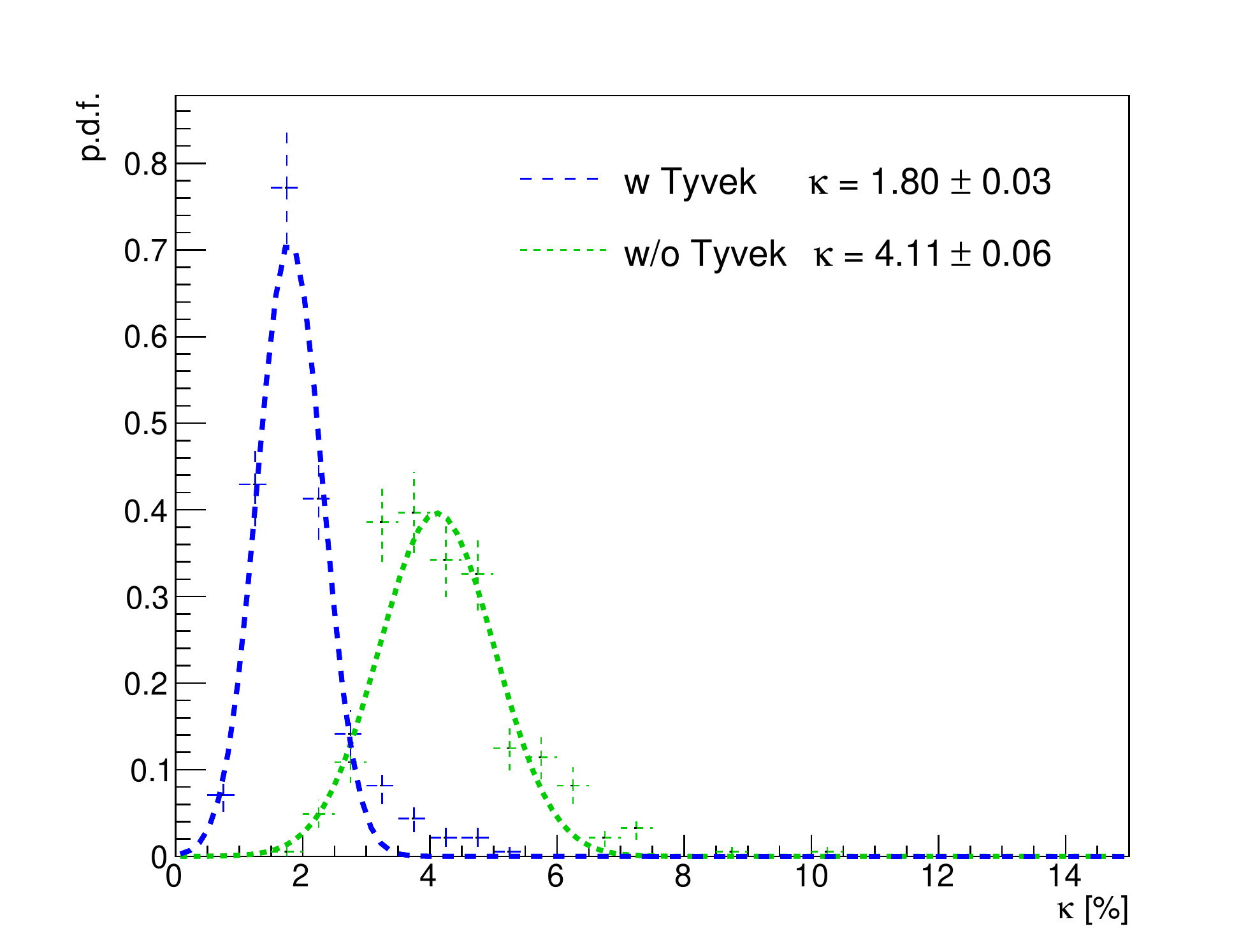}{Measurement of the percentage fraction of light $\kappa$ that flows from main cubes to neighbor crosstalk cubes using equation~\ref{eq:kappa}. The left plot corresponds to the best estimate of $\kappa=2.94\pm0.05$ made using the stopping point of proton events. The right plot shows the optical insulation effect of Tyvek sheets.}{fig:kappa}
,  allows to estimate in 50--60$\%$ the crosstalk reduction achieved thanks to Tyvek. For this measurement, 2.0~GeV/$c$ tracks with a $\mu$/$\pi$ trigger, instead of stopping protons, have been used in order to alleviate the saturation effects.

\subsection{Light Attenuation in WLS Fiber}
\label{sec:atten}

Signal photons are attenuated while traveling inside the WLS fibers. The attenuation process is fit using the equation
\begin{align}
\label{eq:att}
y(d) = LY_0\left(\alpha e^{\frac{-d}{L_S}} + (1-\alpha)e^{\frac{-d}{L_L}}\right),
\end{align}
where $LY_0$ is the unattenuated light yield, $\alpha$ is a weighting factor, and $L_S$ and $L_L$ are respectively short and long attenuation constants and $d$ is the distance from the center of the cube with the main energy deposit to the side of the detector the MPPC detecting the signal is on, including a $2.3mm$ offset. $L_L$ is known to be 4 m from the manufacturer specifications and $L_S$ depends on the specific length of the fiber. The measurement of the light attenuation in 24~cm and 8~cm long fibers is presented in
Fig.~\ref{fig:att_XZ}.

\insertgraph{0.5}{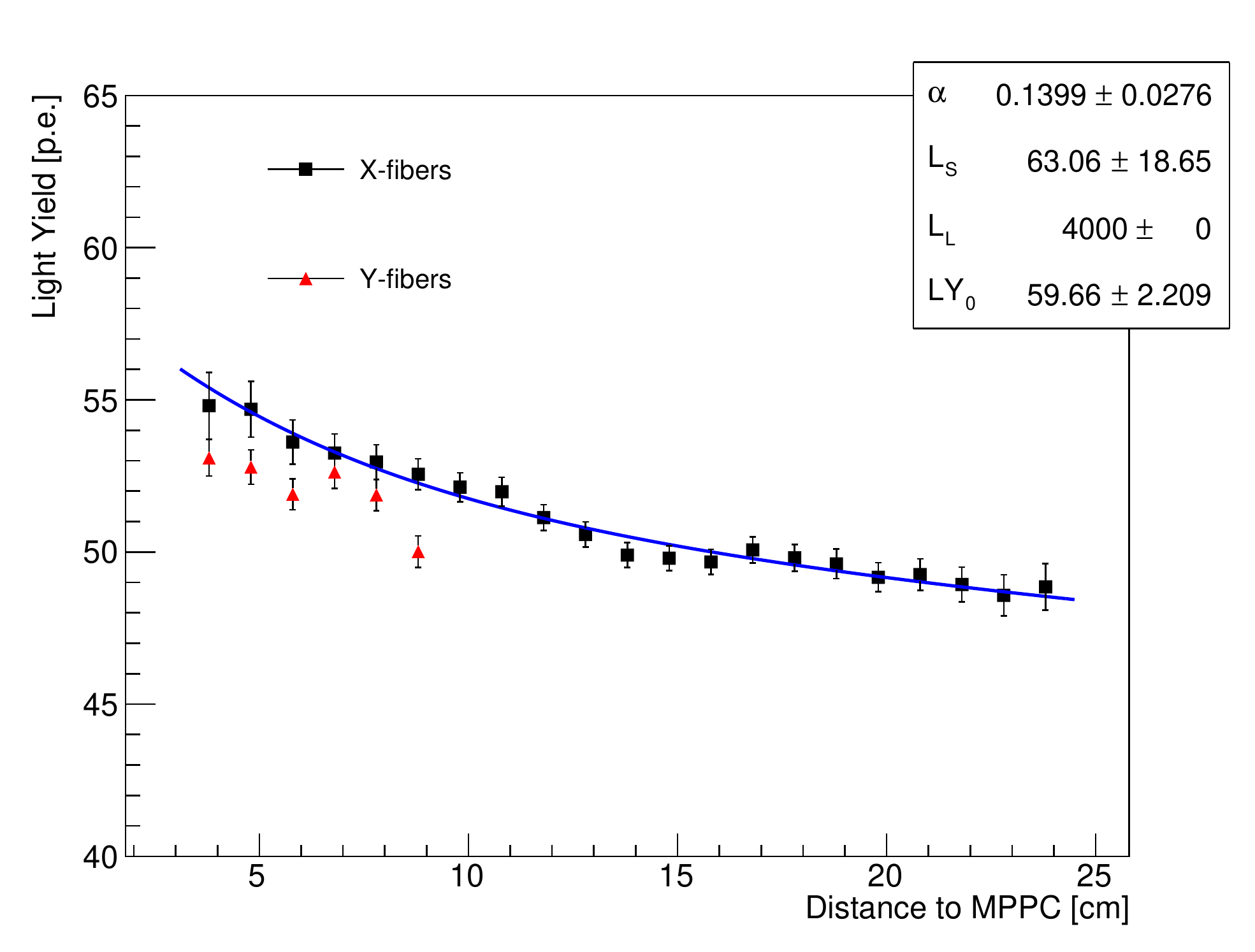}{Light attenuation in 8~cm and 24~cm long fibers. The line shows the fit to $x$-fibers data using Eq.~\ref{eq:att}. The $L_L$ parameter is fixed to 4~m according to manufacturer specifications.}{fig:att_XZ}

The data in Fig.~\ref{fig:att_XZ} are obtained by selecting straight tracks parallel to $z$-fibers so that $x$ and $y$ are known for all hits. The measurement is done by averaging the data of all channels in the $x$-$z$ ($y$-$z$) plane for each specific $y$-distance ($x$-distance). The fit results show the best parameters to be $LY_0=59.7\pm2.2$, $L_S=63.1\pm18.6$, and $\alpha=0.14\pm0.03$.

This method can be used in the future detector to characterize the attenuation with the fibers in-place using cosmic tracks. With high statistics, a measurement of the attenuation in individuals fibers might be possible.

\subsection{Cube Response}
 
A study on the CERN beam test data was performed to check the individual response of cubes. Uniformity in light yield across the volume of a single cube is necessary for reliable particle identification. Potential failures in the WLS fiber and SiPM or presence of impurities in the scintillator could introduce a source of non-uniformity. Hence, checking for inconsistent signals is an excellent screening for faulty hardware. Moreover, analysis tools used for this study are the same as those for the determination of hit position, the first step towards event reconstruction.

For this study, the cube from which a signal originated had to be identified. Studying the hit time of an individual channel in one projection is not sufficient to provide information about the cube position along the WLS fiber. There is a spread in photon arrival time at the photosensor due mostly to processes local to the energy deposition site such as the scintillation process in the cube and the wavelength conversion process in the WLS fiber. There is also a time delay due to light propagation down the WLS fiber. These time-dependent processes, along with the processing of the photosensor's electrical signal by the CITIROC preamplifier and shaper, exclude the possibility of resolving the hit position along the fiber based on the hit profile of a single channel with a resolution better than 10~cm. It is therefore impossible to resolve the main cube hit from adjacent cubes that might collect a fraction of photons from optical crosstalk. That type of distinction can only be made by studying the coincidence of hits in two or more projections.

0.8 GeV/$c$ muons were used in the study. Straight tracks were selected to minimize differences in track length through each cube. The average light yields for a number of cubes with the SuperFGD prototype are shown in Fig.~\ref{fig:cube_light_yield_atten}. 
\inserttwographs{0.495}{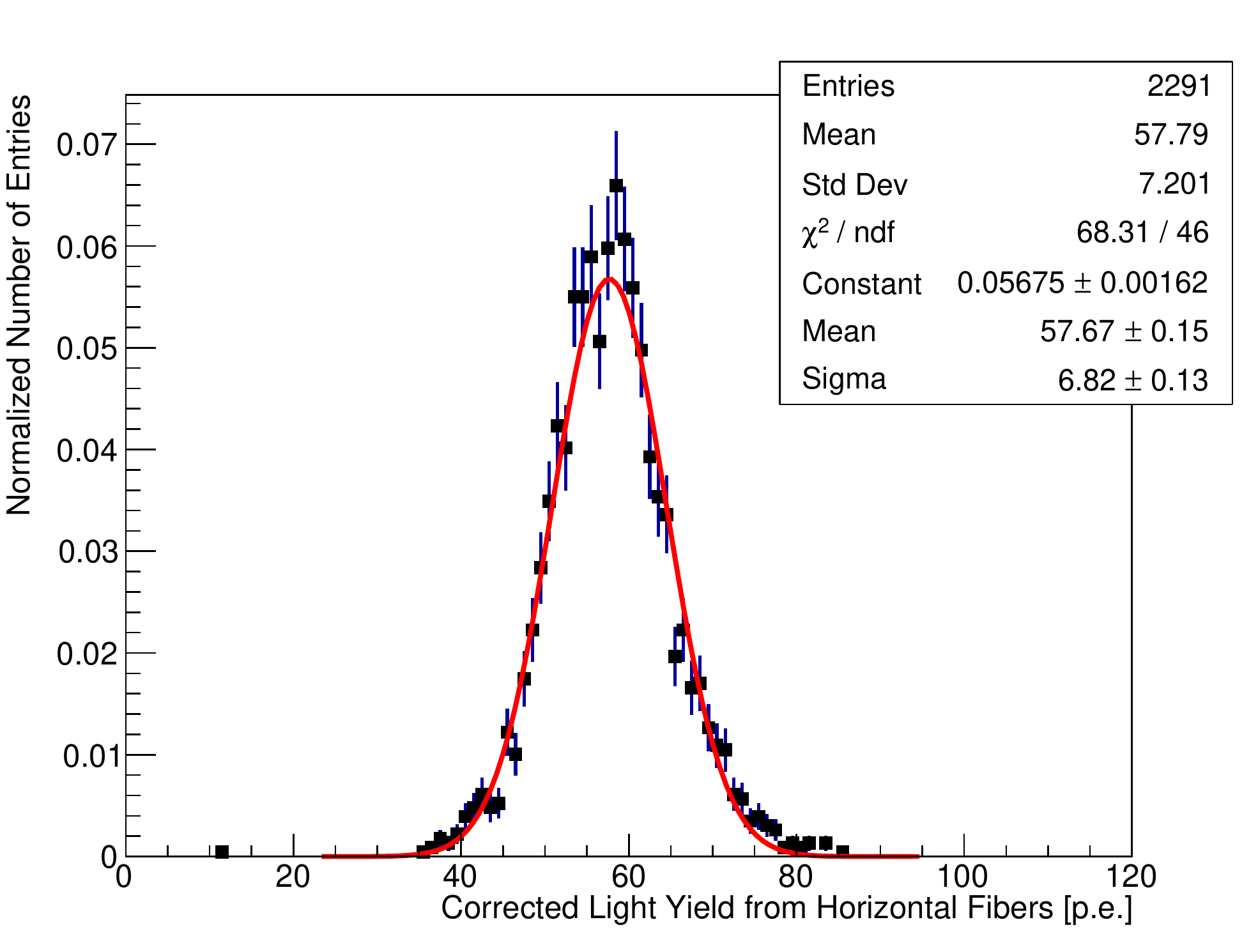}{0.495}{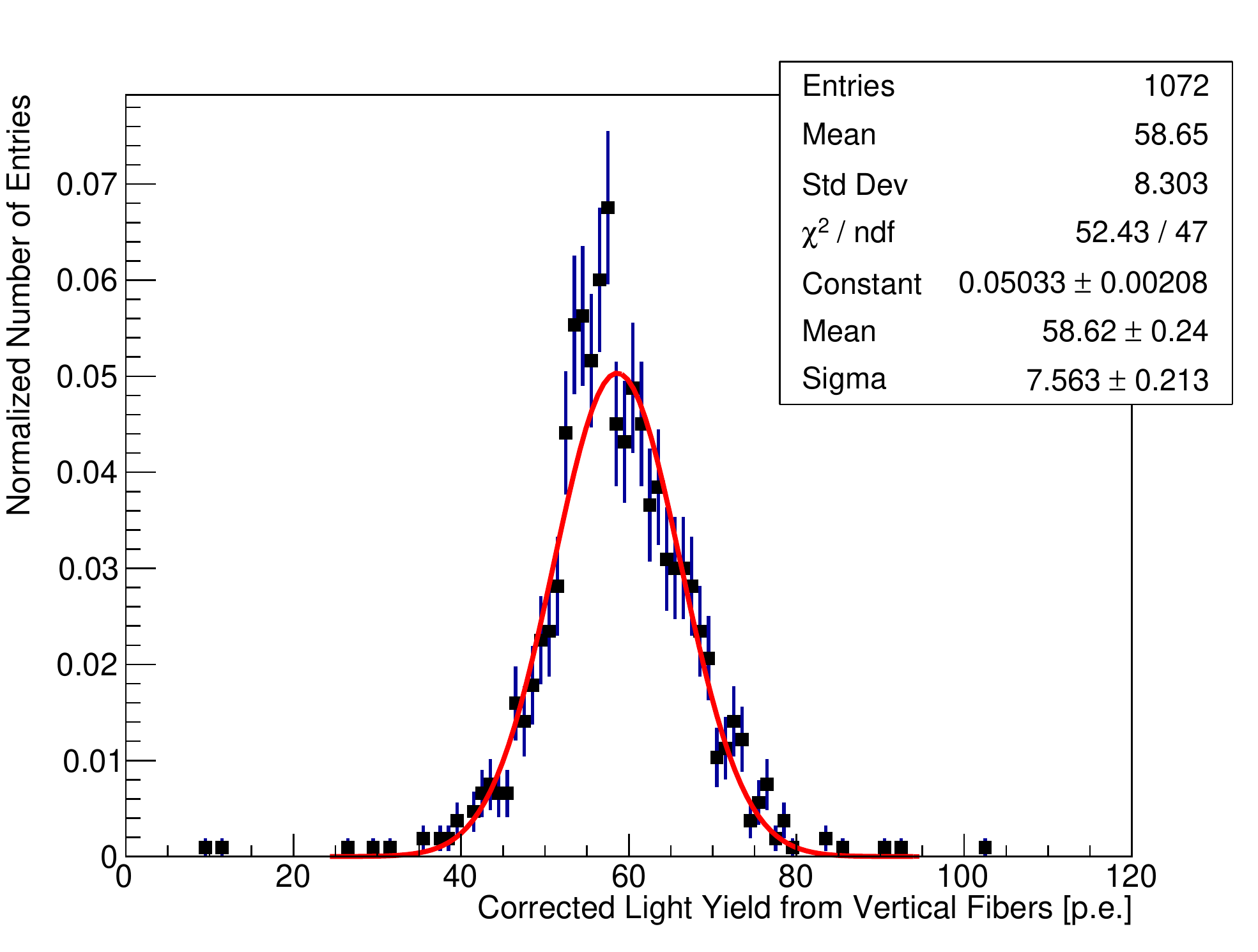}{Left: cube light yields for 2,291 cubes read out by horizontal fibers. Right: cube light yields for 1,072 cubes read out by vertical fibers. Signal attenuation is corrected for. The mean light yields along vertical and horizontal fibers are very similar after the corrections.}{fig:cube_light_yield_atten} Each light yield has been corrected for light attenuation using Eq.~\ref{eq:att} and the parameters quoted in Sec.~\ref{sec:atten}. The light yields have also been separated by fiber orientation: vertical (8~cm) and horizontal (24~cm). The distance from the cube where the signal originated to the MPPC that detected the signal was found by analyzing individual $z$-layers of the detector and finding the hit with the largest amplitude in the $x-z$ and $y-z$ plane. We assume the signal originated at the cube with the $x$, $y$ and $z$ coordinates indicated by these two hits. By selecting the hits with largest amplitude for each $z$-layer, we eliminated crosstalk signals from this analysis.

After attenuation corrections, average light value for both the horizontal and vertical fibers were 58~p.e.\ and 59~p.e.\ respectively, with standard deviations of 7~p.e\ and 8~p.e. No corrections were made to account for the crosstalk picked up by the fibers or the effect of the Tyvek sheets between horizontal cube layers. As such, we would expect the average value of the corrected light yields for cube signals read out by vertical fibers to be slightly lower than those read by horizontal fibers, as the Tyvek sheets will block crosstalk passing into other cubes along the vertical fiber. We do not see this with our fitted means, though there are some overly populated bins for the vertical fiber histograms between 50--60~p.e., which are not accounted for in the Gaussian fit.

\subsection{Time Resolution}

The time resolution of a channel is the time interval around a single time measurement from a channel that we are confident the measured event occurred in. This value is influenced by several factors, the dominant ones being the time variation in the scintillation and wavelength shifting processes and the readout electronics response. 

In order to measure the time resolution of a channel, one can compare hit times in the channel to other hit times from the event. Here, an event is defined as a particle passing through the detector. For a valid comparison, a time reference consistent in position with respect to the detector is used. Standard practice is to use the trigger time as a reference, however, the trigger used for the beam test had a significantly coarser time resolution than the detector's channels and is unsuitable as a reference. Instead, for each event, the hit time in the 24$^\mathrm{th}$ $z$ layer was used. The analyzed data was taken with the beam directed along the $z$-axis, so after crosstalk is removed there should only be one hit in the $z=24$ layer. If there was more than one hit in this layer, then the hit with the highest charge deposit was used. Other time reference candidates, such as the average event hit time and the earliest event hit time, were tested, but it was found that using a time reference from a fixed layer in the detector helped avoid variation from MPPC type and FEB to FEB differences.

To ensure consistent hit times, the time-walk effect was corrected for in this study. This refers to a dependence of registered hit time on the hit amplitude as it crosses the applied hit threshold, with smaller signals being recorded later than larger signals. It was corrected for by fitting the relationship between hit charge and hit time since trigger for electron beam data with the SuperFGD Prototype.

The method to measure a channel's time resolution was as follows: 
\begin{itemize}
 \setlength\itemsep{0mm}
 \item events were selected so that only muons traveling along the $z$-axis were used in the analysis;
 \item in each selected event, every measured hit time was corrected for time-walk and travel time along the WLS fibers;
 \item a histogram of ($T_\mathrm{ref} - T_\mathrm{hit}$), where $T_\mathrm{ref}$ is the reference time and $T_\mathrm{hit}$ is the hit time measured in the channel being analyzed, was built up for each channel using all the selected events;
 \item a Gaussian distribution was fitted to the histogram if there were more than 250 entries;
 \item the standard deviation of the fitted Gaussian was then taken to be the combined time resolutions of the time reference and the channel;
 \item to find the channel time resolution, it was assumed the time reference had the same time resolution as the channel, so the standard deviation was divided by $\sqrt{2}$.

\end{itemize}

Using the above method with data from a 2~GeV $\mu^+$ beam directed along the $z$-axis of the prototype, time resolution values for 854 of 1,728 channels were measured. Channels at the extreme $x$ values in the $x$-$z$ plane were rarely triggered, seeing as the beam was directed down the center of the detector. Hence, not all channels had enough statistics to give a reliable time resolution measurement. The channels that were evaluated mostly give consistent values and are a good indication of the consistent performance of the detector as a whole. 
\begin{figure}
\centering
 \includegraphics[scale=0.6]{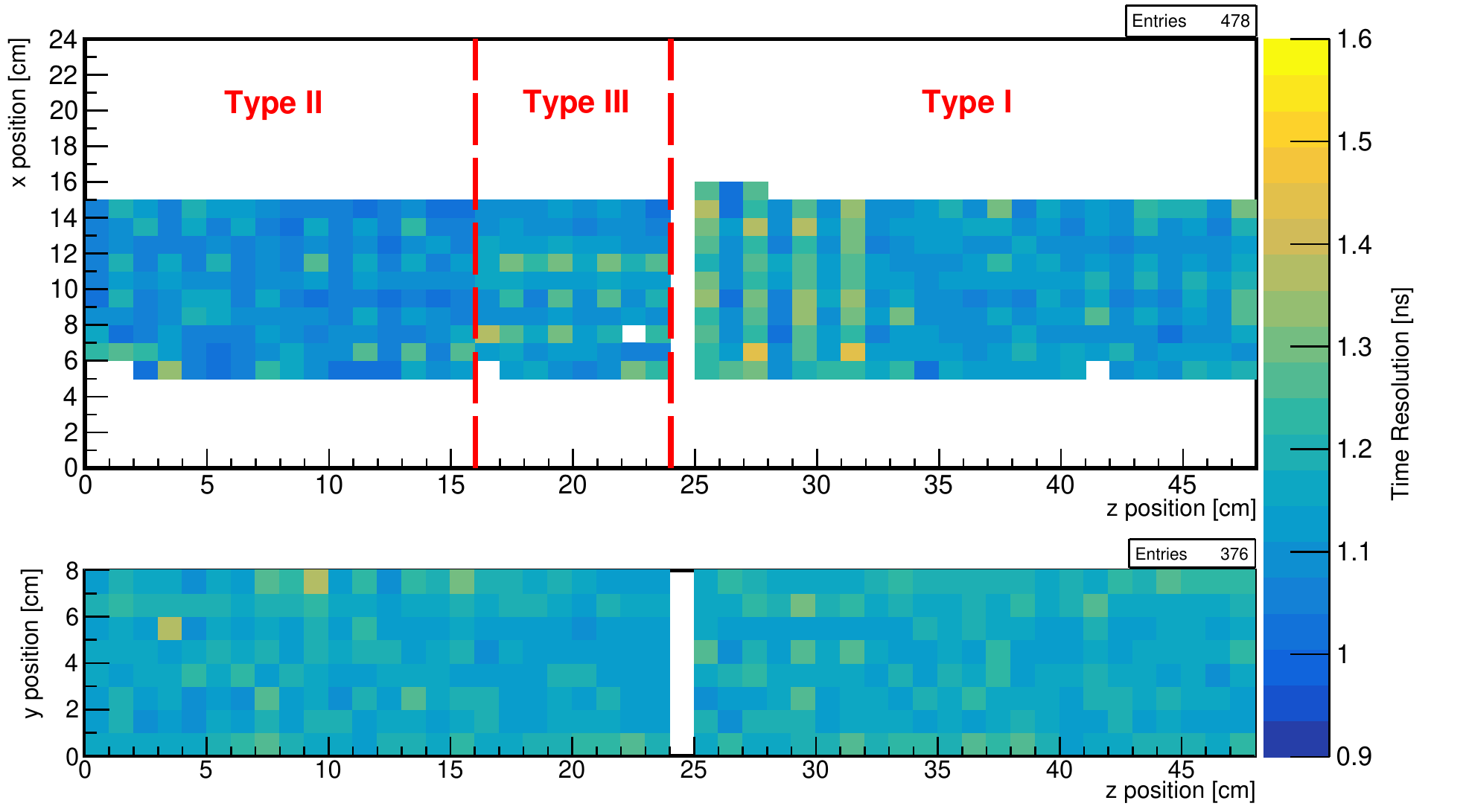}
 \caption{A mapping of measured time resolution values to their corresponding channels. Color corresponds to the time resolution value. The top plot shows channels in the $x$-$z$ plane and the lower plot shows ones in the $z$-$y$ plane. The dashed red lines separate different MPPC types. All channels in the $z$-$y$ plane use Type I MPPCs. There is a noticeable group of outlying values in the $x$-$z$ plane that correspond to channels read out by FEB 18.}
 \label{fig:map}
\end{figure}

\begin{figure}
 \begin{subfigure}{.495\linewidth}
  \centering
  \includegraphics[width=\textwidth]{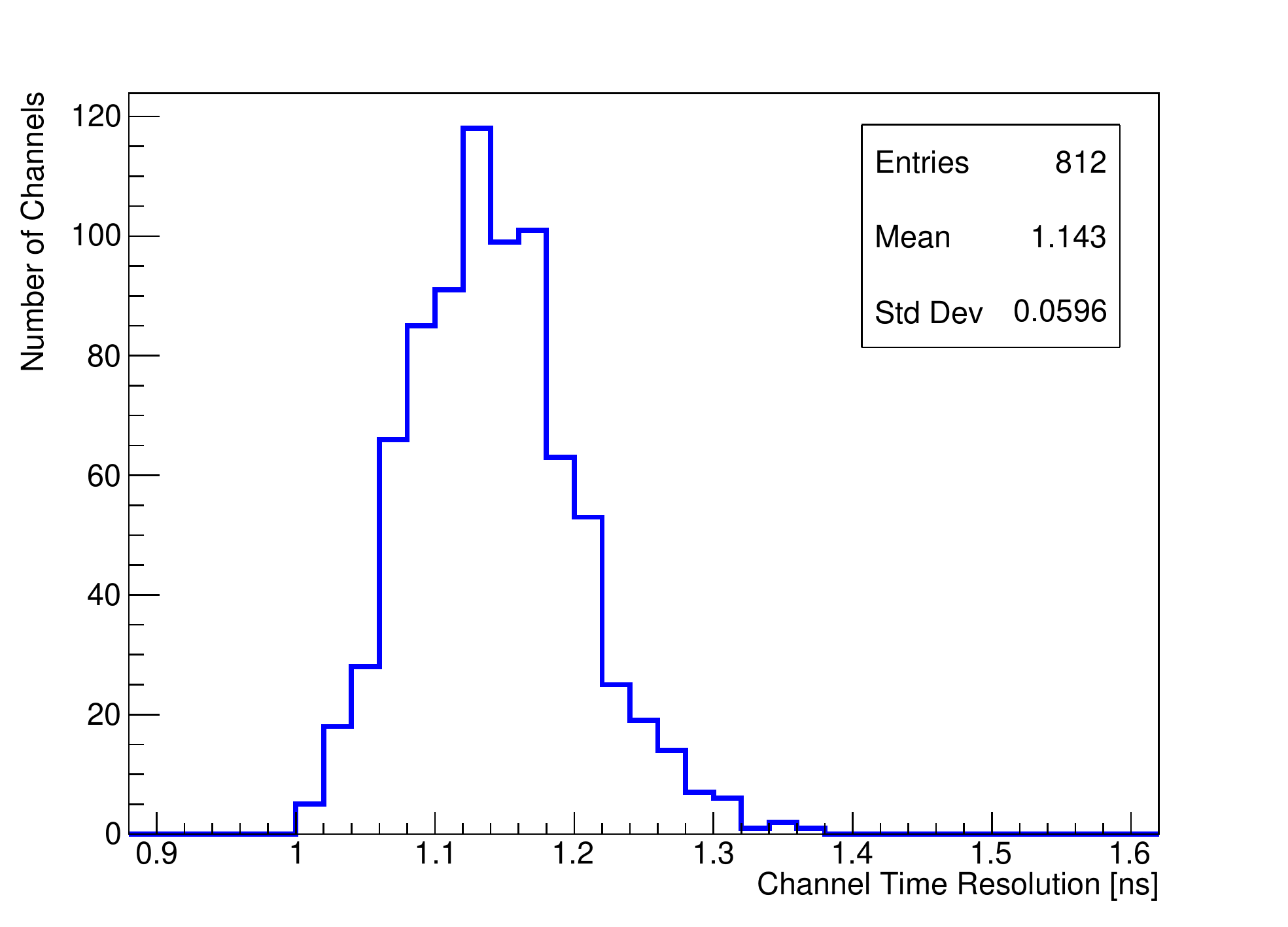}
  \caption{}
  \label{fig:sub1}
 \end{subfigure}
 \begin{subfigure}{.495\linewidth}
  \centering
  \includegraphics[width=\textwidth]{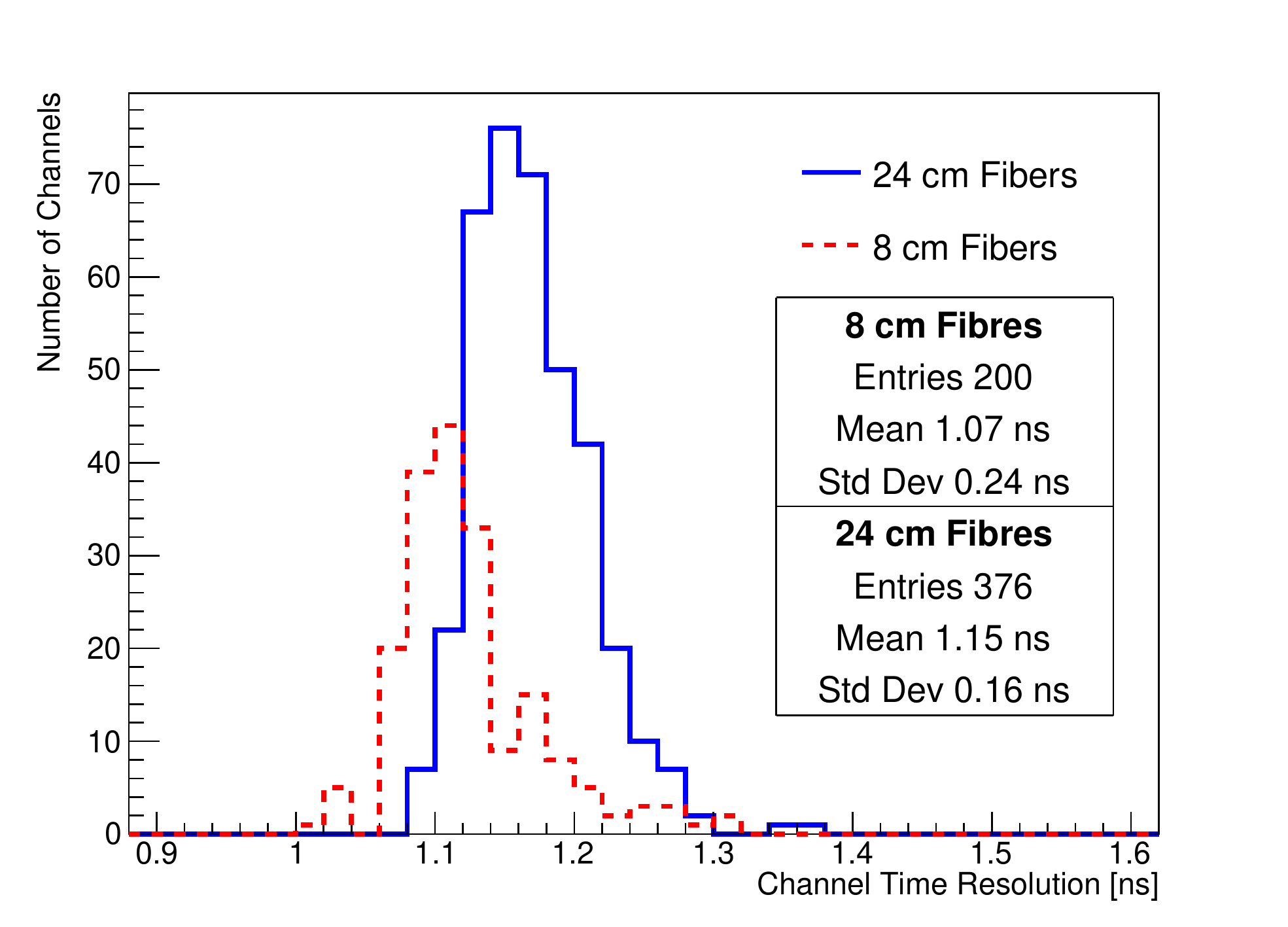}
  \caption{}
  \label{fig:sub2}
 \end{subfigure}\\[1ex]
 \begin{subfigure}{\linewidth}
  \centering
  \includegraphics[width=0.495\textwidth]{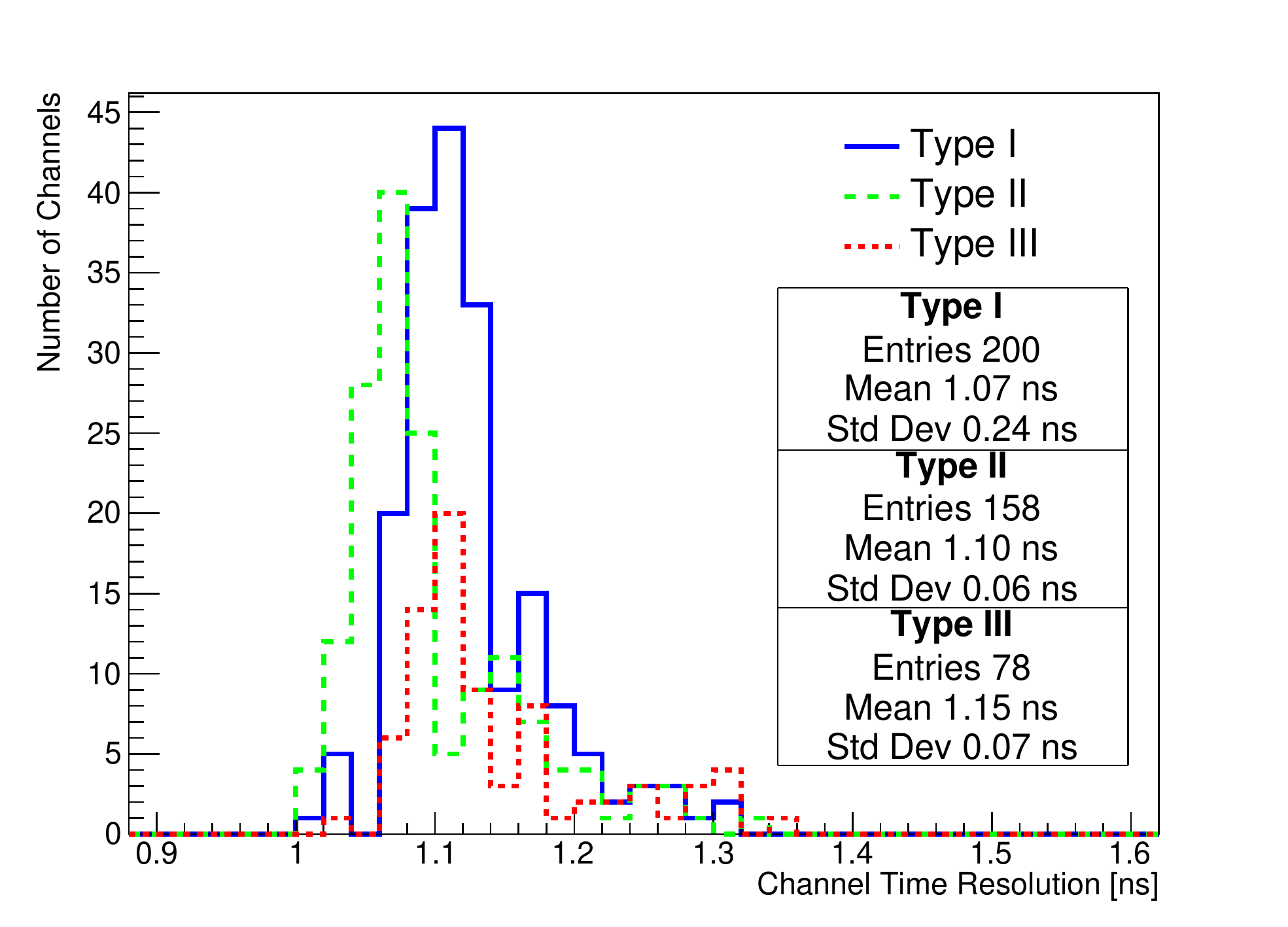}
  \caption{}
  \label{fig:sub3}
 \end{subfigure}
 \caption{Histograms of the measured channel time resolutions. Channels read out by FEB 18 are excluded. (a) shows a single histogram containing all the channels with an associated time resolution. (b) displays two histograms, the blue solid line corresponds to channels read out by short fibers (channels in the $x$-$z$ plane) and the red dashed line corresponds to channels read out by long fibers (channels in the $z$-$y$ plane). Only channels with Type I MPPCs are shown. (c) splits the channels into three by separating them by MPPC type. Green solid, blue dotted and red dashed lines correspond to Type I, Type II, and Type III MPPCs respectively. Only channels with 8~cm fibers are shown.}
 \label{fig:res}
\end{figure}

The channel map in Fig.~\ref{fig:map} shows that, despite most of the channel time resolution values being consistent, there is some visible structure to the $x$-$z$ map between $25 < z < 32$. In this region, channels read off by FEB 18 appear to have larger time resolutions than the average (by about 0.2~ns). The cause of this is as yet unknown, as data from FEB 18 looks otherwise normal. The most likely cause is a different response of FEB 18 to the clock and synchronization signals sent by the Master Clock Board. The channels from FEB 18 are ignored for the remainder of this section.

Fig.~\ref{fig:res} shows distributions of the measured channel time resolutions. Fig.~\ref{fig:sub1} shows the distribution for all the measured channels, except FEB 18 channels. The distribution is Gaussian-like, with a mean value of 1.14~ns and a standard deviation of 0.06~ns. Previous measurements using a 5~GHz sampler with the 125 cube prototype of the SuperFGD yielded an average time resolution of roughly 0.95~ns. Measuring this with such a fast sampler shows that the scintillation and wavelength shifting processes within the detector limit the minimum time resolution to 0.95~ns. Combine this in quadrature with an expected contribution of 0.7~ns from the readout electronics of the SuperFGD prototype and we predict a time resolution of roughly 1.1~ns for individual channels. This is within one standard deviation of the average value measured: $1.14 \pm 0.06$~ns.

An investigation into whether differences between the detector's readout channels affected their time resolution is shown in Figs.~\ref{fig:sub2} and \ref{fig:sub3}. The first plot shows histograms of the time resolutions for channels that use Type I MPPCs with 8~cm fibers (channels in the $x$-$z$ plane) and 24~cm fibers (in the $z$-$y$ plane). There is a noticeable difference between the two fiber lengths, despite fiber length being corrected for in the analysis. Other factors that could cause this difference in time resolution measurements are the Tyvek layers between horizontal cube layers that could affect the timings observed in orthogonal planes, and also the synchronization of FEBs.

Fig.~\ref{fig:sub3} shows channel time resolution values for each different MPPC type, but only for channels with 8~cm fibers. The mean value of the Type I MPPCs is lowest at 1.07~ns, and the mean for the Type III MPPCs is the highest at 1.15~ns. However, these distributions have tails on their falling edge, with Type II MPPCs having the largest tail. The peak of the Type II MPPC distribution is the lowest at 1.08~ns, whereas the Type I and III MPPCs have peaks around 1.1~ns. It's unclear whether the tails of the distributions are physical or a statistical artifact. Assuming they are a statistical feature and just using the peaks of the distributions, the Type I and III MPPCs have similar time resolution values and the Type II MPPCs give slightly better time resolutions.

As a remark, this study has been made by using only muons at a single momentum value with an ionization close to that of a MIP. Given that the time resolution is dependent on the time of arrival of the first detected photons tracks of different momemtum or particle type, such as protons, may produce more light and hence the time resolution may improve for such particles. 


\section{Physics Studies}
\label{physicsStudies}

\subsection{Simulations of the SuperFGD}
Simulations for the SuperFGD were created within the upgraded ND280 framework. The geometry of the SuperFGD detector is simulated in GEANT4~\cite{AGOSTINELLI2003250} by creating replicas of the $10\times10\times10$~mm$^3$ cubes.  

In order to simulate some aspects of the detector response, several effects were taken into account and applied to the GEANT4 output. These effects include the response of the plastic scintillator (e.g. saturation effects), the response of the fibers (e.g. attenuation), and the response of the MPPCs. The parameters used to introduce these effects were optimized and validated using the beam tests of the SuperFGD prototypes.

\subsection{Stopping Proton Response}
Stopping proton studies provide a great tool to help understand the response of the detector for large energy depositions. At the stopping point, the proton deposits a significant amount of energy in a small volume. This large value of d$E$/d$x$ (here $E$ is the deposited energy and $x$ is the distance traveled) can be used to study crosstalk, MPPC and scintillator saturation, and quenching effects. 

In this section, we show how we isolate a sample of stopping protons using several event selection criteria. The sample is then analyzed to extract some of the characteristics of stopping protons, such as their range in the detector and their d$E$/d$x$ curve. These results are then compared with simulations of events in the SuperFGD Prototype using a proton particle gun.

A data sample of protons with 800~MeV/$c$ beam momentum and 0.2~T magnetic field was selected from the beam test runs. When taking into account the presence of the TPC and other elements in front of the prototype during the beam test, this beam momentum is reduced. Therefore, in order to perform the comparison with simulations, a lower momentum of 750~MeV/$c$ was chosen for the Monte Carlo (MC).

In order to select a sample of stopping proton events, there are several selection criteria that must be considered. A stopping proton must be completely contained within the volume of the SuperFGD. It is expected to deposit a large amount of energy at its stopping point, some of which may leak to neighboring cubes in the form of crosstalk. To make the analysis easier, events with multiple tracks are eliminated.

The event selection criteria applied in this study, to both data and MC, are as follows:

\begin{enumerate}
    \item Stopping cut: No energy deposition in the last layer of the detector along the $z$-axis,
    \item Maximum energy cut: The maximum energy deposited in an event (in one fiber) must be above 125 p.e.,
    \item Single straight track cut: For the front ($x$-$y$) view, the standard deviation of the distribution of the position of hits in both $x$ and $y$ must be below 1.0 cm, 
    \item Track gaps cut: Allow a maximum of 2 layers with zero deposited energy along the proton's track, and
    \item Containment cut: No energy deposition in the outer layers of the detector along the X and Y axes.
\end{enumerate}

We observed that, among these cuts, the stopping and containment cuts reject the highest fraction of events. Consequently, the low statistics are mainly due to the hard conditions on the outer layers of the detector, especially on the first and last layers in the $y$-axis which constitute 25$\%$ of the eight $y$ layers.

Fig.~\ref{fig:protonevent} shows an example of an event display for a stopping proton event that has passed the event selection cuts. The three views show the proton track and the large energy deposition at the stopping point, as well as the optical crosstalk on each side of the proton path. By comparing the top and side views, it is clear that the amount of horizontal crosstalk is significantly larger than the vertical crosstalk. This is mostly due to the Tyvek reflector layers used between the horizontal planes of cubes in the prototype, which limit the amount of light leaking vertically to other cubes. This agrees with the conclusions from Sec.~\ref{subsec:crosstalk}.

\insertgraph{1.0}{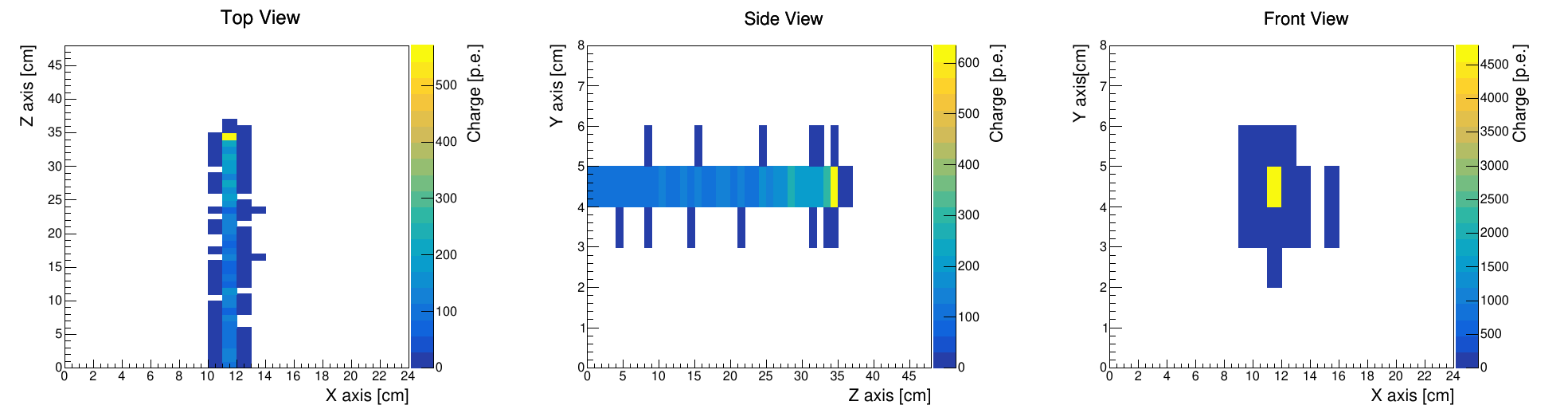}{A stopping proton event from the beam test data [800 MeV/c]. Each plot shows charge readouts from one of the three readout planes.}{fig:protonevent}

\begin{figure}
    \begin{subfigure}{.5\textwidth}
    \centering
    
     \includegraphics[width=.9\linewidth]{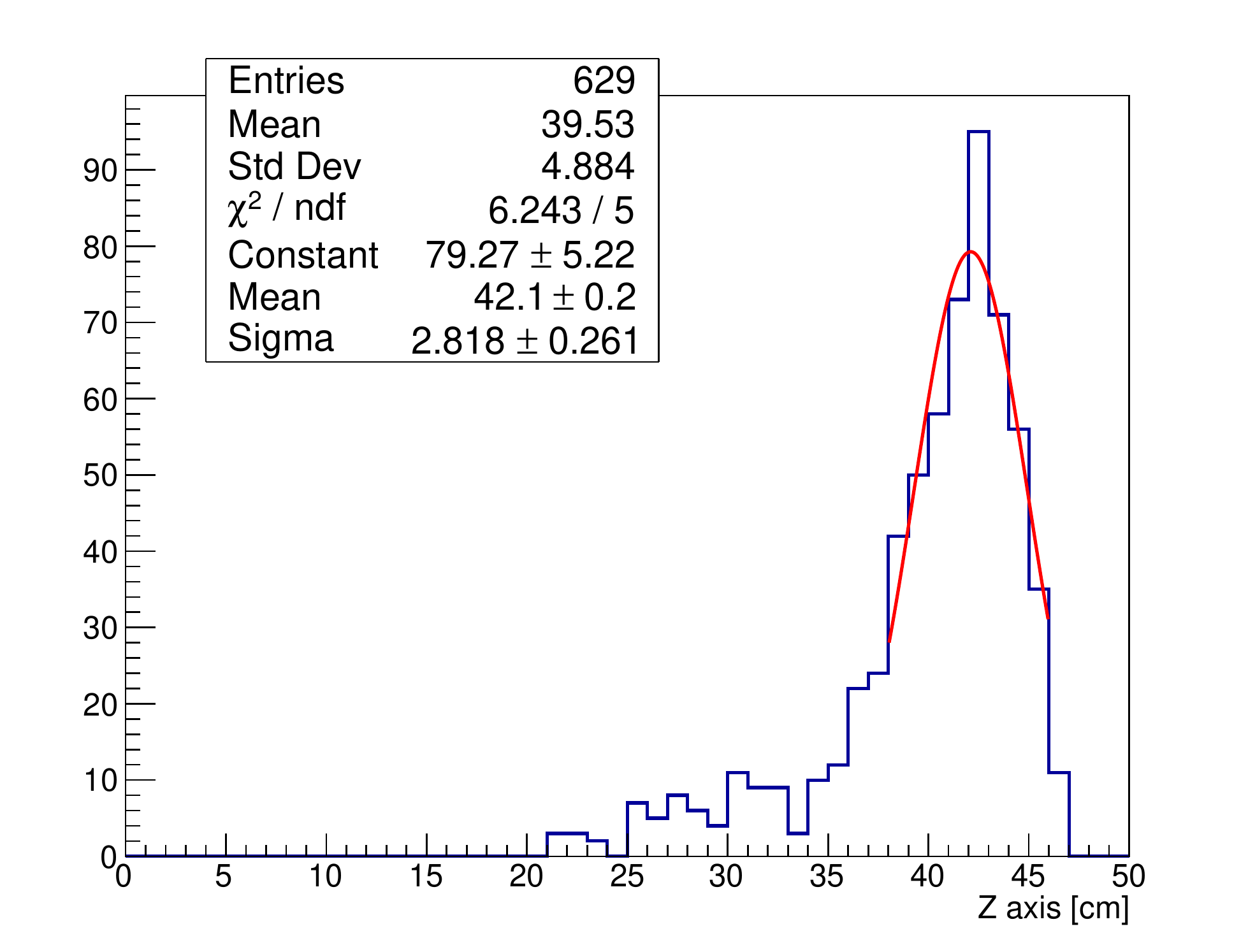}
     \caption{}
    \end{subfigure}%
    \begin{subfigure}{.5\textwidth}
    \centering
     
     \includegraphics[width=.9\linewidth]{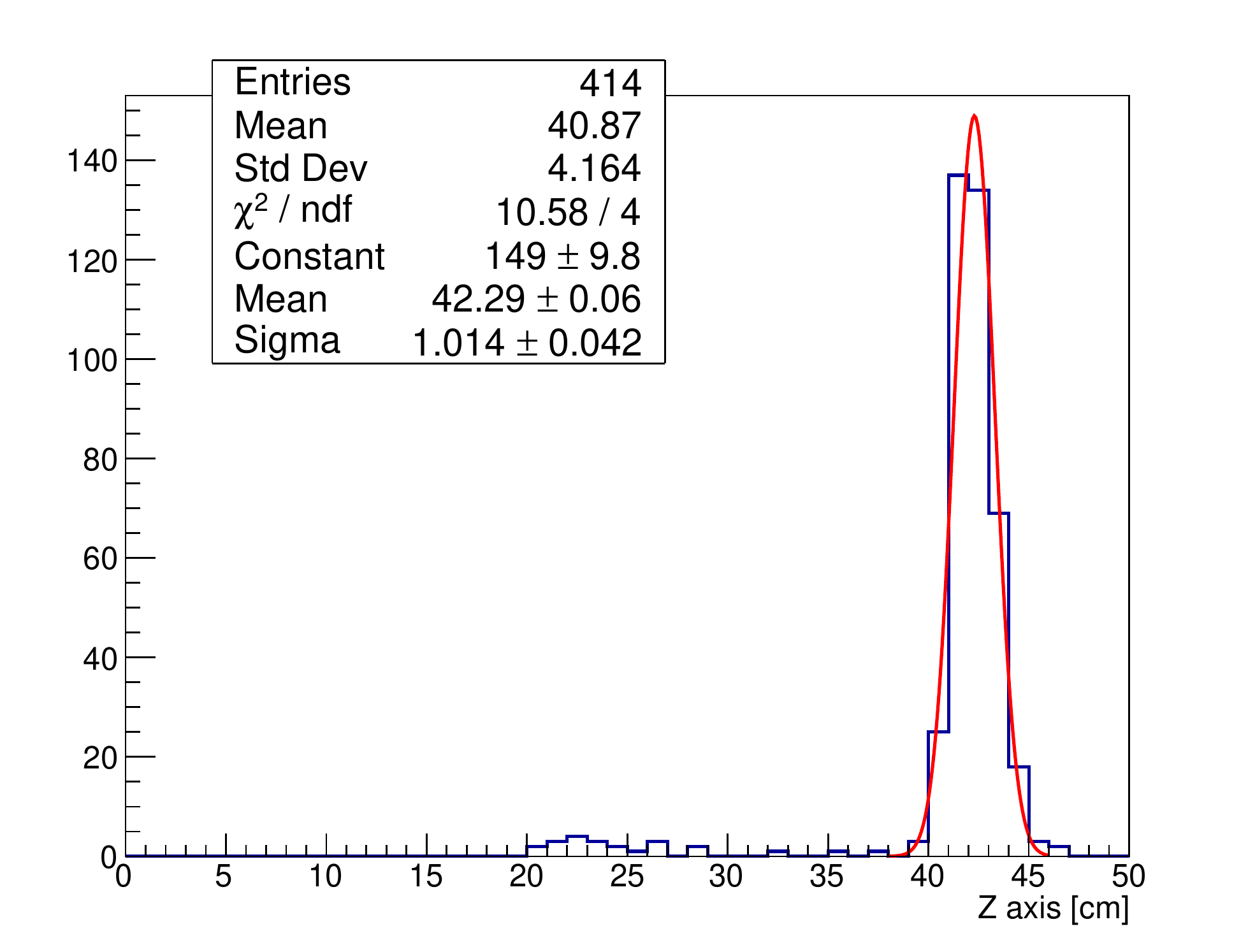}
    \caption{}
    \end{subfigure}
    \caption{The range distribution for the beam test sample [800 MeV/c] (a) and the MC sample [750 MeV/c] (b).}
    \label{fig:Prange}
    \end{figure}

By utilizing the characteristic behavior of protons to deposit the largest amount of energy before they stop, we can locate the stopping point of the proton in each event and use it to plot the range of the sample, the range being the distance the proton has traveled in the detector. Fig.~\ref{fig:Prange} shows the range distributions for the data and MC samples, with mean values of $42.1\pm0.2$~cm and $42.29\pm0.06$~cm respectively.

The wider range distribution in the data can be attributed to the momentum dispersion of the beam which was not simulated in the MC. The dispersion of the T9 beam can be estimated using the results above for the beam test data and the simulation sample, assuming that:
\begin{equation}
    \sigma_\mathrm{Data}^2 = \sigma_\mathrm{Beam}^2 + \sigma_\mathrm{MC}^2, 
\end{equation}
where $\sigma_\mathrm{Data}=2.818$~cm is the variation in proton track length from data and $\sigma_\mathrm{MC}=1.014$~cm is the variation in proton track length from MC (from Fig.~\ref{fig:Prange}). The beam dispersion, $\delta$, is given by:
\begin{equation}
    \delta = \cfrac{\sigma_\mathrm{Beam}}{L} = \cfrac{2.629}{42.1} = 0.062,
\end{equation}
where $L$ is the average range of the protons in the detector. For an 800~MeV/$c$ beam, a 6.2$\%$ dispersion corresponds to a window of $\pm$49.6~MeV/$c$.

The d$E$/d$z$ curve of an event can be constructed using the total energy deposited in each 1~cm layer of the detector along its $z$-axis. Since the vertical fibers in the SuperFGD Prototype are connected to three different types of MPPCs with different gain, the top ($x$-$z$) view is not used in this study. The following d$E$/d$z$ analysis is performed using the side view obtained from the horizontal fibers along the $x$-axis.

\begin{figure}[hbt!]
 \begin{center}
  
  \includegraphics[width=0.6\textwidth]{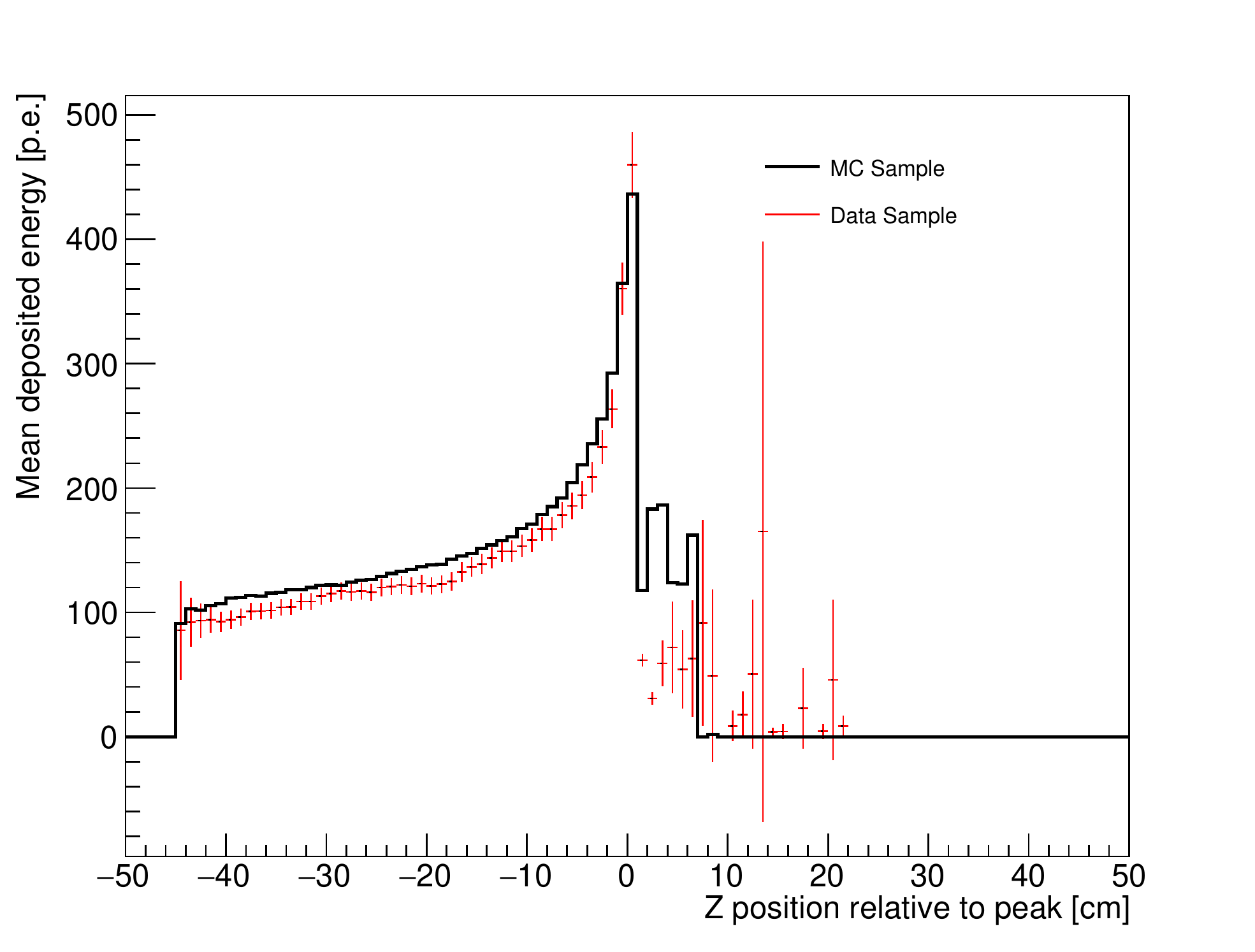}
  \caption{d$E$/d$z$ curves for a 750~MeV/c MC simulation and the 800~MeV/c beam test data. The error bars in the data sample reflect the number of statistics at each point.}
  \label{fig:dedx}
 \end{center}
\end{figure}

The total d$E$/d$z$ curve of the sample is created from the sum of d$E$/d$z$ curves of all selected events by locating the stopping point of the proton and using it as a reference point to construct all points before and after it. Each bin in the curve is then normalized to the number of times it has been filled such that differences in the range of the protons between different events are taken into account. The d$E$/d$z$ curves of the beam test and MC samples are shown in Fig.~\ref{fig:dedx}.

According to these d$E$/d$z$ curves, the data and MC samples show a good agreement up to the stopping point of the proton. Both curves display a plateau region with mean energy deposition of around 100~p.e.\ in the beginning and middle of the proton track. This energy is twice as large as that deposited by a MIP, which is consistent with a proton of 700-800 MeV/c momentum. The plateau region is followed by a large energy deposition of around 450~p.e.\ towards the end of the track where the proton is expected to stop. 

In both cases, the peak is followed by a distribution of non-zero energy deposition mainly caused by interactions at the end of the track that cause scatterings or release secondaries. However, these events are rare as shown by the large statistical fluctuations.


\subsection{Response to Different Particle Types}

To study the particle identification potential of this technology, the detector signals for different track types are compared in this section. The particle trigger information described in Sec.~\ref{subsec:triggers} provided the particle type on an event-by-event basis, with trigger samples including a sample of a mix of muons and pions, and a sample of protons. The light yield for these different samples at a known beam momentum of 0.8~GeV/$c$ is shown in Fig.~\ref{fig:trigg_peaks}. It is computed using MPPC hits from the $y$-$z$ plane only as all these MPPCs are of the same type, Type I. The $x$-fibers are also parallel to the Tyvek sheets, such that $y$-$z$ plane measurements are very similar to the final SuperFGD design.

To measure the average light yield per unit length, the light yield is recorded in each $z$-layer in the $y$-$z$ plane. The range is computed by finding the distance between the centers of the cubes with highest light yield in the first and last $z$-layers of the particle track. We obtain the average d$E$/d$x$ expressed in p.e./cm by dividing the sum of the light yield in all $z$-layers by the range of the track. This is shown in Fig.~\ref{fig:trigg_peaks}.

\insertgraph{0.49}{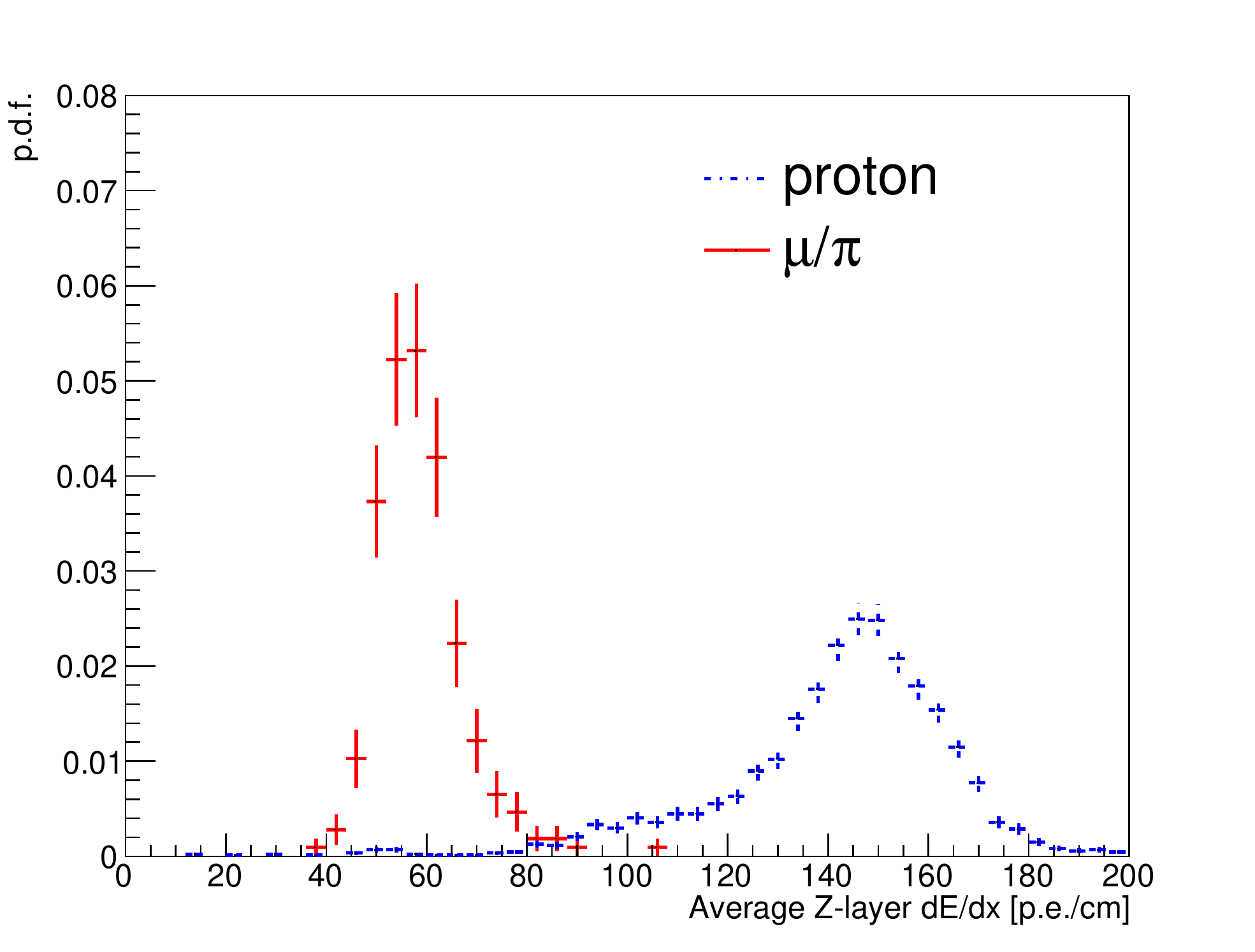}{Average energy deposit per unit length for different particle triggers. Data conditions: 0.2~T, 0.8~GeV/$c$.}{fig:trigg_peaks}

Additional information about the particle track can be extracted by studying the rate at which the ionization changes along its range. This is measured by computing the average d$E$/d$x$ in each $z$-layer as shown in Fig.~\ref{fig:bragg_diff}. For tracks of similar momentum, the detector shows a very clear detector response for different particle types.

\insertgraph{0.49}{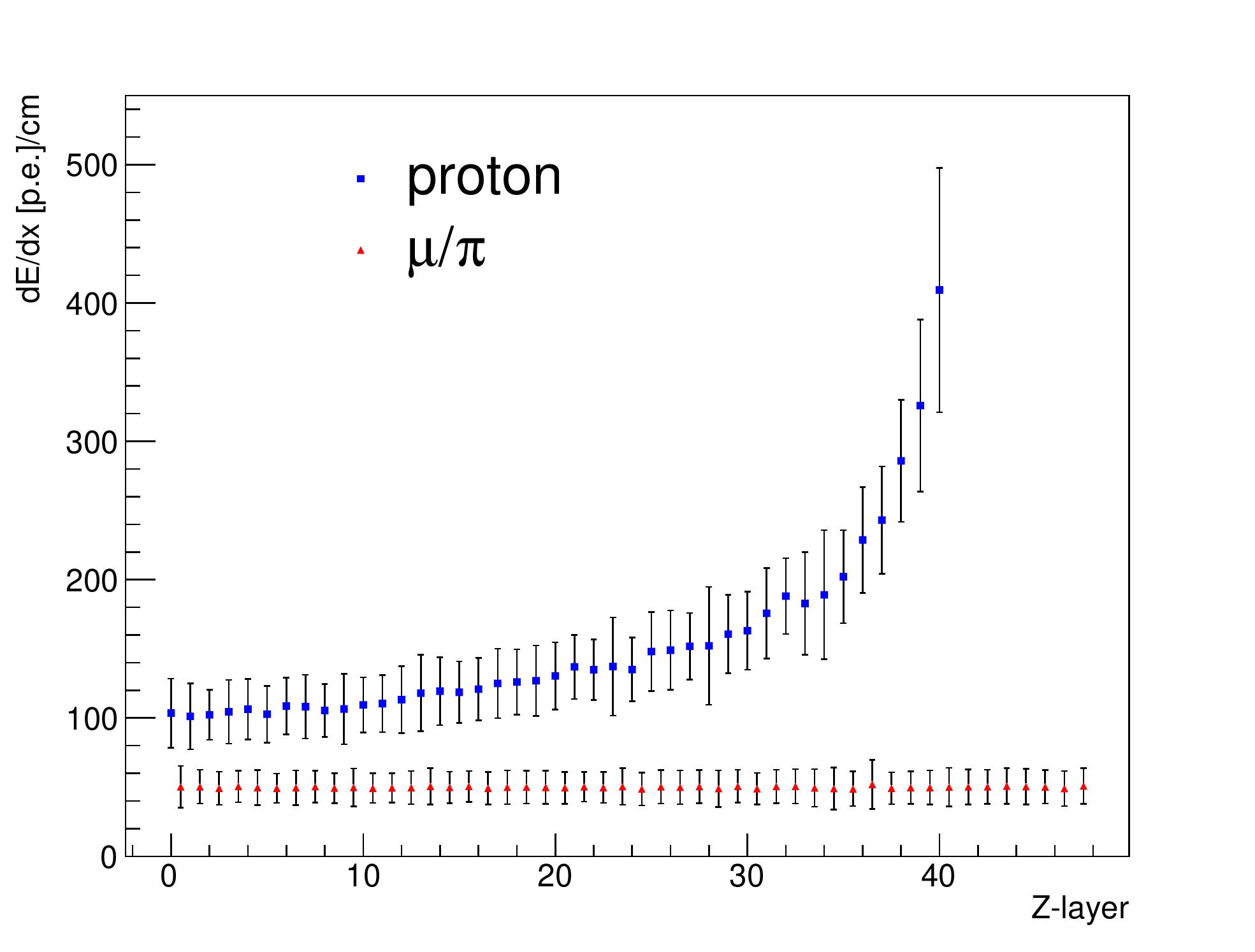}{The d$E$/d$x$ measurement per $z$-layer for different particle triggers. Data conditions: 0.2~T, 0.8~GeV/$c$.}{fig:bragg_diff}

\subsection{dE/dx Resolution}
In the case of muons or pions with a ionization close to a MIP, the d$E$/d$x$ is almost constant. In this way, the d$E$/d$x$ measurement in each layer constitutes a estimator of the true d$E$/d$x$ of the particle. Therefore, combining the measurements of different layers a more accurate d$E$/d$x$ measurement can be achieved. To quantify this improvement, the d$E$/d$x$ resolution is computed using only N of the 48 layers of the prototype. To perform this measurement,  the N layers are chosen at random without repetition. The result is presented in Fig.~\ref{fig:dedx_res}.
\insertgraph{0.49}{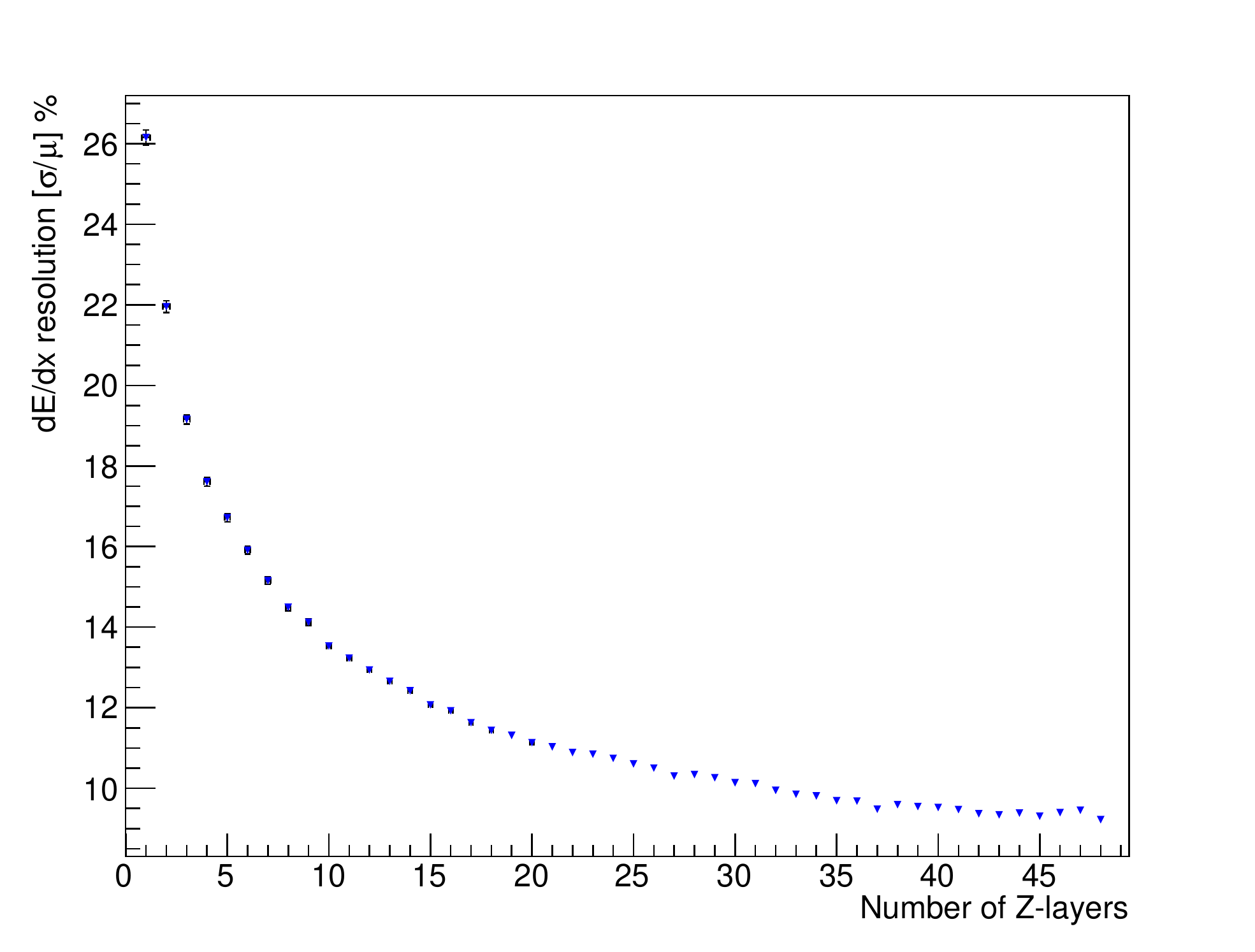}{The d$E$/d$x$ resolution vs number of layers crossed by track (each of them being 1cm-thick), for a quasi-constant d$E$/d$x$. Data conditions: 0.2~T, 0.8~GeV/$c$, $\mu$/$\pi$ trigger).}{fig:dedx_res}
In the context of T2K this performance is relevant. For T2K flux, many outgoing muons and pions have an ionization close to a MIP. Those tracks may travel for some centimeters before escaping the detector such that a precise dE/dx estimate at SuperFGD might help with the particle identification (PID). For context, a ND280 HA-TPC prototype, designed to perform PID using d$E$/d$x$ measurements showed 10-11$\%$ d$E$/d$x$ resolution for 30~cm $\mu/\pi$ tracks of 0.8~GeV/c momentum \cite{attie2019performances}. Therefore, although performing PID is not the primary goal of SuperFGD for this type of tracks, its performance shows that it might be a valuable asset for this task.

\subsection{Electron-Gamma Separation}
An event display of a photon interacting in the SuperFGD Prototype is shown in Fig.~\ref{fig:gamma_conversion}. Such events were extremely rare. The event display illustrates the capacity of this detector for resolving $e^+, e^-$ tracks emerging from the photon interaction vertex, provided both tracks are long enough and diverge, as is the case in this example due to the applied magnetic field.

\insertgraph{0.99}{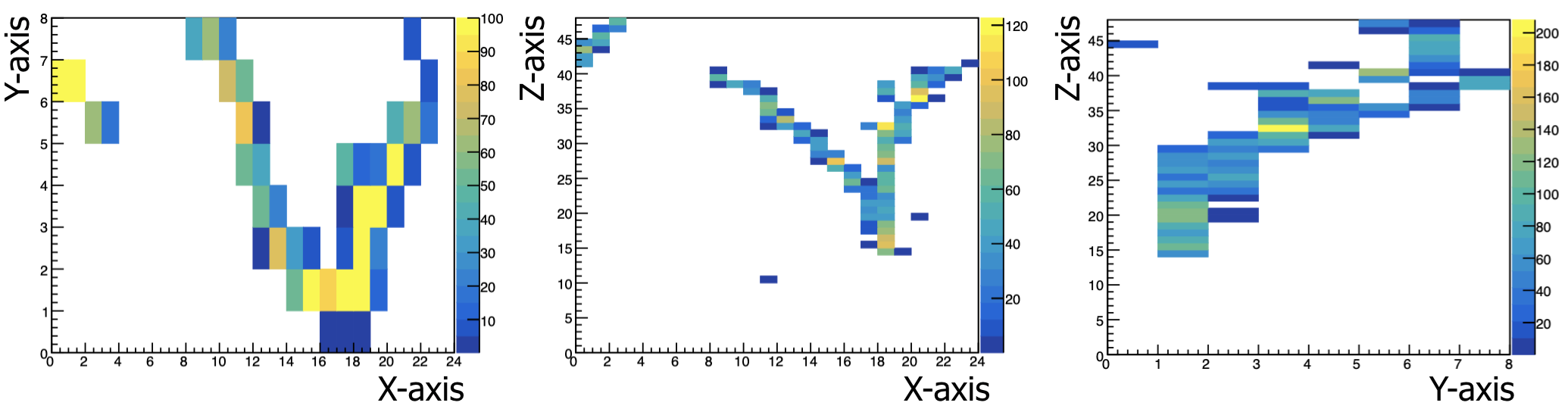}{Event display of a gamma conversion event.}{fig:gamma_conversion}


\section{Conclusion and Outlook}
A campaign of tests at the CERN-PS T9 beamline provided the opportunity for detailed studies of a new detector concept based on 3D readout of scintillator cubes with WLS fibers and photosensors, with a dedicated prototype. The preassembly of the SuperFGD Prototype's 9,216 cubes using the ``fishing line" method was validated, confirming the feasibility of such a scheme for much larger detectors. The detector was instrumented with readout electronics that form the baseline for a new design to be deployed on the full SuperFGD detector at T2K. 

Careful calibration of the SuperFGD Prototype photosensors and readout electronics was undertaken to ensure good hit amplitude resolution for MIPs using the CITIROC HG analogue signal path, and to fully exploit the dynamic range provided by the LG analogue signal path. The use of ToT signals ensured a reliable workaround of the HG and LG signals hit multiplicity limitations and deadtime. The detector operated throughout the beam test campaign in an essentially deadtime-free mode.

The CERN-PS T9 beamline provided charged particles of various types in the momentum range $\pm400$~MeV/$c$ to $\pm8$~GeV/$c$. The large number of recorded events permitted a detailed study of the detector response. The raw light yield measured by the readout channels was 53.7~p.e.\ and 51.0~p.e.\ for events recorded on channels equipped with Type I MPPCs and the 8~cm and 24~cm fibers respectively. Optical crosstalk was studied in detail, using several hundred 8~cm and 24~cm optical fibers. The expected leakage of light from the cube producing scintillation light to adjacent cubes is at the level of 2.94\% per side.

Coincidence matching of hits in two projections led to a determination of the response of individual cubes. Corrections for the attenuation of optical signals in the WLS fibers were applied, leading to similar light yields per MIP from cubes of 58~p.e. ($\sigma=7$ p.e.) and 59~p.e. ($\sigma=8$ p.e.) from horizontal (24~cm long) and vertical (8~cm long) fibers respectively. 

The time resolution of the detector was measured to be 1.14~ns for a single channel. This is inherently limited by the scintillation process and wavelength conversion time constants. Improvements in time resolution for a single event can be obtained by considering several channels on one cube, and several cubes along a track. The recovery time, the time required for a readout channel to be ready to receive a second hit, was estimated to be of order 200~ns, which is sufficient to record a Michel electron event following muon decay in a cube.

With the charged particle beam tuned to 0.8~GeV/$c$, a small sample of protons stopping in the detector bulk was collected. Topologies, light yield, and d$E$/d$x$ curves are reported, showing good agreement with simulations. Saturation effects were indirectly observed, with instances of individual $z$-fibers collecting light from up to 48 cubes for a single event. This confirmed the wide dynamic range of the electronics and photosensor combination, which easily accommodates a proton energy deposition event at the Bragg peak (400~p.e.), a factor 4 below the saturation point.

A detailed study of the energy deposition profiles highlights possibilities for particle identification based on d$E$/d$x$. Muon/pion, and proton hit amplitude samples are clearly distinct for a given momentum. The d$E$/d$x$ resolution for a single cube expressed as $\sigma/\mu$ of 25\% is sufficient to resolve one MIP from two MIPs equivalent events such as two co-linear tracks through the same set of cubes. When the hit amplitude is constant along several layers, as it is the case for MIP-like tracks, the information from different layers may be combined to improve the d$E$/d$x$ resolution up to  10$\%$.

Studies of the SuperFGD Prototype reported here demonstrate its applicability for near detector physics at the T2K ND280 detector complex. Improved angular acceptance, lower detection thresholds, better particle identification capabilities, and clearer resolution around neutrino interaction vertices will contribute to lowering systematic uncertainties in future oscillation analyses at T2K.

\section*{Acknowledgments}
We thank CERN for hosting the beam tests and the CERN Neutrino Platform for contributions to the readout electronics. We acknowledge the support of the MEXT/JSPS KAKENHI Grant No. JP26247034 and JP16H06288, Japan; the National Science Centre (NCN, Grant No. UMO-2018/30/E/ST2/00441 and UMO-2016/21/B/ST2/01092) and Ministry of Science and Higher Education (Grant No. DIR/WK/2017/05), Poland; Ministerio de Economía y Competitividad (SEIDI-MINECO) under Grants No. FPA2016-77347-C2-2-P and No. SEV-2016-0588, Spain; Swiss National Foundation (Grant No. 200021$\_$85012 and 20FL21$\_$186178), Switzerland; the Fondation Ernst et Lucie Schmidheiny, Switzerland; the Science and Technology Facilities Council, UK; RFBR grants $\#$19-32-90100, $\#$19-32-90101, JSPS-RFBR grant $\#$20-52-50010, Russia; and the DOE, USA.

\bibliographystyle{unsrt}
\bibliography{bibliography}

\begin{thebibliography}{10}

\bibitem{Amaudruz:2012esa}
P.A. Amaudruz et~al.
\newblock {The T2K Fine-Grained Detectors}.
\newblock {\em Nucl. Instrum. Meth. A}, 696:1--31, 2012.

\bibitem{Abe:2011ks}
K.~Abe et~al.
\newblock {The T2K Experiment}.
\newblock {\em Nucl. Instrum. Meth.}, A659:106--135, 2011.

\bibitem{Abe:Nature}
K.~Abe et~al.
\newblock {Constraint on the Matter–Antimatter Symmetry-Violating Phase in
  Neutrino Oscillations}.
\newblock {\em Nature}, 580:339--344, 2020.

\bibitem{Michael:2008bc}
D.G. Michael et~al.
\newblock {The Magnetized steel and scintillator calorimeters of the MINOS
  experiment}.
\newblock {\em Nucl. Instrum. Meth. A}, 596:190--228, 2008.

\bibitem{Aliaga:2013uqz}
L.~Aliaga et~al.
\newblock {Design, Calibration, and Performance of the MINERvA Detector}.
\newblock {\em Nucl. Instrum. Meth. A}, 743:130--159, 2014.

\bibitem{Abe:2019whr}
K.~Abe et~al.
\newblock {T2K ND280 Upgrade - Technical Design Report}, 2019.
\newblock arXiv:1901.03750.

\bibitem{Friend:2017oav}
M.~Friend.
\newblock {J-PARC Accelerator and Neutrino Beamline Upgrade programme}.
\newblock {\em J. Phys. Conf. Ser.}, 888(1):012042, 2017.

\bibitem{Abe:2019fux}
K.~Abe et~al.
\newblock {J-PARC Neutrino Beamline Upgrade Technical Design Report}, 2019.
\newblock arXiv:1908.05141.

\bibitem{Sgalaberna:2017khy}
A.~Blondel et~al.
\newblock {A Fully Active Fine Grained Detector with Three Readout Views}.
\newblock {\em JINST}, 13(02):P02006, 2018.

\bibitem{attie2019performances}
D.~Atti{\'e} et~al.
\newblock {Performances of a resistive MicroMegas module for the Time
  Projection Chambers of the T2K Near Detector upgrade}.
\newblock {\em Nucl. Instrum. Meth.}, A957, 2019.

\bibitem{Munteanu:2019llq}
L.~Munteanu, S.~Suvorov, S.~Dolan, D.~Sgalaberna, S.~Bolognesi, S.~Manly,
  G.~Yang, C.~Giganti, K.~Iwamoto, and C.~Jesús-Valls.
\newblock {A new method for an improved anti-neutrino energy reconstruction
  with charged-current interactions in next-generation detectors}.
\newblock {\em Phys. Rev. D}, 101(9):092003, 2020.

\bibitem{Mineev:nim}
O.~Mineev et~al.
\newblock {Beam Test Results of 3D Fine-grained Scintillator Detector Prototype
  for a T2K ND280 Neutrino Active Target}.
\newblock {\em Nucl. Instrum. Meth.}, A923:134--138, 2019.

\bibitem{Kudenko:2001qj}
{\relax Yu}.~G. Kudenko, L.~S. Littenberg, V.~A. Mayatsky, O.~V. Mineev, and
  N.~V. Ershov.
\newblock {Extruded plastic counters with WLS fiber readout}.
\newblock {\em Nucl. Instrum. Meth.}, A469:340--346, 2001.

\bibitem{Kuraray}
{Kuraray Wavelength Shifting Fibers}.
\newblock \url{http://kuraraypsf.jp/psf/ws.html}.

\bibitem{Mineev:jinst}
O.~Mineev, {\relax Yu}.~Kudenko, {\relax Yu}.~Musienko, I.~Polyansky, and
  N.~Yershov.
\newblock {Scintillator detectors with long WLS fibers and multi-pixel
  photodiodes}.
\newblock {\em JINST}, 6:P12004, 2011.

\bibitem{Noah:2017ati}
M.~Antonova et~al.
\newblock {The Baby MIND spectrometer for the J-PARC T59(WAGASCI) experiment}.
\newblock {\em PoS}, EPS-HEP2017:508, 2017.

\bibitem{Noah:2266598}
Etam Noah et~al.
\newblock {Readout Scheme for the Baby-MIND Detector}.
\newblock {\em PoS}, PhotoDet2015:031. 12 p, 2016.

\bibitem{Basille:2019mcp}
O.~Basille et~al.
\newblock {Baby MIND Readout Electronics Architecture for Accelerator Neutrino
  Particle Physics Detectors Employing Silicon Photomultipliers}.
\newblock {\em JPS Conf. Proc.}, 27:011011, 2019.

\bibitem{7753500}
G.~M. {Mitev}, L.~T. {Tsankov}, M.~G. {Mitev}, and E.~N. {Messomo}.
\newblock Light pulse generator for multi-element scintillation detectors
  testing.
\newblock In {\em 2016 XXV International Scientific Conference Electronics
  (ET)}, pages 1--4, 2016.

\bibitem{AGOSTINELLI2003250}
S.~Agostinelli et~al.
\newblock Geant4--a simulation toolkit.
\newblock {\em Nucl. Instrum. Meth.}, 506(3):250 -- 303, 2003.

\end{thebibliography}

\end{document}